\documentclass[12pt]{article}
\pdfoutput=1
\usepackage{jheppub}
\usepackage{ytableau}
\usepackage{comment}
\usepackage[vcentermath]{youngtab}
\usepackage{tikz}

\newcommand\be{\begin{equation}}
\newcommand\ee{\end{equation}}

\newcommand\Tr{\mathrm{Tr}}

\newcommand{\q}{\mathsf{q}}
\renewcommand{\t}{\mathsf{t}}

\preprint{
RUP-25-13
}

\title{S-duality of boundary lines in $\mathcal{N}=4$ SYM theories and supersymmetric indices}
\abstract{
We analyze the supersymmetric defect indices of $\mathcal{N}=4$ super Yang Mills theories 
which are simultaneously decorated by the BPS line operators and the boundary conditions. 
We demonstrate that the two-point functions of the boundary 't Hooft lines of magnetic charges associated with the minuscule representations 
in the presence of the regular Nahm pole boundary conditions can be obtained 
by applying the Higgsing prescription to the half-indices of the Dirichlet boundary conditions. 
Accordingly, we find precise matching of the indices for pairs of the S-dual configurations 
with the Wilson lines and Neumann boundary conditions and those with the 't Hooft lines and the regular Nahm pole boundary conditions. 
Alternatively, we analytically compute the indices by means of the inner product of the Macdonald polynomials to find the exact closed-form expressions. 
}
\author[a]{Yasuyuki Hatsuda}
\author[b]{and Tadashi Okazaki}

\emailAdd{yhatsuda@rikkyo.ac.jp, tokazaki@seu.edu.cn}
\affiliation[a]{Department of Physics, Rikkyo University, Toshima, Tokyo 171-8501, Japan}
\affiliation[b]{
Shing-Tung Yau Center of Southeast University,\\
Yifu Architecture Building, No.2 Sipailou, Xuanwu district, Nanjing, Jiangsu, 210096, China}
\begin{document}
\maketitle

\section{Introduction and summary}
\label{sec_intro}

In this paper we study the BPS boundary line operators, that is the BPS line operators living on the boundary of 4d $\mathcal{N}=4$ super Yang-Mills (SYM) theories. 
We analyze the line defect half-indices 
that decorate the half-indices \cite{Dimofte:2011py,Gang:2012yr} counting the BPS local operators supported at certain junctions of the line operators and boundaries. 
We focus on the one-sided case with gauge theories living in a half-space obeying the half-BPS boundary conditions \cite{Gaiotto:2008sa}. 
When the gauge field satisfies the Neumann boundary condition, the BPS Wilson lines \cite{Wilson:1974sk,Maldacena:1998im,Rey:1998ik} can be introduced at a boundary. 
When the Dirichlet boundary condition is imposed on the gauge field, 
a singular configuration of scalar fields in the adjoint representation is allowed as a solution to Nahm's equation, 
which is referred to as the Nahm pole boundary condition \cite{Nahm:1979yw,Diaconescu:1996rk,Gaiotto:2008sa}. 
The singular configuration of the Nahm pole boundary condition can be generalized in the presence of the 't Hooft line operator \cite{tHooft:1977nqb,Kapustin:2005py,Kapustin:2006pk,Gaiotto:2011nm,Witten:2011zz}. 
The Neumann boundary condition in $\mathcal{N}=4$ SYM theory of gauge group $G$ is conjectured to be dual to 
the regular Nahm pole boundary condition in $\mathcal{N}=4$ SYM theory of the Langlands dual gauge group $G^{\vee}$ \cite{Gaiotto:2008ak}. 
Besides, the Wilson line in $\mathcal{N}=4$ SYM of $G$ is dual to the 't Hooft line  
in $\mathcal{N}=4$ SYM of $G^{\vee}$ \cite{Kapustin:2005py,Gomis:2009ir,Aharony:2013hda}. 
Therefore, it is expected that 
the boundary Wilson line with the Neumann boundary condition in $\mathcal{N}=4$ SYM of $G$
is dual to the boundary 't Hooft line with the regular Nahm pole boundary condition in $\mathcal{N}=4$ SYM of $G^{\vee}$. 

In spite of the fact that 
the line defect half-indices of the Neumann boundary condition together with the boundary Wilson lines 
can be evaluated by integral representations \cite{Dimofte:2011py,Gang:2012yr}, 
little is known about the exact formulas for these line defect half-indices and dual descriptions in the literature. 
We show that the exact closed-form expressions for the line defect half-indices can be obtained by employing the Higgsing procedure proposed in \cite{Gaiotto:2019jvo}. 
\footnote{Also see \cite{Gaiotto:2012xa} for the Higgsing manipulation of the supersymmetric indices.} 
The key idea is that the Nahm pole boundary conditions with the 't Hooft lines 
can be viewed as a certain deformed Dirichlet boundary conditions  
upon turning on a non-trivial nilpotent and holomorphic vevs for the scalar fields \cite{Gaiotto:2011nm,Witten:2011zz}. 
We obtain from the Dirichlet half-indices \cite{Gaiotto:2019jvo,Okazaki:2019ony} the exact closed-form expressions for the two-point functions of the boundary 't Hooft lines 
of magnetic charges corresponding to the minuscule representations \cite{MR2109105} in the presence of the regular Nahm pole boundary conditions. 
Accordingly, we find strong evidence of S-duality of the configurations of 4d $\mathcal{N}=4$ SYM theories in the presence of both line operators and boundaries 
as precise matching of the two-point functions of the boundary Wilson lines with the Neumann boundary conditions  
and the two-point functions of the boundary 't Hooft lines with the regular Nahm pole boundary conditions. 
Our results generalize the identities of the half-indices \cite{Gaiotto:2019jvo,Hatsuda:2024lcc}, 
which demonstrate S-duality of the Neumann boundary conditions and the Nahm pole boundary conditions \cite{Gaiotto:2008ak} 
and those of the line defect indices \cite{Gang:2012yr,Hatsuda:2025jze} without boundary conditions, 
which result from S-duality of the Wilson lines and the 't Hooft lines \cite{Kapustin:2005py,Aharony:2013hda}. 

Moreover, we observe that 
the integral representations for the line defect half-indices of the Neumann boundary condition together with the boundary Wilson lines 
can be exactly computed using the inner product of the Macdonald polynomials \cite{MR1354144,MR1976581,MR1314036,MR1354956} 
that appears in a framework of the double affine Hecke algebra (DAHA) \cite{MR1185831,MR1314036,MR1358032,MR1354956,MR1613515,MR1768938,MR1715325,MR1792347,MR1411136}. 
Such calculations reproduce the results obtained from the Higgsing prescription and also allow us to find the exact closed-form formulas for the line defect half-indices 
even for the case where the dual 't Hooft lines are associated with the non-minuscule modules. 

\subsection{Structure}
The organization of the paper is as follows. 
In section \ref{sec_LineHindex} we review the configurations in $\mathcal{N}=4$ SYM theories with line operators and boundaries. 
We introduce the line defect half-indices, the supersymmetric indices which are decorated by the line operators and the boundaries for 4d $\mathcal{N}=4$ SYM theories. 
In section \ref{sec_uN} we analyze the line defect half-indices for $U(N)$ SYM theories. 
The two-point function of the boundary Wilson lines in the rank-$k$ antisymmetric representation 
is shown to agree with that of the dual boundary 't Hooft lines. 
In section \ref{sec_so2N+1} the line defect half-indices of the Wilson lines for SYM theories based on the gauge algebra $\mathfrak{so}(2N+1)$ 
and those of the dual 't Hooft lines for the theories with the gauge algebra $\mathfrak{usp}(2N)$ are examined. 
In section \ref{sec_usp2N} the opposite types of the line defect half-indices in these theories are investigated. 
In section \ref{sec_so2N} we present the line defect half-indices in SYM theories with the gauge algebra $\mathfrak{so}(2N)$. 
In Appendix \ref{app_ch} we list the character formulas of the irreducible representations for the classical Lie algebras. 

\subsection{Future works}
\begin{itemize}

\item It would be interesting to extend our results to the cases with more general boundary conditions. 
In particular, more general dualities of boundary conditions, including the Dirichlet b.c., enriched Neumann b.c., and interfaces are confirmed in \cite{Gaiotto:2019jvo,Okazaki:2019ony,Hatsuda:2024lcc} as precise matching pairs of half-indices. 
It is tempting to include the line operators in the latter setup to analyze the line defect half-indices. 
We hope to report our results in the upcoming works. 

\item Extension of our results to the cases with more general line operators should be also interesting. 
For the 't Hooft lines of magnetic charges associated with the non-minuscule representations, 
one encounters the monopole bubbling effects \cite{Kapustin:2006pk}. 
It would be intriguing to analyze the line defect half-indices for such 't Hooft lines 
by taking into account the monopole bubbling index \cite{Ito:2011ea} (also see \cite{Brennan:2018yuj,Brennan:2018rcn,Hayashi:2019rpw,Hayashi:2020ofu}). 

\item According to the brane construction in Type IIB string theory, 
the coexistence of the line operators and the boundary conditions in $\mathcal{N}=4$ SYM theories can have the gravity dual as discussed in 
\cite{Nagasaki:2011ue,Estes:2012nx,Nagasaki:2013hwa,deLeeuw:2016vgp,Aguilera-Damia:2016bqv,Coccia:2021lpp,Karch:2022rvr,Bergman:2022otk}. 
Hence the line defect half-indices are expected to encode the spectra of the gravity dual geometries. 
In the absence of the boundary, the large $N$ limits of the line defect indices of Wilson lines in various representations for SYM theories  
are examined in \cite{Gang:2012yr,Drukker:2015spa,Hatsuda:2023iwi,Hatsuda:2023imp,Hatsuda:2023iof,Hatsuda:2025jze}. 
Besides, they admit non-trivial giant graviton expansions \cite{Imamura:2024lkw,Imamura:2024pgp,Beccaria:2024oif,Beccaria:2024dxi,Hatsuda:2024uwt,Imamura:2024zvw,Beccaria:2024lbt}. 
We hope to report the detailed analysis of the large $N$ limits and the giant graviton expansions of the line defect half-indices. 

\item Alternative interpretation of the matrix integrals of the Wilson line defect half-indices associated with the Macdonald polynomials 
appear in the study of the BPS defect operators in 3d gauge theories \cite{Okazaki:2023kpq}. 
It would be interesting to investigate the 3d gauge theoretical interpretation.  

\end{itemize}

\section{Line defect half-indices}
\label{sec_LineHindex}

\subsection{Neumann and Nahm pole b.c.}
The half-BPS boundary conditions in 4d $\mathcal{N}=4$ SYM theories are classified by Gaiotto and Witten \cite{Gaiotto:2008sa}. 
They preserve 3d $\mathcal{N}=4$ supersymmetry and $SU(2)_C\times SU(2)_H$ R-symmetry. 
Consequently, six adjoint scalar fields in 4d $\mathcal{N}=4$ SYM theorie split into two scalar fields $\vec{X}$ and $\vec{Y}$  
acted on by $SU(2)_C$ $\times$ $SU(2)_H$ as $(\mathbf{1},\mathbf{3})$ and $(\mathbf{3},\mathbf{1})$. 
Suppose that the theories are supported along the directions $x^0$, $x^1$, $x^2$ and $x^6$ and that there is a boundary at $x^6=0$. 
The half-BPS boundary conditions at $x^6=0$ can be obtained by finding the solutions to the vanishing of normal component of supercurrent. 
The two basic ones are
\begin{align}
\label{NeuBC}
&
\textrm{Neumann b.c. $\mathcal{N}$: }\qquad 
F_{6\mu}|_{\partial}=0, \quad D_{\mu}\vec{X}|_{\partial}=0, \quad D_{6}\vec{Y}|_{\partial}=0, \\
\label{NahmBC}
&
\textrm{Nahm pole b.c. $\textrm{Nahm}$: }\qquad 
F_{\mu\nu}|_{\partial}=0, \quad D_{6}\vec{X}+\vec{X}\times \vec{X}|_{\partial}=0, \quad D_{\mu}\vec{Y}|_{\partial}=0,  
\end{align}
where $\mu,\nu$ $=$ $0,1,2$. 
The first equations (\ref{NeuBC}) define the Neumann boundary conditions for which gauge group $G$ is completely preserved at the boundary 
as the gauge field obeys the Neumann boundary condition. 
They set the scalar fields $\vec{X}$ to zero or some fixed values, 
whereas the scalar fields $\vec{Y}$ can fluctuate at the boundary. 
The second equations (\ref{NahmBC}) are identified with the Nahm pole boundary conditions. 
As the gauge field is subject to the Dirichlet boundary condition, gauge group $G$ is broken at the boundary. 
The scalar fields $\vec{X}$ obeys Nahm's equation. 
The solutions involve the Neumann boundary condition, for which the scalar fields $\vec{X}$ can fluctuate, 
however, more generally they also admit singular profile at $x^6=0$
\begin{align}
\label{NahmPole}
\vec{X}(x^6)&=\frac{\vec{\mathfrak{t}}}{x^6}, 
\end{align}
where $\vec{\mathfrak{t}}$ $=$ $(\mathfrak{t}_1,\mathfrak{t}_2,\mathfrak{t}_3)$ is a triplet element 
of the Lie algebra $\mathfrak{g}$ obeying the commutation relation of the Lie algebra $\mathfrak{su}(2)$. 
The choice of $\vec{\mathfrak{t}}$ is labeled by a homomorphism of Lie algebras 
\begin{align}
\label{hom_su2}
\rho:\quad \mathfrak{su}(2) \rightarrow \mathfrak{g}. 
\end{align}
The homomorphism describes how the fundamental representation $V$ of $\mathfrak{g}$ 
decomposes as a direct sum of the $n_{i}$ $=$ $2j_{i}+1$ dimensional irreducible spin-$j_i$ representations $\mathcal{R}_{j_i}$ of $\mathfrak{su}(2)$ 
\begin{align}
V=\bigoplus_{i=1}^{s} \mathcal{R}_{j_i}. 
\end{align} 
For $\mathfrak{g}=\mathsf{u}(N)$, possible homomorphisms (\ref{hom_su2}) correspond to partitions of $N$ 
\begin{align}
N=n_1+\cdots+n_{s}
\end{align}
or Young diagrams $\lambda$ with $N$ boxes and $s$ rows. 
For $\mathfrak{g}=\mathfrak{so}(N)$, they are described by partitions of $N$ with even multiplicity of even $n_i$ due to the reality condition of the fundamental representation. 
For $\mathfrak{g}=\mathfrak{usp}(2N)$, they are given by partitions of $N$ with even multiplicity of odd $n_i$ as the fundamental representation is pseudo-real. 
As the boundary condition preserves 3d $\mathcal{N}=4$ supersymmetry, 
the moduli space $\mathcal{M}$ of solutions of Nahm pole boundary conditions at $x^6=0$ and the value of $\vec{X}$ at $x^6=\infty$ is hyperK\"{a}hler. 
So it is useful to describe $\mathcal{M}$ as complex manifold by taking the complexified fields 
\begin{align}
\mathcal{X}&=X_1+iX_2,& 
\mathcal{A}&=A_6+iX_3. 
\end{align}
Then Nahm's equation in (\ref{NahmBC}) is given by a single holomorphic equation \cite{Gaiotto:2008sa}
\begin{align}
\frac{\mathcal{D}\mathcal{X}}{\mathcal{D}x^6}&=0, 
\end{align}
where $\frac{\mathcal{D}\mathcal{X}}{dx^6}$ $=$ $\frac{d}{dx^6}+[\mathcal{A},\mathcal{X}]$. 
The solution (\ref{NahmPole}) takes the form 
\begin{align}
\mathcal{X}&=\frac{\mathfrak{t}_1+i\mathfrak{t}_2}{x^6},& 
\mathcal{A}&=\frac{i \mathfrak{t}_3}{x^6}, 
\end{align}
where $1/2(\mathfrak{t}_1+i\mathfrak{t}_2)$ $=$ $e$ and $\mathfrak{t}_3$ $=$ $h$ are the raising operator  
and the diagonal operator in the $\mathfrak{sl}(2)$ Lie algebra 
so that the complex scalar field $\mathcal{X}$ acquires the nilpotent vev. 
According to the Jacobson-Morozov theorem \cite{MR49882,MR125180,MR114875}, 
any nilpotent element $e$ in a complex semi-simple Lie algebra 
$\mathfrak{g}_{\mathbb{C}}$ $=$ $\mathfrak{g}\times_{\mathbb{R}}\mathbb{C}$ can be extended to $\mathfrak{sl}(2)$-triple $\{e,f,h\}$ 
so that there is a homomorphism of the Lie algebra $\mathfrak{sl}(2)$ $\rightarrow$ $\mathfrak{g}_{\mathbb{C}}$, corresponding to (\ref{hom_su2}). 
Therefore a choice of the homomorphism $\rho$ given by (\ref{hom_su2}) is equivalent to a choice of nilpotent conjugacy class in $\mathfrak{g}_{\mathbb{C}}$. 

The basic case is a principal embedding of $\mathfrak{su}(2)$, 
which corresponds to the condition that $\mathcal{X}$ is a regular \footnote{An element of Lie algebra is called regular if the commutant $C$, 
the subalgebra that commutes with it, has the minimum possible dimension. } nilpotent element of $\mathfrak{g}_{\mathbb{C}}$. 
For $\mathfrak{g}_{\mathbb{C}}$ $=$ $\mathfrak{gl}(N)$, 
a nilpotent element is simply an $N\times N$ nilpotent matrix. 
According to the Jordan decomposition theorem, it can be conjugate to a Jordan canonical form
\begin{align}
\mathcal{X}&\sim 
\left(
\begin{matrix}
0&*&\quad &\quad &\quad \\
&0&*&\ &\  \\
&&0&*&\  \\
&&&0&*\\
&&&&0\\
\end{matrix}
\right). 
\end{align}
Here $\sim$ indicates that it can be written as the matrix form up to regular terms and $x^6$ dependence. 
The elements labeled by $*$ are all $1$ or $0$. 
They determine the conjugacy class of nilpotent elements of $\mathfrak{sl}(2)$ up to permutations of blocks. 
The basic case is realized by a principle embedding $\rho$, which can be labeled by a partition with a single row 
involving the $\mathfrak{su}(2)$-module of maximal dimension. 
It corresponds to a single Jordan block with all elements $*$ are equal to $1$. 
It can be identified with the regular nilpotent conjugacy class of maximal dimension. 
Therefore the corresponding boundary condition is called the regular Nahm pole boundary condition. 
Another extreme case is an embedding $\rho$, which can be labeled by a partition with a single column 
involving the trivial representation of the $\mathfrak{su}(2)$. 
It is obtained by setting all elements $*$ to zero. 
Such a trivial Nahm pole boundary condition is identified with the Dirichlet boundary condition. 
With the Dirichlet boundary condition, there exists the global symmetry 
as the commutant $C$ of $\rho(\mathfrak{su}(2))$ in $G$. 
It is argued \cite{Gaiotto:2008ak} that 
under the action of S-duality \cite{Montonen:1977sn}, 
the Neumann b.c. (\ref{NeuBC}) of 4d $\mathcal{N}=4$ SYM theory of gauge group $G$ 
maps to the regular Nahm pole b.c. (\ref{NahmBC}) of 4d $\mathcal{N}=4$ SYM theory of the Langlands dual gauge group $G^{\vee}$. 

\subsection{Boundary Wilson and 't Hooft lines}
Here we further consider the \textit{boundary line operators} which are localized at the boundary $x^6=0$. 
In the presence of the half-BPS Neumann b.c. (\ref{NeuBC}), 
we can introduce an electric Wilson line characterized by the representation $\mathcal{R}$ of gauge group $G$ preserved at the boundary. 
As in the construction of the supersymmetric Wilson lines in the bulk \cite{Maldacena:1998im,Rey:1998ik}, 
in order to preserve supersymmetry, we need to replace the connection in the holonomy matrix with 
a modified connection comprising an ordinary connection and scalar fields as it can be annihilated by supercharges. 
Consider a boundary Wilson line of the form 
\begin{align}
\label{bdy_W}
W_{\mathcal{R}}&={\Tr}_{\mathcal{R}} P \exp\left[
\oint i (A_{\mu}\dot{x}^{\mu}+\sqrt{-\dot{x}^2}Y) ds
\right], 
\end{align}
where $Y:=Y^3$ is the scalar filed inside the triplet $\vec{Y}$ obeying the Neumann boundary condition in (\ref{NeuBC}) 
and $x^{\mu}$, $\mu=0,1,2$ parametrize the 3d boundary. 
While it preserves the $SU(2)_H$ R-symmetry, it manifestly breaks the $SU(2)_C$ R-symmetry down to $U(1)_C$. 
It can preserve 1d $\mathcal{N}=4$ supersymmetry, i.e. four supercharges. 

When one has the Nahm pole b.c. (\ref{NahmBC}), the gauge group $G$ is broken. 
As $\vec{Y}$ obeys the Dirichlet boundary condition, one cannot define the modified connection 
to be used to construct supersymmetric electric Wilson line as in (\ref{bdy_W}). 
Instead, it admits non-trivial 't Hooft line operator \cite{tHooft:1977nqb,Kapustin:2005py,Gaiotto:2011nm,Witten:2011zz}. 
The 't Hooft line operator is associated with the Dirac monopole singularity 
and it carries magnetic charge $B$ as an element of the cocharacter lattice of gauge group $G$
or equivalently a weight of $\mathfrak{g}^{\vee}$, the Langlands dual of $\mathfrak{g}=\mathrm{Lie}(G)$, 
corresponding to an embedding of $U(1)$ into gauge group $G$. 
The supersymmetric 't Hooft line in the bulk can be constructed by requiring the scalar field $X:=X^3$ inside the triplet $\vec{X}$ 
to have a certain type of singularities along the line. 
When we introduce the 't Hooft line at $z=x_1+ix_2=0$ in the presence of the Nahm pole b.c. (\ref{NahmBC}), 
the nilpotent matrix is deformed in such a way that 
the configuration of $\mathcal{X}$ $=$ $X^1+iX^2$ is holomorphic in $z$ $=$ $x_1+ix_2$ and nilpotent \cite{Gaiotto:2011nm,Witten:2011zz}. 
For example, for $\mathfrak{g}=\mathfrak{su}(2)$, 
any nilpotent $\mathcal{X}(z)$ is conjugate to 
\begin{align}
\mathcal{X}(z)
&\sim \left(
\begin{matrix}
0&z^m\\
0&0\\
\end{matrix}
\right)
\end{align}
up to the residual gauge transformation. 
More generally, for $\mathfrak{sl}(N,\mathbb{C})$ 
a solution is given by 
\begin{align}
\mathcal{X}(z)&\sim 
\left(
\begin{matrix}
0&z^{m_1}&&&\textrm{}\\
&0&z^{m_2}&& \\
&&0&z^{m_{N-2}}& \\
&&&0&z^{m_{N-1}}\\
&&&&0\\
\end{matrix}
\right), 
\end{align}
where $m_i$ are non-negative integers which encode the magnetic charge carried by the 't Hooft line. 
The boundary 't Hooft line can preserve 1d $\mathcal{N}=4$ supersymmetry and the $SU(2)_C$ R-symmetry, 
whereas it breaks down the $SU(2)_H$ R-symmetry. 
Similarly, for other gauge groups, the nilpotent matrices need to be deformed due to the 't Hooft lines appropriately. 
For example, the 't Hooft lines in $USp(2N)/\mathbb{Z}_2$ and $SO(2N)/\mathbb{Z}_2$ gauge theories may carry the fractional magnetic charges (see (\ref{Table_duallines})). 

It is conjectured that S-duality of $\mathcal{N}=4$ SYM theory maps the Wilson line of certain representation $\mathcal{R}$ of $G$ 
to the 't Hooft line of certain magnetic charge $B$ for the Langlands dual $G^{\vee}$ \cite{Kapustin:2005py,Aharony:2013hda}. 
Accordingly, the following dualities of boundary line operators in $\mathcal{N}=4$ SYM theories are conjectured: 
\begin{align}
\label{dual_bdyline}
&
\textrm{boundary Wilson line of $\mathcal{R}$ with Neumann b.c. of $G$ SYM theory}
\nonumber\\
&\leftrightarrow 
\textrm{boundary 't Hooft line of $B$ with Nahm pole b.c. of $G^{\vee}$ SYM theory}. 
\end{align}
The complexified coupling constant $\tau$ $=$ $\frac{\theta}{2\pi}+\frac{4\pi i}{g^2}$ of $G$ SYM theory, 
where $\theta$ is the theta angle and $g$ is the gauge coupling, 
is related to the coupling constant $\tau^{\vee}$ of $G^{\vee}$ SYM theory by
\begin{align}
\tau^{\vee}=-\frac{1}{n_{\mathfrak{g}}\tau}, 
\end{align}
where $n_{\mathfrak{g}}$ is the ratio of the length squares of the long and short roots. 
For simply-laced $\mathfrak{g}$ we have $n_{\mathfrak{g}}=1$. 
For non-simply-laced $\mathfrak{g}$ we have 
$n_{\mathfrak{so}(2N+1)}$ $=$ $n_{\mathfrak{usp}(2N)}$ $=$ $n_{F_4}$ $=$ $2$ and $n_{G_2}$ $=$ $3$. 
In general, the magnetic charge $B$ carried by the 't Hooft line can be lowered by the smooth monopoles surrounding the singular monopole. 
This phenomenon is called the monopole bubbling effect \cite{Kapustin:2006pk}. 
In the cases with the 't Hooft lines associated with the minuscule representation \cite{MR2109105}
\footnote{
There exist many ways of defining minuscule representations. 
See \cite{MR2109105} for six equivalent conditions. 
}, for which all of their weights lie in a single Weyl group orbit there is no monopole bubbling effect. 
For the case with gauge algebra $\mathfrak{g}$ being the classical Lie algebra, 
dualities of the Wilson lines of representations $\mathcal{R}$ 
and the 't Hooft lines of magnetic charges $B$ associated with the minuscule representations are given as follows: 
\begin{align}
\label{Table_duallines}
\begin{array}{c|c||c|c}
G&\mathcal{R}&G^{\vee}&B \\ \hline 
U(N)&(1^k)&U(N)&(1^{k},0^{N-k}) \\
Spin(2N+1)&\textrm{spinor}&USp(2N)/\mathbb{Z}_2&(\frac12^N) \\
USp(2N)&(1)&SO(2N+1)&(1,0^{N-1}) \\
Spin(2N)&\textrm{(anti)chiral spinor}&SO(2N)/\mathbb{Z}_2&(\frac12^N) \\ 
SO(2N)&(1)&SO(2N)&(1,0^{N-1}) \\ 
O(2N)&(1)&O(2N)&(1,0^{N-1}) \\ 
\end{array}
\end{align}
Here the notation $(\lambda_1,\cdots, \lambda_N)$ in $\mathcal{R}$ stands for the partition associated with the highest weight $\lambda$ of the representation $\mathcal{R}$. 
In the following sections, we demonstrate the dualities (\ref{dual_bdyline}) of the boundary line operators 
involving the 't Hooft lines associated with the minuscule representations 
as precise matching pairs of line defect half-indices. 

\subsection{Brane construction}
The configuration of the boundary line operators in 4d $\mathcal{N}=4$ SYM theories can be realized in Type IIB string theory 
in a similar manner as the brane construction of the line operators in 3d $\mathcal{N}=4$ gauge theories \cite{Assel:2015oxa}. 
We consider a stack of $N$ D3-branes supported along the directions $0126$. 
For $\mathrm{Lie}(G)$ $=$ $\mathfrak{g}$ $=$ $\mathfrak{so}(2N+1)$, $\mathfrak{usp}(2N)$ and $\mathfrak{so}(2N)$, 
the theories can be constructed by introducing an additional O3-plane. 
There are four kinds of O3-planes with two $\mathbb{Z}_2$-valued discrete torsions $\theta_{RR}$ and $\theta_{NS}$ 
for the Ramond-Ramond (RR) and Neveu-Schwarz (NS) $2$-forms. 
Four distinct types of O3-planes and SYM theories are given by 
\begin{align}
\label{O3_SYM}
\begin{array}{c|c|c|c|c}
&SO(2N+1)&USp(2N)&O(2N)&USp(2N)'\\ \hline 
\theta_{RR}&1/2&0&0&1/2 \\
\theta_{NS}&0&1/2&0&1/2 \\
\textrm{D3-brane charge}&1/4&1/4&-1/4&1/4 \\
\textrm{orientifold}&\widetilde{\textrm{O3}}^{-}&\textrm{O3}^+&\textrm{O3}^{-}&\widetilde{\textrm{O3}}^+ \\
\textrm{$S$ operation}&\textrm{O3}^+&\widetilde{\textrm{O3}}^{-}&\textrm{O3}^{-}&\widetilde{\textrm{O3}}^+ \\
\end{array}.
\end{align}
The half-BPS boundary conditions in $\mathcal{N}=4$ SYM theories can be 
constructed by introducing $5$-branes which are localized at $x^6=0$ 
in such a way that D3-branes terminate on them \cite{Hanany:1996ie,Gaiotto:2008ak}. 
In the absence of the O3-plane, one finds the Neumann b.c. (\ref{NeuBC}) for $U(N)$ SYM theory 
from a single NS5-brane along the $012345$ directions and the Nahm pole b.c. (\ref{NahmBC}) from a single D5-brane along the $012789$ directions. 
When the O3-plane interact with these $5$-branes, these $5$-branes are recognized as half-NS5- and D5-branes \cite{Gaiotto:2008ak}. 
The fluctuations of the D3-branes along the $345$ directions and the $789$ directions 
are described by the scalar fields $\vec{Y}$ and $\vec{X}$ respectively. 

Furthermore, one can construct the boundary line operators in $\mathcal{N}=4$ SYM theories 
by introducing strings which are localized at $x^6=0$. 
In the presence of the NS5-brane, the boundary Wilson line in the fundamental representation with the Neumann b.c. 
can be described by a fundamental string along the $05$ directions in a similar manner as the fundamental Wilson line in the bulk \cite{Maldacena:1998im,Rey:1998ik}. 
When $k$ fundamental strings terminate on another D5-brane along the $012789$ directions that is separated from the boundary D5-brane in the $x^5$ direction at one end 
and on $N$ D3-branes at the other end, we find the boundary Wilson line in the rank-$k$ symmetric representation. 
On the other hand, when $k$ fundamental strings are stretched in the $x^5$ directions between the D3-branes and an additional D5-brane along the $034789$ directions, which we denote by $\widetilde{\textrm{D5}}$, 
we obtain the boundary Wilson line in the rank-$k$ antisymmetric representation. 
This generalizes the construction of the antisymmetric Wilson line in the bulk \cite{Yamaguchi:2006tq}. 
For orthogonal gauge groups, there is the Wilson line in the spinor representation. 
It corresponds to a \textit{fat string}, a wrapped D5-brane at an orientifold fixed point which is conserved modulo two due to the topological restrictions 
of the vanishing $\theta_{NS}$ in the presence of $\widetilde{\textrm{O3}}^{-}$- or $\textrm{O3}^-$-plane \cite{Witten:1998xy}. 
The brane construction of the boundary Wilson line operators with the Neumann b.c. (\ref{NeuBC}) is summarized as follows: 
\begin{align}
\begin{array}{c|cccccccccc}
&0&1&2&3&4&5&6&7&8&9 \\ \hline
\textrm{$N$ D3}&\circ&\circ&\circ&&&&\circ&&& \\
\textrm{O3}&\circ&\circ&\circ&&&&\circ&&& \\
\textrm{NS5}&\circ&\circ&\circ&\circ&\circ&\circ&&&& \\
\textrm{D5}&\circ&\circ&\circ&&&&&\circ&\circ&\circ \\
\textrm{F1}&\circ&&&&&\circ&&&& \\
\widetilde{\textrm{D5}}&\circ&&&\circ&\circ&&&\circ&\circ&\circ \\
\end{array}
\end{align}

The S-dual brane construction of the boundary 't Hooft lines can be found by considering the $S$ transformation of the $SL(2,\mathbb{Z})$ action of Type IIB string theory. 
The D3-branes are invariant, however, the NS5-branes, the D5-branes and the fundamental strings map to the D5-branes, the NS5-branes and the D1-branes respectively, 
while the O3-planes transform as in (\ref{O3_SYM}). 
The brane configuration of the boundary 't Hooft line operators with the Nahm pole b.c. (\ref{NahmBC}) is
\begin{align}
\begin{array}{c|cccccccccc}
&0&1&2&3&4&5&6&7&8&9 \\ \hline
\textrm{$N$ D3}&\circ&\circ&\circ&&&&\circ&&& \\
\textrm{O3}&\circ&\circ&\circ&&&&\circ&&& \\
\textrm{NS5}&\circ&\circ&\circ&\circ&\circ&\circ&&&& \\
\textrm{D5}&\circ&\circ&\circ&&&&&\circ&\circ&\circ \\
\textrm{D1}&\circ&&&&&&&&&\circ \\
\widetilde{\textrm{NS5}}&\circ&&&\circ&\circ&\circ&&\circ&\circ& \\
\end{array}
\end{align}
For example, 
the configuration consisting of a D1-brane along the $09$ directions together with $N$ D3-branes and a single D5-brane at $x^6=0$ 
is obtained from the fundamental string along the $05$ directions as well as $N$ D3-branes and a single NS5-brane at $x^6=0$ under the S-transformation. 
Hence it describes the boundary 't Hooft line of magnetic charge $B=(1,0^{N-1})$ with the Nahm pole b.c.

\subsection{Line defect half-indices}
The half-indices can count the boundary BPS local operators for the half-BPS boundary conditions $\mathcal{B}$ in 4d $\mathcal{N}=4$ SYM theory of gauge group $G$. 
They can be defined as the trace over the cohomology $\mathcal{H}$ of the supercharges 
which are contained in the subalgebra of the 3d $\mathcal{N}=4$ superalgebra \cite{Gaiotto:2019jvo} 
\footnote{Here we use the convention and definition in \cite{Gaiotto:2019jvo,Okazaki:2019ony}. 
See \cite{Dimofte:2011py,Gang:2012yr} for the other conventions. }
\begin{align}
\label{hindex_def}
\mathbb{II}_{\mathcal{B}}^{\textrm{4d $G$}}(t,z;q)
&={\Tr}_{\mathcal{H}}(-1)^{F} q^{J+\frac{H+C}{4}}t^{H-C}z^f, 
\end{align}
where $F$ is the Fermion number operator, $J$ spin, 
$H,C$ the Cartan generators of the two $SU(2)_H$ and $SU(2)_C$ factors of the R-symmetry group and $f$ the Cartans of the other global symmetries. 

The half-indices can be expressed in terms of the $q$-shifted factorial. 
Let us define 
\begin{align}
\label{qpoch_def}
(a;q)_{0}&:=1,\qquad
(a;q)_{n}:=\prod_{k=0}^{n-1}(1-aq^{k}),\qquad 
(q)_{n}:=\prod_{k=1}^{n}(1-q^{k}),
\nonumber \\
(a;q)_{\infty}&:=\prod_{k=0}^{\infty}(1-aq^{k}),\qquad 
(q)_{\infty}:=\prod_{k=1}^{\infty} (1-q^k), 
\end{align}
with $a, q \in \mathbb{C}$ and $|q|<1$. 
For simplicity we also use the following notations: 
\begin{align}
(x^{\pm};q)_{n}&:=(x;q)_{n}(x^{-1};q)_{n}, \\
(x_1,\cdots,x_k;q)_{n}&:=(x_1;q)_{n}\cdots (x_k;q)_{n}. 
\end{align}

The local operators in 4d $\mathcal{N}=4$ SYM theory of gauge group $G$ which contribute to the index are 
\begin{align}
\label{4dn4_ch}
\begin{array}{c|cccc}
&\partial^{n}X&\partial^{n}Y&\partial^{n}\lambda&\partial^{n}\overline{\lambda} \\ \hline
G&\textrm{adj}&\textrm{adj}&\textrm{adj}&\textrm{adj} \\
U(1)_{J}&n&n&n+\frac12&n+\frac12 \\
U(1)_{C}&0&2&+&+ \\
U(1)_{H}&2&0&+&+  \\
\textrm{fugacity}
&q^{n+\frac12}t^{2}e^{\alpha}
&q^{n+\frac12}t^{-2}e^{\alpha}
&-q^{n+1}e^{\alpha}
&-q^{n+1}e^{\alpha} \\
\end{array}. 
\end{align}
Here $e^{\alpha}$ is associated with gauge fugacities $s$, 
where $\alpha$ stand for elements of a root system $R$ of the Lie algebra $\mathfrak{g}$ for gauge group $G$. 
In the case with the Neumann b.c. (\ref{NeuBC}) and for the Dirichlet b.c. the calculation of the half-indices is straightforward. 
The Neumann half-index can be computed by collecting the contributions obeying the Neumann boundary conditions 
and projecting these to $G$-invariants by performing the integration over the gauge fugacities $s$. 
On the other hand, the Dirichlet half-index can be obtained by projecting out the contributions which cannot freely fluctuate at the boundary 
and identifying the gauge fugacities $s$ with the fugacities of the boundary global symmetry \cite{Gaiotto:2019jvo}. 
For the Nahm pole b.c. (\ref{NahmBC}), the half-index can be obtained by employing the Higgsing manipulation to the Dirichlet half-index \cite{Gaiotto:2019jvo,Hatsuda:2024lcc}. 
The basic idea is to view the regular Nahm pole b.c. as the deformed Dirichlet b.c. 
due to the nilpotent vev for the adjoint scalar fields $\mathcal{X}$ 
so that we specialize the global fugacities of the Dirichlet half-index and removing the index 
\begin{align}
\mathcal{I}^{\textrm{3d HM}}(t,x;q)
&=\frac{(q^{\frac34}t^{-1}x^{\mp};q)_{\infty}}
{(q^{\frac14}tx^{\mp};q)_{\infty}}
\end{align}
for the 3d hypermultiplet. 
As a consequence of S-duality, one finds matching pairs of the Neumann half-indices and the Nahm pole half-indices 
(see \cite{Gaiotto:2019jvo,Hatsuda:2024lcc} and the discussion below). 

The half-indices can be decorated by introducing the boundary line operators 
so that they can enumerate the BPS boundary local operators appearing at the junction of the boundary lines \cite{Cordova:2016uwk}. 
We refer to them as \textit{line defect half-indices} or briefly \textit{line defect correlation functions}. 
When the line operators wrap the $S^1$ circle and sit at points on a great circle on the boundary of the hemisphere with the $S^2$ topology, 
the line defect half-indices have no spatial dependence \cite{Cordova:2016uwk}. 
In the following sections we investigate the line defect half-indices for the boundary Wilson lines with the Neumann b.c. (\ref{NeuBC}) 
and those for the boundary 't Hooft lines with the regular Nahm pole b.c. (\ref{NahmBC}). 
For the boundary Wilson lines in the representations $\mathcal{R}_i$, $i=1,\cdots, k$ with the Neumann b.c. (\ref{NeuBC}), 
the line defect half-indices can be evaluated by the matrix integral over the gauge fugacities $s$ associated with $e^{\alpha}$,\footnote{More precisely, if $\{ \varepsilon_i\}$ is an orthonormal basis of the Euclidian space to expand $\alpha$, we define $s_i=e^{\varepsilon_i}$.} 
\begin{align}
\label{G_Wintegral}
\langle W_{\mathcal{R}_1}\cdots W_{\mathcal{R}_k}\rangle_{\mathcal{N}}^{G}
=\frac{1}{|W_R|}
\oint ds\prod_{\alpha\in R} 
\frac{(e^{\alpha};q)_{\infty}}
{(q^{\frac12}t^{-2}e^{\alpha};q)_{\infty}}
\prod_{i=1}^{k}\chi^{\mathfrak{g}}_{\mathcal{R}_i}(s). 
\end{align}
Here $|W_R|$ is the order of Weyl group $W_R$ associated with the root system $R$ of gauge group $G$. 
The numerator encodes contributions from the fermionic modes $\partial^n\overline{\lambda}$ as well as the Vandermonde determinant, 
while the denominator stems from the bosonic excitations $\partial^n Y$ which arise in the presence of the Neumann b.c. (\ref{NeuBC}).  
$\chi^{\mathfrak{g}}_{\mathcal{R}_i}(s)$ is the character of the representation $\mathcal{R}_i$ of gauge algebra $\mathfrak{g}$. 
It is given by the Weyl character formula (see Appendix \ref{app_ch} and \cite{MR1153249} for the details). 
In particular, two-point function of a pair of the boundary Wilson lines in the irreducible representation $\mathcal{R}$ 
and its conjugate $\overline{\mathcal{R}}$ can form a straight line in the flat space under the conformal map 
so that the configuration can preserve a one-dimensional superconformal symmetry \cite{Cordova:2016uwk}. 
 
Since the half-indices are protected along the RG-flow in such a way that they are independent of gauge coupling, 
the dualities (\ref{dual_bdyline}) would imply that 
the two-point functions of the boundary Wilson lines for SYM theory of gauge group $G$ with the Neumann b.c. (\ref{NeuBC}) 
are related to the two-point function of the dual boundary 't Hooft lines for SYM theory of gauge group $G^{\vee}$ with the Nahm pole b.c. (\ref{NahmBC}) 
under the transformation $t$ $\rightarrow$ $t^{-1}$. 
In the following, we propose that the two-point functions of a pair of the boundary 't Hooft lines of magnetic charge 
associated with the minuscule representation that is free from the monopole bubbling 
can be obtained from the Dirichlet half-index by performing the Higgsing procedure \cite{Gaiotto:2019jvo}. 
We confirm matching pairs of the two-point functions of the Wilson line defect half-indices and those of the 't Hooft line defect half-indices 
as strong evidence of our proposal and the dualities (\ref{dual_bdyline}) of the boundary lines in $\mathcal{N}=4$ SYM theories.

We also point out that the matrix integral (\ref{G_Wintegral}) is related to the 
inner product of Macdonald polynomials associated with the root system $R$ \cite{MR1354144,MR1976581,MR1314036,MR1354956}. 
In fact, the Schur indices and the interface half-indices for $\mathcal{N}=4$ $U(N)$ SYM theories are exactly computed in 
\cite{Hatsuda:2025mvj} in terms of the Macdonald polynomials. 
Let $P_\lambda(s;\q,\t)$ be the Macdonald polynomials associated with $R$. Here we introduce a notation of parameters
\begin{align}
\q=q,\qquad \t=q^{\frac{1}{2}}t^{-2},
\end{align}
for the Macdonald polynomials. The Macdonald polynomials are orthogonal symmetric polynomials with respect to the following inner product:
\begin{align}
\langle f, g \rangle=\frac{1}{|W_R|} \oint ds\, w_R(s; \q,\t) f(s) \overline{g(s)},
\end{align}
where $\overline{g(s)}=g(s^{-1})$.
The weight function is given by
\begin{align}
w_R(s; \q,\t)=\prod_{\alpha \in R} \frac{(e^{\alpha};\q)_\infty}{(\t e^{\alpha};\q)_\infty}.
\end{align}

According to \cite{MR1976581}, the norm of the Macdonald polynomial is given by
\begin{align}
\langle P_\lambda, P_\lambda \rangle=\prod_{\alpha \in R^+}
\frac{(\t^{(\rho, \alpha^\vee)}\q^{(\lambda, \alpha^\vee)};\q)_\infty (\t^{(\rho, \alpha^\vee)}\q^{(\lambda, \alpha^\vee)+1};\q)_\infty}{(\t^{(\rho, \alpha^\vee)+1}\q^{(\lambda, \alpha^\vee)};\q)_\infty (\t^{(\rho, \alpha^\vee)-1}\q^{(\lambda, \alpha^\vee)+1};\q)_\infty},
\end{align}
where $R^+$ is the set of the positive roots, and
\begin{align}
\alpha^\vee=\frac{2\alpha}{(\alpha, \alpha)},\qquad
\rho=\frac{1}{2}\sum_{\alpha \in R^+} \alpha.
\end{align}
It is more useful to rewrite it as the normalized inner product
\begin{align}
\frac{\langle P_\lambda, P_\lambda \rangle}{\langle 1, 1 \rangle}=\prod_{\alpha \in R^+}
\frac{(\t^{(\rho, \alpha^\vee)+1};\q)_{(\lambda, \alpha^\vee)} (\t^{(\rho, \alpha^\vee)-1}\q ;\q)_{(\lambda, \alpha^\vee)}}{(\t^{(\rho, \alpha^\vee)};\q)_{(\lambda, \alpha^\vee)} (\t^{(\rho, \alpha^\vee)}\q;\q)_{(\lambda, \alpha^\vee)}},
\end{align}
and
\begin{align}
\langle 1, 1 \rangle=\prod_{\alpha \in R^+}
\frac{(\t^{(\rho, \alpha^\vee)};\q)_\infty (\t^{(\rho, \alpha^\vee)}\q;\q)_\infty}{(\t^{(\rho, \alpha^\vee)+1};\q)_\infty (\t^{(\rho, \alpha^\vee)-1}\q;\q)_\infty}.
\end{align}
We will use these results to evaluate the matrix integral (\ref{G_Wintegral}) analytically.

\section{$\mathfrak{u}(N)$}
\label{sec_uN}

\subsection{$\mathfrak{u}(2)$}
We begin with $\mathcal{N}=4$ $U(2)$ gauge theory. 
The half-index of the Neumann b.c. for $U(2)$ gauge theory is given by
\begin{align}
\label{u2N}
\mathbb{II}_{\mathcal{N}}^{\textrm{4d $\mathcal{N}=4$ $U(2)$}}(t;q)
&=\frac12 \frac{(q)_{\infty}^2}{(q^{\frac12}t^{-2};q)_{\infty}^2}
\oint \prod_{i=1}^2 \frac{ds_i}{2\pi is_i} 
\prod_{i\neq j} \frac{\left( \frac{s_i}{s_j};q \right)_{\infty}}{\left(q^{\frac12}t^{-2}\frac{s_i}{s_j};q \right)_{\infty}}. 
\end{align}
Under the transformation $t$ $\rightarrow$ $t^{-1}$, 
it is equal to the $U(2)$ Nahm pole half-index 
\begin{align}
\label{u2Nahm}
\mathbb{II}_{\mathrm{Nahm}}^{\textrm{4d $\mathcal{N}=4$ $U(2)$}}(t;q)
&=\frac{(q)_{\infty} (q^{\frac32}t^{2};q)_{\infty}}
{(q^{\frac12}t^{2};q)_{\infty}(qt^{4};q)_{\infty}}. 
\end{align}
As observed in \cite{Gaiotto:2019jvo}, 
it is obtained from the Dirichlet half-index
\begin{align}
\label{u2D}
\mathbb{II}_{\mathcal{D}}^{\textrm{4d $\mathcal{N}=4$ $U(2)$}}(t,x_1,x_2;q)
&=
\frac{(q)_{\infty}^2}{(q^{\frac12}t^2;q)_{\infty}^2} 
\frac{(q\frac{x_1}{x_2};q)_{\infty}(q\frac{x_2}{x_1};q)_{\infty}}
{(q^{\frac12}t^2 \frac{x_1}{x_2};q)_{\infty} (q^{\frac12}t^2 \frac{x_2}{x_1};q)_{\infty}}
\end{align}
by employing the Higgsing procedure. 
By specializing the global fugacities as
\begin{align}
\label{u2spe_Nahm}
x_1&=q^{\frac14}t, \qquad 
x_2=q^{\frac34}t^3, 
\end{align}
the Dirichlet half-index (\ref{u2D}) is factorized as
\begin{align}
&
\mathbb{II}_{\mathcal{D}}^{\textrm{4d $\mathcal{N}=4$ $U(2)$}}
\left(t,x_1=q^{\frac14}t,x_2=q^{\frac34}t^3;q\right)
\nonumber\\
&=\mathbb{II}_{\mathrm{Nahm}}^{\textrm{4d $\mathcal{N}=4$ $U(2)$}}(t;q)
\mathcal{I}^{\textrm{3d HM}}(t,x=q^{\frac14}t;q), 
\end{align}
and stripping off the decoupled 3d matter indices,  
which are expected to be contributed along the RG flow after turning on the nilpotent vev for the adjoint scalar fields, 
we get the desirable regular Nahm pole half-index (\ref{u2Nahm}). 
Notice that the same result can be obtained by considering the Higgsing manipulation with different specializations as 
it only depends on the ratio $x_1/x_2$. 
In particular, we can take 
\begin{align}
x_1&=q^{\frac54}t, \qquad 
x_2=q^{\frac74}t^3. 
\end{align}
As we will see, this can be viewed as the Higgsing for the two-point function of the 't Hooft lines of magnetic charge $B=(1,1)$ with the regular Nahm pole b.c. 
However, since it is dual to the trivial Wilson line for $U(2)$ gauge theory, 
it leads to the Nahm pole half-index (\ref{u2Nahm}) without any insertion of lines. 

\subsubsection{Fundamental Wilson line}
Let us study the line defect half-indices of the Wilson lines in the fundamental representation of $U(2)$ gauge group. 
According to the Gauss law, the one-point function of the boundary Wilson line in the fundamental representation with Neumann b.c. vanishes. 
On the other hand, the two-point function of the boundary (anti)fundamental Wilson lines with the Neumann b.c. is non-trivial. 
It can be evaluated by the matrix integral of the form 
\begin{align}
\label{u2N_W1}
&
\langle W_{\tiny \yng(1)} W_{\overline{\tiny \yng(1)}}\rangle_{\mathcal{N}}^{\textrm{4d $\mathcal{N}=4$ $U(2)$}}(t;q)
\nonumber\\
&=\frac12 \frac{(q)_{\infty}^2}{(q^{\frac12}t^{-2};q)_{\infty}^2}
\oint \prod_{i=1}^2 \frac{ds_i}{2\pi is_i} 
\prod_{i\neq j} \frac{\left( \frac{s_i}{s_j};q \right)_{\infty}}{\left(q^{\frac12}t^{-2}\frac{s_i}{s_j};q \right)_{\infty}}
(s_1+s_2)(s_1^{-1}+s_2^{-1}). 
\end{align}
The boundary Wilson line in the fundamental representation with the Neumann boundary condition 
is conjecturally dual to the boundary 't Hooft line of magnetic charge $B=(1,0)$ with the Nahm pole boundary condition. 
Here we demonstrate that the Higgsed Dirichlet half-index leads to the two-point function of the dual 't Hooft lines with the Nahm pole b.c. which is equivalent to (\ref{u2N_W1}), 
as discussed in \cite{Gaiotto:2019jvo}. 
By identifying the global fugacities 
\begin{align}
x_1&=q^{\frac14}t,\qquad 
x_2=q^{\frac74}t^3, 
\end{align}
the Dirichlet half-index (\ref{u2D}) is factorized as
\begin{align}
&
\mathbb{II}_{\mathcal{D}}^{\textrm{4d $\mathcal{N}=4$ $U(2)$}}
\left(t,x_1=q^{\frac14}t,x_2=q^{\frac74}t^3;q\right)
\nonumber\\
&=
\langle T_{(1,0)}T_{(1,0)}\rangle_{\textrm{Nahm}}^{\textrm{4d $\mathcal{N}=4$ $U(2)$}}(t;q)
\mathcal{I}^{\textrm{3d HM}}(t,x=q^{\frac54}t;q), 
\end{align}
where 
\begin{align}
\label{u2Nahm_T1}
\langle T_{(1,0)}T_{(1,0)}\rangle_{\textrm{Nahm}}^{\textrm{4d $\mathcal{N}=4$ $U(2)$}}(t;q)&=
\frac{(q)_{\infty}^2 (q^{\frac32}t^{2};q)_{\infty}(q^{\frac52}t^{2};q)_{\infty}}
{(q^{\frac12}t^{2};q)_{\infty}^2(q^2;q)_{\infty}(q^2t^{4};q)_{\infty}}
\end{align}
is identified with the two-point function of the boundary 't Hooft lines with the Nahm pole b.c. for $U(2)$ gauge theory. 
In fact, one can check that the two expressions (\ref{u2N_W1}) and (\ref{u2Nahm_T1}) agree with each other upon 
the transformation $t$ $\rightarrow$ $t^{-1}$
\begin{align}
\langle W_{\tiny \yng(1)} W_{\overline{\tiny \yng(1)}}\rangle_{\mathcal{N}}^{\textrm{4d $\mathcal{N}=4$ $U(2)$}}(t;q)
&=\langle T_{(1,0)}T_{(1,0)}\rangle_{\textrm{Nahm}}^{\textrm{4d $\mathcal{N}=4$ $U(2)$}}(t^{-1};q). 
\end{align}

\subsubsection{Symmetric Wilson lines}
The line defect half-indices of the boundary Wilson lines which 
have the magnetic duals associated with the non-minuscule representations are rather cumbersome, 
however, we find that they also admit several closed-form expressions. 
Let us consider the boundary Wilson line in the symmetric representation. 
While the one-point function of the boundary Wilson line in the rank-$2$ symmetric representation with the Neumann b.c. vanishes, 
the two-point function is non-trivial. 
We find that the two-point function can be written as
\begin{align}
\label{u2N_Wsym2}
&
\langle W_{\tiny \yng(2)} W_{\overline{\tiny \yng(2)}}\rangle_{\mathcal{N}}^{\textrm{4d $\mathcal{N}=4$ $U(2)$}}(t;q)
\nonumber\\
&=3\frac{(q)_{\infty}(q^{\frac32}t^{-2};q)_{\infty}}
{(q^{\frac12}t^{-2};q)_{\infty}(qt^{-4};q)_{\infty}}
-2\frac{(q^2;q)_{\infty}^2(q^{\frac52}t^{-2};q)_{\infty}}
{(qt^{-4};q)_{\infty}(q^{\frac32}t^{-2};q)_{\infty}(q^3;q)_{\infty}}
\nonumber\\
&-q^{\frac12}t^{-2}
\frac{(q^{\frac12}t^2;q)_{\infty}(q)_{\infty}(q^3;q)_{\infty}(q^{\frac72}t^{-2};q)_{\infty}(q^4;q)_{\infty}}
{(qt^{-4};q)_{\infty}(q^{\frac32}t^{-2};q)_{\infty}(q^{\frac32}t^2;q)_{\infty}(q^2;q)_{\infty}(q^5;q)_{\infty}}. 
\end{align}
To gain more insight, 
we explore the closed-form expressions for the two-point function of the boundary Wilson line in the rank-$3$ symmetric representation. 
We find that they are given by
\begin{align}
\label{u2N_Wsym3}
&
\langle W_{\tiny \yng(3)} W_{\overline{\tiny \yng(3)}}\rangle_{\mathcal{N}}^{\textrm{4d $\mathcal{N}=4$ $U(2)$}}(t;q)
\nonumber\\
&=4\frac{(q)_{\infty}(q^{\frac32}t^{-2};q)_{\infty}}
{(q^{\frac12}t^{-2};q)_{\infty}(qt^{-4};q)_{\infty}}
-3\frac{(q^2;q)_{\infty}^2(q^{\frac52}t^{-2};q)_{\infty}}
{(qt^{-4};q)_{\infty}(q^{\frac32}t^{-2};q)_{\infty}(q^3;q)_{\infty}}
\nonumber\\
&-2q^{\frac12}t^{-2}
\frac{(q^{\frac12}t^2;q)_{\infty}(q)_{\infty}(q^3;q)_{\infty}(q^{\frac72}t^{-2};q)_{\infty}(q^4;q)_{\infty}}
{(qt^{-4};q)_{\infty}(q^{\frac32}t^{-2};q)_{\infty}(q^{\frac32}t^2;q)_{\infty}(q^2;q)_{\infty}(q^5;q)_{\infty}}
\nonumber\\
&-qt^{-4}
\frac{(q^{\frac12}t^2;q)_{\infty}(q)_{\infty}(q^4;q)_{\infty}(q^{\frac92}t^{-2};q)_{\infty}(q^6;q)_{\infty}}
{(qt^{-4};q)_{\infty}(q^{\frac32}t^{-2};q)_{\infty}(q^{\frac52}t^2;q)_{\infty}(q^3;q)_{\infty}(q^7;q)_{\infty}}. 
\end{align}
Now it is straightforward to generalize the results to the two-point function of the boundary Wilson lines in the rank-$k$ symmetric representation. 
We propose that it is given by
\begin{align}
\label{u2N_Wsymk}
&
\langle W_{(k)} W_{\overline{(k)}}\rangle_{\mathcal{N}}^{\textrm{4d $\mathcal{N}=4$ $U(2)$}}(t;q)
=
(k+1)\frac{(q)_{\infty}(q^{\frac32}t^{-2};q)_{\infty}}
{(q^{\frac12}t^{-2};q)_{\infty}(qt^{-4};q)_{\infty}}
\nonumber\\
&-\frac{(q^{\frac12}t^2;q)_{\infty}(q)_{\infty}}
{(qt^{-4};q)_{\infty}(q^{\frac32}t^{-2};q)_{\infty}}
\sum_{i=1}^{k}
(k-i+1)q^{\frac{i-1}{2}}t^{-2(2i-1)}
(1+q^i)
\frac{(q^{\frac{2i+3}{2}}t^{-2};q)_{\infty}}
{(q^{\frac{2i-1}{2}}t^{2};q)_{\infty}}. 
\end{align}
This expression is reminiscent of the formula for the two-point function of the Wilson lines 
in the rank-$k$ symmetric representation without boundary that is obtained via the Fermi-gas method \cite{Hatsuda:2023iwi} 
\begin{align}
&
\langle W_{(k)}W_{\overline{(k)}}\rangle^{\textrm{4d $\mathcal{N}=4$ $U(2)$}}(q)
\nonumber\\
&=(k+1)\mathcal{I}^{U(2)}(t;q)
-q^{-1}t^4 \sum_{i=1}^{k}(k-i+1)
\frac{t^{2i}-t^{-2i}}{q^{\frac{i}{2}}-q^{-\frac{i}{2}}}
P_1
\left[
\begin{matrix}
q^{-1}t^4\\
1
\end{matrix}
\right](\zeta,\tau), 
\end{align}
where $P_1
\left[
\begin{matrix}
\theta \\
\phi
\end{matrix}
\right](\zeta,\tau)$ is the twisted Weierstrass function with $q^{1/2}t^{-2}=e^{2\pi i\zeta}$ and $q=e^{2\pi i\tau}$ 
(see \cite{Hatsuda:2023iwi} for the details). 
In the unflavored limit $t\rightarrow 1$, the line defect half-index (\ref{u2N_Wsymk}) reduces to 
\begin{align}
\label{u2N_Wsymk_uflavor}
&
\langle W_{(k)} W_{\overline{(k)}}\rangle_{\mathcal{N}}^{\textrm{4d $\mathcal{N}=4$ $U(2)$}}(q)
\nonumber\\
&=\frac{k+1}{1-q^{\frac12}}
+
(1-q^{\frac12})
\sum_{i=1}^k (k-i+1)q^{\frac{i-1}{2}}\frac{(1+q^i)}
{(1-q^{\frac{2i-1}{2}})(1-q^{\frac{2i+1}{2}})}. 
\end{align}
We observe that in the large representation limit $k\rightarrow \infty$ 
the unflavored two-point function (\ref{u2N_Wsymk_uflavor}) can be written as
\begin{align}
\langle W_{(k=\infty)} W_{\overline{(k=\infty)}}\rangle_{\mathcal{N}}^{\textrm{4d $\mathcal{N}=4$ $U(2)$}}(q)
&=\sum_{n=1}^{\infty} \frac{q^{\frac{n-1}{2}}}{1-q^{n-\frac12}}
\nonumber\\
&=q^{-\frac12}\sum_{n=1}^{\infty} \sigma_0(2n-1)q^{\frac{n}{2}}, 
\end{align}
where
\begin{align}
\sigma_k(n)&=\sum_{d|n}d^{k}
\end{align}
is the sum of divisors function. 
It can be worthwhile to compare the large representation limit of the 
unflavored two-point function of the two-point function of the symmetric Wilson lines without boundary. 
It is given by \cite{Hatsuda:2023iwi}
\begin{align}
\langle W_{(k=\infty)} W_{\overline{(k=\infty)}}\rangle^{\textrm{4d $\mathcal{N}=4$ $U(2)$}}(q)
&=\sum_{n=1}^{\infty}\frac{n^2 q^{\frac{n-1}{2}}}{1-q^n}
\nonumber\\
&=q^{-\frac12}\sum_{n=1}^{\infty}
\frac{\sigma_2(2n)-\sigma_2(n)}{4}q^{\frac{n}{2}}. 
\end{align}

\subsection{$\mathfrak{u}(3)$}
Next consider $U(3)$ gauge theory. 
The half-index of Neumann boundary condition is evaluated as 
\begin{align}
\label{u3N}
\mathbb{II}_{\mathcal{N}}^{\textrm{4d $\mathcal{N}=4$ $U(3)$}}(t;q)
&=\frac{1}{3!}
\frac{(q)_{\infty}^3}{(q^{\frac12}t^{-2};q)_{\infty}^3}
\oint 
\prod_{i=1}^3 \frac{ds_i}{2\pi is_i}
\prod_{i\neq j} \frac{\left( \frac{s_i}{s_j};q \right)_{\infty}}{\left(q^{\frac12}t^{-2}\frac{s_i}{s_j};q \right)_{\infty}}. 
\end{align}
On the other hand, 
the half-index of the regular Nahm pole boundary condition for $U(3)$ gauge theory reads \cite{Gaiotto:2019jvo}
\begin{align}
\label{u3Nahm}
\mathbb{II}_{\textrm{Nahm}}^{\textrm{4d $\mathcal{N}=4$ $U(3)$}}(t;q)
&=\frac{(q)_{\infty}(q^{\frac32}t^{2};q)_{\infty}(q^2t^{4};q)_{\infty}}
{(q^{\frac12}t^{2};q)_{\infty}(qt^{4};q)_{\infty}(q^{\frac32}t^{6};q)_{\infty}}. 
\end{align}
As discussed in \cite{Gaiotto:2019jvo}, 
the two expressions (\ref{u3N}) and (\ref{u3Nahm}) become equivalent 
by flipping $t$ to $t^{-1}$ 
as a consequence of S-duality of the Neumann b.c. and the Nahm pole b.c. 
It is shown in \cite{Gaiotto:2019jvo} that the Nahm pole half-index (\ref{u3Nahm}) is obtained from the Dirichlet half-index 
\begin{align}
\label{u3D}
&
\mathbb{II}_{\mathcal{D}}^{\textrm{4d $\mathcal{N}=4$ $U(3)$}}(t,x_1,x_2,x_3;q)
\nonumber\\
&=
\frac{(q)_{\infty}^3}{(q^{\frac12}t^2;q)_{\infty}^3}
\frac{(q\frac{x_1}{x_2};q)_{\infty}(q\frac{x_1}{x_3};q)_{\infty}(q\frac{x_2}{x_1};q)_{\infty}(q\frac{x_2}{x_3};q)_{\infty}
(q\frac{x_3}{x_1};q)_{\infty}(q\frac{x_3}{x_2};q)_{\infty}}
{
(q^{\frac12}t^2\frac{x_1}{x_2};q)_{\infty}(q^{\frac12}t^2\frac{x_1}{x_3};q)_{\infty}
(q^{\frac12}\frac{x_2}{x_1};q)_{\infty}(q^{\frac12}t^2\frac{x_2}{x_3};q)_{\infty}
(q^{\frac12}t^2\frac{x_3}{x_1};q)_{\infty}(q^{\frac12}t^2\frac{x_3}{x_2};q)_{\infty}
}
\end{align}
by applying the Higgsing manipulation in such a way that 
the global fugacities are specialized as
\begin{align}
x_1&=q^{\frac14}t,\qquad 
x_2=q^{\frac34}t^3,\qquad 
x_3=q^{\frac54}t^5. 
\end{align}
Consequently, the Higgsed Dirichlet half-index takes the form
\begin{align}
&
\mathbb{II}_{\mathcal{D}}^{\textrm{4d $\mathcal{N}=4$ $U(3)$}}
\left(t,x_1=q^{\frac14}t,x_2=q^{\frac34}t^3,x_3=q^{\frac54}t^5;q\right)
\nonumber\\
&=\mathbb{II}_{\textrm{Nahm}}^{\textrm{4d $\mathcal{N}=4$ $U(3)$}}(t;q)
\mathcal{I}^{\textrm{3d HM}}(t,x=q^{\frac14}t;q)
\mathcal{I}^{\textrm{3d HM}}(t,x=q^{\frac34}t^3;q). 
\end{align}
In this case, the Higgsed Dirichlet half-index only depends on the ratio $x_{i+1}/x_{i}$ 
and therefore the specialization is not unique. 
For example, we can also get the same Higgsed Dirichlet half-index by taking
\begin{align}
x_1&=q^{\frac54}t,\qquad 
x_2=q^{\frac74}t^3,\qquad 
x_3=q^{\frac94}t^5. 
\end{align}
As we will see, this corresponds to the Higgsed Dirichlet half-index for the two-point function of the 't Hooft lines of magnetic charge $B=(1,1,1)$ with the Nahm pole b.c. 
For $U(3)$ gauge theory, it is trivial so that the two-point function is equivalent to the Nahm pole half-index (\ref{u3Nahm}) with empty line. 

\subsubsection{Fundamental Wilson line}
Consider the boundary Wilson line in the fundamental representation of $U(3)$ gauge group. 
One finds non-trivial line defect half-indices of the fundamental Wilson lines with two or more insertions. 
The two-point function of the boundary Wilson lines in the (anti)fundamental representation with the Neumann b.c. is evaluated as
\begin{align}
&
\label{u3N_W1}
\langle W_{\tiny \yng(1)} W_{\overline{\tiny \yng(1)}}\rangle_{\mathcal{N}}^{\textrm{4d $\mathcal{N}=4$ $U(3)$}}(t;q)
\nonumber\\
&=\frac{1}{3!} \frac{(q)_{\infty}^3}{(q^{\frac12}t^{-2};q)_{\infty}^3}
\oint \prod_{i=1}^3 \frac{ds_i}{2\pi is_i} 
\prod_{i\neq j} \frac{\left( \frac{s_i}{s_j};q \right)_{\infty}}{\left(q^{\frac12}t^{-2}\frac{s_i}{s_j};q \right)_{\infty}}
(s_1+s_2+s_3)(s_1^{-1}+s_2^{-1}+s_3^{-1}). 
\end{align}
Again the boundary Wilson line in the fundamental representation with the Neumann b.c. 
is conjecturally dual to the boundary 't Hooft line of magnetic charge $B=(1,0,0)$ with the Nahm pole boundary condition. 
To perform the Higgsing procedure, 
we specialize the global fugacities of the Dirichlet half-index (\ref{u3D}) as
\begin{align}
x_1&=q^{\frac14}t,\qquad 
x_2=q^{\frac34}t^3, \qquad
x_3=q^{\frac94}t^5. 
\end{align}
Then we get
\begin{align}
&
\mathbb{II}_{\mathcal{D}}^{\textrm{4d $\mathcal{N}=4$ $U(3)$}}
\left(t,x_1=q^{\frac14}t,x_2=q^{\frac34}t^3,x_3=q^{\frac94}t^5;q\right)
\nonumber\\
&=\langle T_{(1,0,0)}T_{(1,0,0)}\rangle_{\textrm{Nahm}}^{\textrm{4d $\mathcal{N}=4$ $U(3)$}}(t;q)
\nonumber\\
&\times 
\mathcal{I}^{\textrm{3d HM}}(t,x=q^{\frac14}t;q)
\mathcal{I}^{\textrm{3d HM}}(t,x=q^{\frac54}t;q)
\mathcal{I}^{\textrm{3d HM}}(t,x=q^{\frac74}t^3;q). 
\end{align}
Here there are three deformed hypermultiplet indices with non-standard R-symmetry assignment. 
After stripping off them, we find the two-point function of the boundary 't Hooft line of magnetic charge $B=(1,0,0)$ with the Nahm pole boundary condition of the form 
\begin{align}
\label{u3Nahm_T1}
\langle T_{(1,0,0)}T_{(1,0,0)}\rangle_{\textrm{Nahm}}^{\textrm{4d $\mathcal{N}=4$ $U(3)$}}(t;q)
&=\frac{(q)_{\infty}^2 (q^{\frac32}t^2;q)_{\infty}^2 (q^3t^{4};q)_{\infty}}
{(q^{\frac12}t^{2};q)_{\infty}^2 (qt^{4};q)_{\infty} (q^2;q)_{\infty} (q^{\frac52}t^{6};q)_{\infty}}. 
\end{align}
It follows that 
the two-point functions (\ref{u3N_W1}) and (\ref{u3Nahm_T1}) agree with each other under $t$ $\rightarrow$ $t^{-1}$
\begin{align}
\langle W_{\tiny \yng(1)} W_{\overline{\tiny \yng(1)}}\rangle_{\mathcal{N}}^{\textrm{4d $\mathcal{N}=4$ $U(3)$}}(t;q)
&=\langle T_{(1,0,0)}T_{(1,0,0)}\rangle_{\textrm{Nahm}}^{\textrm{4d $\mathcal{N}=4$ $U(3)$}}(t^{-1};q). 
\end{align}
Note that the result (\ref{u3Nahm_T1}) can be also obtained from the Higgsing process with the following specialization: 
\begin{align}
\label{u3spe_NahmT11}
x_1&=q^{\frac14}t,\qquad 
x_2=q^{\frac74}t^3, \qquad
x_3=q^{\frac{9}{4}}t^5.
\end{align}
This can be regarded as the Higgsed Dirichlet half-index for the two-point function of the 't Hooft lines of magnetic charge $B=(1,1,0)$ with the Nahm pole b.c., 
which is dual to the rank-$2$ antisymmetric Wilson lines with the Neumann b.c. 
For $U(3)$ gauge theory, it is equivalent to the 't Hooft line of magnetic charge $B=(1,0,0)$ or the fundamental Wilson line 
so that the Higgsed Dirichlet half-index triggered by the specialization (\ref{u3spe_NahmT11}) simply generates the two-point function (\ref{u3Nahm_T1}). 

\subsubsection{Symmetric Wilson lines}
Although the boundary Wilson lines in the symmetric representation is not simply obtained via the Higgsing procedure, 
they also possess beautiful closed-form expressions for the line defect half-indices. 
While the one-point function of the boundary Wilson line in the rank-$2$ symmetric representation of $U(3)$ gauge group vanishes, the two-point function is non-trivial. 
We find that it can be written as
\begin{align}
\label{u3N_Wsym2}
&
\langle W_{\tiny \yng(2)} W_{\overline{\tiny \yng(2)}}\rangle_{\mathcal{N}}^{\textrm{4d $\mathcal{N}=4$ $U(3)$}}(t;q)
\nonumber\\
&=\frac{(q)_{\infty}^2(q^{\frac32}t^{-2};q)_{\infty}^2(q^3t^{-4};q)_{\infty}}
{(q^{\frac52}t^{-2};q)_{\infty}^2(qt^{-4};q)_{\infty}(q^2;q)_{\infty}(q^{\frac52}t^{-6};q)_{\infty}}
\nonumber\\
&+qt^{-4}\frac{(q^{\frac12}t^2;q)_{\infty}(q)_{\infty}^2(q^{\frac32}t^{-2};q)_{\infty}(q^{\frac52}t^{-2};q)_{\infty}(q^4;q)_{\infty}(q^4t^{-4};q)_{\infty}}
{(q^{\frac12}t^{-2};q)_{\infty}^2(qt^{-4};q)_{\infty}(q^{\frac32}t^{2};q)_{\infty}(q^{\frac52}t^{-6};q)_{\infty}(q^3;q)_{\infty}(q^4;q)_{\infty}}. 
\end{align}
The first term is the two-point function (\ref{u3N_W1}) of the boundary Wilson lines in the fundamental representation. 
The second term is identified with the three-point function of the boundary Wilson lines carrying charges $+1$, $+1$ and $-2$
\begin{align}
&
\langle W_{1} W_{1} W_{-2} \rangle_{\mathcal{N}}^{\textrm{4d $\mathcal{N}=4$ $U(3)$}}(t;q)
\nonumber\\
&=qt^{-4}\frac{(q^{\frac12}t^2;q)_{\infty}(q)_{\infty}^2(q^{\frac32}t^{-2};q)_{\infty}(q^{\frac52}t^{-2};q)_{\infty}(q^4;q)_{\infty}(q^4t^{-4};q)_{\infty}}
{(q^{\frac12}t^{-2};q)_{\infty}^2(qt^{-4};q)_{\infty}(q^{\frac32}t^{2};q)_{\infty}(q^{\frac52}t^{-6};q)_{\infty}(q^3;q)_{\infty}(q^4;q)_{\infty}}. 
\end{align}

\subsection{$\mathfrak{u}(N)$}
Now we would like to propose general results for $U(N)$ gauge theories. 
We begin by recalling the ordinary half-indices without any line operator. 
The half-index of the Neumann b.c. (\ref{NeuBC}) of $U(N)$ SYM theory is given by
\begin{align}
\label{uNN}
\mathbb{II}_{\mathcal{N}}^{\textrm{4d $\mathcal{N}=4$ $U(N)$}}(t;q)
&=\frac{1}{N!}
\frac{(q)_{\infty}^N}{(q^{\frac12}t^{-2};q)_{\infty}^N}
\oint 
\prod_{i=1}^N \frac{ds_i}{2\pi is_i}
\prod_{i\neq j} \frac{\left( \frac{s_i}{s_j};q \right)_{\infty}}{\left(q^{\frac12}t^{-2}\frac{s_i}{s_j};q \right)_{\infty}}. 
\end{align}
On the other hand, the half-index of the regular Nahm pole boundary condition in $U(N)$ gauge theory is \cite{Gaiotto:2019jvo}
\begin{align}
\label{uNNahm}
\mathbb{II}_{\textrm{Nahm}}^{\textrm{4d $\mathcal{N}=4$ $U(N)$}}(t;q)
&=\prod_{k=1}^{N}
\frac{(q^{\frac{1+k}{2}}t^{2(k-1)};q)_{\infty}}
{(q^{\frac{k}{2}}t^{2k};q)_{\infty}}. 
\end{align} 
The Neumann half-index (\ref{uNN}) and the Nahm pole half-index (\ref{uNNahm}) agree with each other under $t$ $\rightarrow$ $t^{-1}$ \cite{Gaiotto:2019jvo}. 
As the Nahm pole b.c. can be viewed as the deformed Dirichlet b.c. with a regular nilpotent vev for the scalar fields, 
the Nahm pole half-index (\ref{uNNahm}) follows from the half-index of the Dirichlet b.c. with the form \cite{Gaiotto:2019jvo}
\begin{align}
\label{uND}
\mathbb{II}_{\mathcal{D}}^{\textrm{4d $\mathcal{N}=4$ $U(N)$}}(t,x_i;q)
&=
\frac{(q)_{\infty}^{N}}{(q^{\frac12}t^2;q)_{\infty}^N}
\prod_{i\neq j}
\frac{(q\frac{x_i}{x_j};q)_{\infty}}
{(q^{\frac12}t^2 \frac{x_i}{x_j};q)_{\infty}}
\end{align}
by means of the Higgsing procedure with the specialized global fugacities
\begin{align}
x_i&=q^{\frac{2i-1}{4}}t^{2i-1}. 
\end{align}
The Higgsed Dirichlet half-index is given by
\begin{align}
&
\mathbb{II}_{\mathcal{D}}^{\textrm{4d $\mathcal{N}=4$ $U(N)$}}
\left(t,x_i=q^{\frac{2i-1}{4}}t^{2i-1};q\right)
\nonumber\\
&=\mathbb{II}_{\textrm{Nahm}}^{\textrm{4d $\mathcal{N}=4$ $U(N)$}}(t;q)
\prod_{i=1}^{N-1}
{\mathcal{I}^{\textrm{3d HM}}(t,x=q^{\frac{2i-1}{4}}t^{2i-1};q)}^{N-i}. 
\end{align}
In this Higgsing process, one finds $N(N-1)/2$ decoupled 3d matter fields. 

\subsubsection{Fundamental Wilson line}
Let us consider line operators in the presence of the Neumann b.c. (\ref{NeuBC}). 
According to the Gauss law constraint at the boundary, there is no non-trivial one-point function of the boundary Wilson line for $U(N)$ gauge theory. 
So the basic example is the two-point function of a pair of the boundary Wilson lines in the (anti)fundamental representation. 
It can be computed as
\begin{align}
\label{uNN_W1}
&
\langle W_{\tiny \yng(1)} W_{\overline{\tiny \yng(1)}}\rangle_{\mathcal{N}}^{\textrm{4d $\mathcal{N}=4$ $U(N)$}}(t;q)
\nonumber\\
&=\frac{1}{N!} \frac{(q)_{\infty}^N}{(q^{\frac12}t^{-2};q)_{\infty}^N}
\oint \prod_{i=1}^N \frac{ds_i}{2\pi is_i} 
\prod_{i\neq j} \frac{\left( \frac{s_i}{s_j};q \right)_{\infty}}{\left(q^{\frac12}t^{-2}\frac{s_i}{s_j};q \right)_{\infty}}
(\sum_i s_i)(\sum_i s_i^{-1}). 
\end{align}
The boundary Wilson line in the fundamental representation with the Neumann b.c. is 
conjecturally dual to the 't Hooft line of magnetic charge $B=(1,0,\cdots,0)$ with the Nahm pole b.c. 
This corresponds to the minuscule representation without any monopole bubbling effect. 
Here we propose that
the two-point function of the dual  't Hooft lines by applying the Higgsing procedure \cite{Gaiotto:2019jvo}. 
Let us specialize the global fugacities of the Dirichlet half-index (\ref{uND}) as
\begin{align}
x_i&=q^{\frac{2i-1}{4}}t^{2i-1},\qquad i=1,\cdots, N-1,\nonumber\\
x_N&=q^{\frac{2N+3}{4}}t^{2N-1}. 
\end{align}
Then we get the Higgsed Dirichlet half-index 
\begin{align}
&
\mathbb{II}_{\mathcal{D}}^{\textrm{4d $\mathcal{N}=4$ $U(N)$}}
\left(t,\left\{x_i=q^{\frac{2i-1}{4}}t^{2i-1}\right\}_{i=1}^{N-1}, x_N=q^{\frac{2N+3}{4}}t^{2N-1};q\right)
\nonumber\\
&=\langle T_{(1,0,\cdots,0)}T_{(1,0,\cdots,0)}\rangle_{\textrm{Nahm}}^{\textrm{4d $\mathcal{N}=4$ $U(N)$}}(t;q)
\nonumber\\
&\times 
\prod_{i=1}^{N-2}
{\mathcal{I}^{\textrm{3d HM}}(t,x=q^{\frac{2i-1}{4}}t^{2i-1};q)}^{N-1-i}
\prod_{i=1}^{N-1}\mathcal{I}^{\textrm{3d HM}}(t,x=q^{\frac{2i+1}{4}}t^{2i-1};q). 
\end{align}
By stripping off the indices of the decoupled 3d matters, 
we find the two-point function of the dual  't Hooft lines of the form 
\begin{align}
\label{uNNahm_T1}
&
\langle T_{(1,0,\cdots,0)}T_{(1,0,\cdots,0)}\rangle_{\textrm{Nahm}}^{\textrm{4d $\mathcal{N}=4$ $U(N)$}}(t;q)
\nonumber\\
&=\frac{(q)_{\infty}(q^{\frac32}t^2;q)_{\infty}(q^{\frac{3+N}{2}}t^{2(N-1)};q)_{\infty}}
{(q^{\frac12}t^2;q)_{\infty}(q^2;q)_{\infty}(q^{\frac{2+N}{2}}t^{2N};q)_{\infty}}
\prod_{k=1}^{N-1}
\frac{(q^{\frac{k+1}{2}}t^{2(k-1)};q)_{\infty}}
{(q^{\frac{k}{2}}t^{2k};q)_{\infty}}. 
\end{align}
For example, for $N=4,5$ we have
\begin{align}
\label{u4Nahm_T1}
&
\langle T_{(1,0,0,0)}T_{(1,0,0,0)}\rangle_{\textrm{Nahm}}^{\textrm{4d $\mathcal{N}=4$ $U(4)$}}(t;q)
\nonumber\\
&=\frac{(q)_{\infty}^2(q^{\frac32}t^2;q)_{\infty}^2(q^2t^4;q)_{\infty}(q^{\frac72}t^6;q)_{\infty}}
{(q^{\frac12}t^2;q)_{\infty}(qt^4;q)_{\infty}(q^{\frac32}t^6;q)_{\infty}(q^2;q)_{\infty}(q^3t^6;q)_{\infty}}, \\
\label{u5Nahm_T1}
&
\langle T_{(1,0,0,0,0)}T_{(1,0,0,0,0)}\rangle_{\textrm{Nahm}}^{\textrm{4d $\mathcal{N}=4$ $U(5)$}}(t;q)
\nonumber\\
&=\frac{(q)_{\infty}^2(q^{\frac32}t^2;q)_{\infty}^2(q^2t^4;q)_{\infty}(q^{\frac52}t^6;q)_{\infty}(q^4t^8;q)_{\infty}}
{(q^{\frac12}t^2;q)_{\infty}(qt^4;q)_{\infty}(q^{\frac32}t^6;q)_{\infty}(q^2;q)_{\infty}(q^2t^8;q)_{\infty}(q^{\frac72}t^{10};q)_{\infty}}. 
\end{align}
As a result of S-duality of line operators, 
we propose that the two-point function (\ref{uNN_W1}) of the boundary Wilson lines in the fundamental representation with the Neumann b.c. 
and the two-point function (\ref{uNNahm_T1}) of the boundary 't Hooft lines of magnetic charge $B$ $=$ $(1,0,\cdots,0)$ with the Nahm pole b.c. 
precisely agree with each other under $t$ $\rightarrow$ $t^{-1}$
\begin{align}
\langle W_{\tiny \yng(1)} W_{\overline{\tiny \yng(1)}}\rangle_{\mathcal{N}}^{\textrm{4d $\mathcal{N}=4$ $U(N)$}}(t;q)
&=
\langle T_{(1,0,\cdots,0)}T_{(1,0,\cdots,0)}\rangle_{\textrm{Nahm}}^{\textrm{4d $\mathcal{N}=4$ $U(N)$}}(t^{-1};q). 
\end{align}

\subsubsection{Antisymmetric Wilson lines}
Provided that $U(N)$ gauge theory is subject to the Neumann b.c., 
one can introduce non-trivial Wilson lines in the rank-$k$ antisymmetric representation for $k\le N$. 
The two-point function of the boundary Wilson lines in the rank-$k$ antisymmetric representation with the Neumann b.c. is given by 
\begin{align}
\label{uNN_Wasymk}
&
\langle W_{(1^k)} W_{\overline{(1^k)}}\rangle_{\mathcal{N}}^{\textrm{4d $\mathcal{N}=4$ $U(N)$}}(t;q)
\nonumber\\
&=\frac{1}{N!} \frac{(q)_{\infty}^N}{(q^{\frac12}t^{-2};q)_{\infty}^N}
\oint \prod_{i=1}^N \frac{ds_i}{2\pi is_i} 
\prod_{i\neq j} \frac{\left( \frac{s_i}{s_j};q \right)_{\infty}}{\left(q^{\frac12}t^{-2}\frac{s_i}{s_j};q \right)_{\infty}}
e_k(s)e_k(s^{-1}),  
\end{align}
where 
\begin{align}
e_k(s)&=\sum_{1\le i_1<\cdots<i_k\le N}
s_{i_1}\cdots s_{i_k}
\end{align}
is the elementary symmetric function. 
The rank-$k$ antisymmetric Wilson line is expected to be dual to the 't Hooft line of magnetic charge $B=(1^k,0^{N-k})$. 
We claim that the two-point function of the dual 't Hooft lines with the Nahm pole b.c. can be obtained 
by applying the Higgsing manipulation to the Dirichlet half-index (\ref{uND}). 
By taking the global fugacities as
\begin{align}
x_i&=q^{\frac{2i-1}{4}}t^{2i-1},\qquad i=1,\cdots, N-k
\nonumber\\
x_j&=q^{\frac{2j+3}{4}}t^{2j-1}, \qquad j=N-k+1,\cdots, N, 
\end{align}
we find that the Dirichlet half-index (\ref{uND}) is factorized as
\begin{align}
&
\mathbb{II}_{\mathcal{D}}^{\textrm{4d $\mathcal{N}=4$ $U(N)$}}
\left(t,\left\{x_i=q^{\frac{2i-1}{4}}t^{2i-1}\right\}_{i=1}^{N-k}, 
\left\{x_j=q^{\frac{2j+3}{4}}t^{2j-1}\right\}_{j=N-k+1}^{N};q\right)
\nonumber\\
&=\langle T_{(1^k,0^{N-k})}T_{(1^k,0^{N-k})}\rangle_{\textrm{Nahm}}^{\textrm{4d $\mathcal{N}=4$ $U(N)$}}(t;q)
\nonumber\\
&\times 
\prod_{i=1}^{k-1}
{\mathcal{I}^{\textrm{3d HM}}(t,x=q^{\frac{2i-1}{4}}t^{2i-1};q)}^{k-i}
\prod_{i=1}^{N-k-1}
{\mathcal{I}^{\textrm{3d HM}}(t,x=q^{\frac{2i-1}{4}}t^{2i-1};q)}^{N-k-i}
\nonumber\\
&\times 
\prod_{i=1}^{k-1}
{\mathcal{I}^{\textrm{3d HM}}(t,x=q^{\frac{2i+k+1}{4}}t^{2i+k-3};q)}^{i}
\prod_{i=k}^{N-k}
{\mathcal{I}^{\textrm{3d HM}}(t,x=q^{\frac{2i+k+1}{4}}t^{2i+k-3};q)}^{k}
\nonumber\\
&\times 
\prod_{i=N-k+1}^{N-1}
{\mathcal{I}^{\textrm{3d HM}}(t,x=q^{\frac{2i+k+1}{4}}t^{2i+k-3};q)}^{N-i}. 
\end{align}
After taking off the indices for the decoupled 3d matters, 
we find the two-point function of the 't Hooft line of magnetic charge $B=(1^k,0^{N-k})$ for $U(N)$ gauge theory with the Nahm pole b.c. with the form 
\begin{align}
\label{uNNahm_T1^k}
&
\langle T_{(1^k,0^{N-k})}T_{(1^k,0^{N-k})}\rangle_{\textrm{Nahm}}^{\textrm{4d $\mathcal{N}=4$ $U(N)$}}(t;q)
\nonumber\\
&=
\prod_{l=1}^{N-k}
\frac{(q^{\frac{l+1}{2}}t^{2(l-1)};q)_{\infty}}
{(q^{\frac{l}{2}}t^{2l};q)_{\infty}}
\prod_{m=1}^{k}
\frac{(q^{\frac{m+1}{2}}t^{2(m-1)};q)_{\infty}}
{(q^{\frac{m}{2}}t^{2m};q)_{\infty}}
\frac{(q^{\frac{m+2}{2}}t^{2m};q)_{\infty}}
{(q^{\frac{m+3}{2}}t^{2(m-1)};q)_{\infty}}
\frac{(q^{\frac{m+N-k+3}{2}}t^{2m+2N-2k-2};q)_{\infty}}
{(q^{\frac{m+N-k+2}{2}}t^{2m+2N-2k};q)_{\infty}}. 
\end{align}
Note that 
the expression reduces to the Nahm pole half-index (\ref{uNNahm}) when $k=0$ and $k=N$ 
while it becomes the two-point function (\ref{uNNahm_T1}) of the 't Hooft lines of magnetic charge $B=(1,0,\cdots,0)$ for $k=1$, as expected. 

For example, for $(N,k)$ $=$ $(4,2)$, $(5,2)$, $(6,2)$ and $(6,3)$, 
we have
\begin{align}
&
\langle T_{(1,1,0,0)}T_{(1,1,0,0)}\rangle_{\textrm{Nahm}}^{\textrm{4d $\mathcal{N}=4$ $U(4)$}}(t;q)
\nonumber\\
&=\frac{(q)_{\infty}^2(q^{\frac32}t^2;q)_{\infty}^3(q^2t^4;q)_{\infty}(q^3t^4;q)_{\infty}(q^{\frac72}t^6;q)_{\infty}}
{(q^{\frac12}t^2;q)_{\infty}^2(qt^4;q)_{\infty}^2(q^2;q)_{\infty}(q^{\frac52}t^2;q)_{\infty}(q^{\frac52}t^6;q)_{\infty}(q^3t^8;q)_{\infty}}, \\
&
\langle T_{(1,1,0,0,0)}T_{(1,1,0,0,0)}\rangle_{\textrm{Nahm}}^{\textrm{4d $\mathcal{N}=4$ $U(5)$}}(t;q)
\nonumber\\
&=\frac{(q)_{\infty}^2(q^{\frac32}t^2;q)_{\infty}^3(q^2t^4;q)_{\infty}^2(q^{\frac72}t^6;q)_{\infty}(q^4t^8;q)_{\infty}}
{(q^{\frac12}t^2;q)_{\infty}^2(qt^4;q)_{\infty}^2(q^{\frac32}t^6;q)_{\infty}(q^2;q)_{\infty}(q^{\frac52}t^2;q)_{\infty}(q^3t^8;q)_{\infty}(q^{\frac72}t^{10};q)_{\infty}}, \\
&
\langle T_{(1,1,0,0,0,0)}T_{(1,1,0,0,0,0)}\rangle_{\textrm{Nahm}}^{\textrm{4d $\mathcal{N}=4$ $U(6)$}}(t;q)
\nonumber\\
&=\frac{(q)_{\infty}^2(q^{\frac32}t^2;q)_{\infty}^3(q^2t^4;q)_{\infty}^2(q^{\frac52}t^6;q)_{\infty}(q^4t^8;q)_{\infty}(q^{\frac92}t^{10};q)_{\infty}}
{(q^{\frac12}t^2;q)_{\infty}^2(qt^4;q)_{\infty}^2(q^{\frac32}t^6;q)_{\infty}(q^2;q)_{\infty}(q^2t^8;q)_{\infty}(q^{\frac52}t^2;q)_{\infty}(q^{\frac72}t^{10};q)_{\infty}(q^4t^{12};q)_{\infty}}, \\
&
\langle T_{(1,1,1,0,0,0)}T_{(1,1,1,0,0,0)}\rangle_{\textrm{Nahm}}^{\textrm{4d $\mathcal{N}=4$ $U(6)$}}(t;q)
\nonumber\\
&=\frac{(q)_{\infty}^2(q^{\frac32}t^2;q)_{\infty}^3(q^2t^4;q)_{\infty}^3(q^{\frac52}t^6;q)_{\infty}(q^{\frac72}t^6;q)_{\infty}(q^4t^8;q)_{\infty}(q^{\frac92}t^{10};q)_{\infty}}
{(q^{\frac12}t^2;q)_{\infty}^2(qt^4;q)_{\infty}^2(q^{\frac32}t^6;q)_{\infty}^2(q^2;q)_{\infty}(q^{\frac52}t^2;q)_{\infty}(q^3t^4;q)_{\infty}(q^3t^8;q)_{\infty}(q^{\frac72}t^{10};q)_{\infty}(q^5t^{12};q)_{\infty}}. 
\end{align}

S-duality of the boundary Wilson line in the rank-$k$ antisymmetric representation of $U(N)$ gauge theory satisfying the Neumann b.c. 
and the boundary 't Hooft line of magnetic charge $B=(1^k,0^{N-k})$ of $U(N)$ gauge theory obeying the Nahm pole b.c. 
would imply that 
\begin{align}
\langle W_{(1^k)} W_{\overline{(1^k)}}\rangle_{\mathcal{N}}^{\textrm{4d $\mathcal{N}=4$ $U(N)$}}(t;q)
&=\langle T_{(1^k,0^{N-k})}T_{(1^k,0^{N-k})}\rangle_{\textrm{Nahm}}^{\textrm{4d $\mathcal{N}=4$ $U(N)$}}(t^{-1};q). 
\end{align}


\subsubsection{Direct evaluation by Macdonald polynomials}
Next we analytically compute the matrix integral (\ref{uNN_Wasymk}) by using the norm formula of the Macdonald polynomials of type $A_{N-1}$. 
In the case of type $A_{N-1}$, the root system is given by
\begin{align}
R&=\{  \varepsilon_i - \varepsilon_j | 1\leq i\ne j \leq N \}, \\
R^+&=\{ \varepsilon_i - \varepsilon_j | 1\leq i<j \leq N \},
\end{align}
where $\varepsilon_i$ ($i=1,2,\dots, N$) are an orthonormal basis of $\mathbb{R}^N$.
Then,
\begin{align}
\rho=\frac{1}{2}\sum_{i=1}^N ( N-2i+1)\varepsilon_i,\qquad \lambda=\sum_{i=1}^N \lambda_i \varepsilon_i,
\end{align}
and
\begin{align}
w_{A_{N-1}}(s; \q,\t)=
\prod_{1\leq i \ne j \leq N} \frac{(s_i/s_j;\q)_\infty}{(\t s_i/s_j;\q)_\infty},
\end{align}
where $s_i=e^{\varepsilon_i}$.

The norm is given by
\begin{align}
\frac{\langle P_\lambda, P_\lambda \rangle}{\langle 1, 1 \rangle}=\prod_{1\leq i<j\leq N} \frac{(\t^{j-i+1};\q)_{\lambda_i-\lambda_j}(\t^{j-i-1}\q;\q)_{\lambda_i-\lambda_j}}{(\t^{j-i};\q)_{\lambda_i-\lambda_j}(\t^{j-i}\q;\q)_{\lambda_i-\lambda_j}},
\end{align}
and also we have
\begin{align}
\langle 1, 1 \rangle=
\prod_{k=1}^N \frac{(\t^{k-1}\q;\q)_\infty}{(\t^{k};\q)_\infty}.
\end{align}
In particular, for $\lambda=(1^k)$, we have $\chi_{(1^k)}=e_k=P_{(1^k)}$. Therefore we immediately find
\begin{align}
\label{uNN_Wasymk_exact}
\frac{\langle W_{(1^k)}W_{\overline{(1^k)}} \rangle_{\mathcal{N}}^{U(N)}}{\mathbb{II}_{\mathcal{N}}^{U(N)}}
=\frac{\langle P_{(1^k)}, P_{(1^k)} \rangle}{\langle 1, 1 \rangle}=\frac{(\q;\t)_k (\t^{N-k+1};\t)_k}{(\t;\t)_k (\t^{N-k}\q;\q)_k}.
\end{align}
We have confirmed that 
the formula (\ref{uNN_Wasymk_exact}) gives rise to the same results as (\ref{uNN_W1}) for $k=1$ and (\ref{uNN_Wasymk}) for general $k$. 
Also it agrees with the expressions (\ref{uNNahm_T1}) and (\ref{uNNahm_T1^k}) obtained from the Higgsing method 
upon the transformation $t$ $\rightarrow t^{-1}$. 

The Higgsing procedure does not simply work for the case involving the non-minuscule module. 
Even for such cases, we can still calculate the line defect half-indices by using the inner product of the Macdonald polynomials. 
Here we give an example for the rank-$2$ symmetric representation. 
Since, for $\lambda=(2)$, we have
\begin{align}
\chi_{(2)}=h_2=P_{(2)}+\frac{\t-\q}{1-\q \t} P_{(1^2)},
\end{align}
we obtain
\begin{equation}
\label{uNN_Wsym}
\begin{aligned}
\frac{\langle W_{(2)}W_{\overline{(2)}} \rangle_{\mathcal{N}}^{U(N)}}{\mathbb{II}_{\mathcal{N}}^{U(N)}}
=\frac{(1-\q)(1-\t^N)}{(1-\t)(1-\q \t)(1-\q \t^{N-1})}
\biggl[ \frac{(1-\q^2)(1-\q\t^N)}{1-\q^2 \t^{N-1}}+\frac{(\t-\q)^2(1-\t^{N-1})}{(1-\t^2)(1-\q \t^{N-2})}\biggr].
\end{aligned}
\end{equation}
One can check that 
the expression (\ref{uNN_Wsym}) reproduces the results (\ref{u2N_Wsym2}) and (\ref{u3N_Wsym2}). 

\section{$\mathfrak{so}(2N+1)$}
\label{sec_so2N+1}

\subsection{$\mathfrak{so}(3)$}
\label{sec_so3}
The half-index of the Neumann b.c. for $SO(3)$ gauge theory takes the form 
\begin{align}
\label{so3N}
\mathbb{II}_{\mathcal{N}}^{\textrm{4d $\mathcal{N}=4$ $SO(3)$}}(t;q)
&=\frac12 \frac{(q)_{\infty}}{(q^{\frac12}t^{-2};q)_{\infty}}
\oint \frac{ds}{2\pi is}
\frac{(s^{\pm};q)_{\infty}}{(q^{\frac12}t^{-2}s^{\pm};q)_{\infty}}. 
\end{align}
Under the transformation $t$ $\rightarrow$ $t^{-1}$, 
it agrees with the $USp(2)$ Nahm pole half-index \cite{Hatsuda:2024lcc}
\begin{align}
\label{usp2Nahm}
\mathbb{II}_{\textrm{Nahm}}^{\textrm{4d $\mathcal{N}=4$ $USp(2)$}}(t;q)
&=\frac{(q^{\frac32}t^{2};q)_{\infty}}{(qt^{4};q)_{\infty}}. 
\end{align}
The half-index of the Dirichlet b.c. for $USp(2)$ gauge theory takes the form
\begin{align}
\label{usp2D}
\mathbb{II}_{\mathcal{D}}^{\textrm{4d $\mathcal{N}=4$ $USp(2)$}}(t,x;q)
&=
\frac{(q)_{\infty}}{(q^{\frac12}t^2;q)_{\infty}}
\frac{(qx^2;q)_{\infty}(qx^{-2};q)_{\infty}}
{(q^{\frac12}t^2x^2;q)_{\infty}(q^{\frac12}t^2x^{-2};q)_{\infty}}. 
\end{align}
By identifying the global fugacity of the $USp(2)$ Dirichlet half-index (\ref{usp2D}) as $x=q^{1/4}t$ and stripping off the 3d index, 
one finds the Nahm pole half-index (\ref{usp2Nahm}) from the Higgsed Dirichlet half-index
\begin{align}
&
\mathbb{II}_{\mathcal{D}}^{\textrm{4d $\mathcal{N}=4$ $USp(2)$}}
\left(t,x=q^{\frac14}t;q\right)
\nonumber\\
&=\mathbb{II}_{\textrm{Nahm}}^{\textrm{4d $\mathcal{N}=4$ $USp(2)$}}(t;q)
\mathcal{I}^{\textrm{3d HM}}(t,x=q^{\frac14}t;q). 
\end{align}

\subsubsection{Spinor Wilson line}
Let us consider the boundary Wilson line in the spinor representation of $Spin(3)$ gauge theory obeying the Neumann b.c. 
While the one-point function vanishes, the two-point function of a pair of the spinor Wilson lines is non-trivial. 
It is evaluated as the matrix integral of the form
\begin{align}
\label{so3N_Wsp}
&
\langle W_{\textrm{sp}} W_{\textrm{sp}}\rangle_{\mathcal{N}}^{\textrm{4d $\mathcal{N}=4$ $Spin(3)$}}(t;q)
\nonumber\\
&=\frac12 \frac{(q)_{\infty}}{(q^{\frac12}t^{-2};q)_{\infty}}
\oint \frac{ds}{2\pi is}
\frac{(s^{\pm};q)_{\infty}}{(q^{\frac12}t^{-2}s^{\pm};q)_{\infty}}
(2+s+s^{-1}). 
\end{align}
The Wilson line in the spinor representation of $\mathcal{N}=4$ $Spin(3)$ gauge theory is conjecturally dual to 
the 't Hooft line of magnetic charge $B=(\frac12)$ in $USp(2)/\mathbb{Z}_2$ gauge theory, 
which corresponds to the minuscule representation. 
In order to derive the two-point function of the dual 't Hooft lines of $USp(2)/\mathbb{Z}_2$ gauge theory with the Nahm pole b.c., 
we apply the Higgsing procedure to the $USp(2)$ Dirichlet half-index (\ref{usp2D}). 
Let us specialize the global fugacity $x$ of the $USp(2)$ Dirichlet half-index as $x$ $=$ $q^{3/4}t$. 
Consequently, we find 
\begin{align}
&
\mathbb{II}_{\mathcal{D}}^{\textrm{4d $\mathcal{N}=4$ $USp(2)$}}
\left(t,x=q^{\frac34}t;q\right)
\nonumber\\
&=\langle T_{(\frac12)}T_{(\frac12)}\rangle_{\textrm{Nahm}}^{\textrm{4d $\mathcal{N}=4$ $USp(2)/\mathbb{Z}_2$}}(t;q)
\mathcal{I}^{\textrm{3d HM}}(t,x=q^{\frac54}t;q). 
\end{align}
By stripping off the index of the 3d matter, 
we obtain the two-point function of the boundary 't Hooft line operators in the $USp(2)/\mathbb{Z}_2$ gauge theory 
with the Nahm pole boundary condition of the form 
\begin{align}
\label{usp2Nahm_T1/2}
\langle T_{(\frac12)}T_{(\frac12)}\rangle_{\textrm{Nahm}}^{\textrm{4d $\mathcal{N}=4$ $USp(2)/\mathbb{Z}_2$}}(t;q)
&=\frac{(q)_{\infty} (q^{\frac32}t^{2};q)_{\infty}}{(q^{\frac12}t^{2};q)_{\infty}(q^2;q)_{\infty}}
\frac{(q^{\frac52}t^{2};q)_{\infty}}{(q^2t^{4};q)_{\infty}}. 
\end{align}
We find that 
the two-point functions (\ref{so3N_Wsp}) and (\ref{usp2Nahm_T1/2}) agree with each other. 
This supports the duality of the boundary Wilson line in the spinor Wilson line for $Spin(3)$ gauge theory 
and the boundary 't Hooft line of magnetic charge $B=(\frac12)$ for $USp(2)/\mathbb{Z}_2$ gauge theory. 

\subsubsection{Fundamental Wilson line}
The Wilson line transforming in the representation other than the fundamental representation is not dual to the 't Hooft line of magnetic charge 
associated with the minuscule representation. 
Nevertheless, we can still find the exact closed-form for the line defect half-indices. 

For the fundamental Wilson line, there is non-trivial one-point function. 
It follows from the closed-form expressions (\ref{usp2Nahm}) and (\ref{usp2Nahm_T1/2}) as well as the group character relation 
that it can be written as 
\begin{align}
\label{so3N_W1_1pt}
\langle W_{\tiny \yng(1)}\rangle_{\mathcal{N}}^{\textrm{4d $\mathcal{N}=4$ $SO(3)$}}(t;q)
&=q^{\frac12}t^{-2}\frac{(q^{\frac12}t^2;q)_{\infty}}{(q^{\frac32}t^2;q)_{\infty}}
\frac{(q^{\frac52}t^{-2};q)_{\infty}}{(qt^{-4};q)_{\infty}}. 
\end{align}
Furthermore, we find the closed-form expressions for the multi-point functions of the fundamental Wilson lines 
in the presence of the Neumann b.c. 
The two-point function is given by
\begin{align}
&
\label{so3N_W1_2pt}
\langle W_{\tiny \yng(1)}W_{\tiny \yng(1)} \rangle_{\mathcal{N}}^{\textrm{4d $\mathcal{N}=4$ $SO(3)$}}(t;q)
\nonumber\\
&=\frac{(q^{\frac32}t^{-2};q)_{\infty}}
{(qt^{-4};q)_{\infty}}
+q^{\frac12}t^{-2}
\frac{(q^{\frac12}t^2;q)_{\infty}(q^{\frac52}t^{-2};q)_{\infty}}
{(qt^{-4};q)_{\infty}(q^{\frac32}t^{2};q)_{\infty}}
+qt^{-4}\frac{(q^{\frac12}t^2;q)_{\infty}(q^{\frac72}t^{-2};q)_{\infty}}
{(q^{\frac52}t^2;q)_{\infty}(qt^{-4};q)_{\infty}}. 
\end{align}
Here the first term is the $SO(3)$ Neumann half-index (\ref{so3N}) 
and the second is the one-point function (\ref{so3N_W1_1pt}). 
The third term turns out to be the one-point function (\ref{so3N_Wsym_1pt}) of the rank-$2$ symmetric Wilson line. 
More generally, the $k$-point function can be expressed as
\begin{align}
\label{so3N_W1_kpt}
\langle \underbrace{W_{\tiny \yng(1)}\cdots W_{\tiny \yng(1)}}_{k} \rangle_{\mathcal{N}}^{\textrm{4d $\mathcal{N}=4$ $SO(3)$}}(t;q)
&=\sum_{i=0}^k 
C_{k,i}
q^{\frac{i}{2}}
t^{-2i}
\frac{(q^{\frac{i}{2}}t^2;q)_{\infty}(q^{\frac{2i+3}{2}}t^{-2};q)_{\infty}}
{(q^{\frac{2i+1}{2}}t^2;q)_{\infty}(qt^{-4};q)_{\infty}}, 
\end{align}
where 
\begin{align}
C_{k,i}&=
T(k,i)-T(k,i-1)
\nonumber\\
&=
\sum_{j=0}^{k}
\left(
\begin{matrix}
k\\
j\\
\end{matrix}
\right)
\left[
\left(
\begin{matrix}
j\\
i-j\\
\end{matrix}
\right)
-
\left(
\begin{matrix}
j\\
i-1-j\\
\end{matrix}
\right)
\right]
\end{align}
is the difference of the trinomial coefficients $T(n,k)$ \cite{euler1805disquitiones} generated by
\begin{align}
(1+x+x^2)^n&=\sum_{k=0}^{2n}T(n,k)x^k. 
\end{align}
The $j$-th term in the expression (\ref{so3N_W1_kpt}) 
is the one-point function (\ref{so3N_Wsym_1pt}) of the boundary Wilson line in the rank-$k$ symmetric representation. 

\subsubsection{Symmetric Wilson lines}
The expression (\ref{so3N_W1_1pt}) for the one-point function of the boundary Wilson line in the fundamental representation 
can be generalized to the symmetric representations. 
We find that the one-point function of the Wilson line in the rank-$k$ symmetric representation can be written as
\begin{align}
\label{so3N_Wsym_1pt}
\langle W_{(k)}\rangle_{\mathcal{N}}^{\textrm{4d $\mathcal{N}=4$ $SO(3)$}}(t;q)
&=q^{\frac{k}{2}}t^{-2k}
\frac{(q^{\frac12}t^2;q)_{\infty}(q^{\frac{2k+3}{2}}t^{-2};q)_{\infty}}
{(q^{\frac{2k+1}{2}}t^2;q)_{\infty}(qt^{-4};q)_{\infty}}. 
\end{align}

\subsection{$\mathfrak{so}(5)$}
Next consider the gauge theory associated with the $\mathfrak{so}(5)$ gauge algebra. 
The $SO(5)$ Neumann half-index reads 
\begin{align}
\label{so5N}
&\mathbb{II}_{\mathcal{N}}^{\textrm{4d $\mathcal{N}=4$ $SO(5)$}}(t;q)
\nonumber\\
&=\frac{1}{8}\frac{(q)_{\infty}^2}{(q^{\frac12}t^{-2})^2}
\oint \prod_{i=1}^2 \frac{ds_i}{2\pi is_i}
\frac{(s_i^{\pm};q)_{\infty}}
{(q^{\frac12}t^{-2}s_i^{\pm};q)_{\infty}}
\prod_{i<j}
\frac{(s_i^{\pm}s_j^{\mp};q)_{\infty}(s_i^{\pm}s_j^{\pm};q)_{\infty}}
{(q^{\frac12}t^{-2}s_i^{\pm}s_j^{\mp};q)_{\infty}(q^{\frac12}t^{-2}s_i^{\pm}s_j^{\pm};q)_{\infty}}. 
\end{align}
The half-index of the dual Nahm pole b.c. in $USp(4)$ gauge theory is given by \cite{Hatsuda:2024lcc}
\begin{align}
\label{usp4Nahm}
\mathbb{II}_{\textrm{Nahm}}^{\textrm{4d $\mathcal{N}=4$ $USp(4)$}}(t;q)
&=\frac{(q^{\frac32}t^{2};q)_{\infty} (q^{\frac52}t^{6};q)_{\infty}}
{(qt^{4};q)_{\infty}(q^2t^{8};q)_{\infty}}. 
\end{align}
As discussed in \cite{Hatsuda:2024lcc}, 
the $SO(5)$ Neumann half-index (\ref{so5N}) is shown to be identical to the $USp(4)$ Nahm pole half-index (\ref{usp4Nahm}) after flipping $t$ to $t^{-1}$. 
The half-index of the Dirichlet b.c. for $USp(4)$ gauge theory reads
\begin{align}
\label{usp4D}
&
\mathbb{II}_{\mathcal{D}}^{\textrm{4d $\mathcal{N}=4$ $USp(4)$}}(t,x_1,x_2;q)
\nonumber\\
&=\frac{(q)_{\infty}^2}{(q^{\frac12}t^2;q)_{\infty}^2}
\prod_{i=1}^2
\frac{(qx_i^{\pm 2};q)_{\infty}}
{(q^{\frac12}t^2 x_i^{\pm 2};q)_{\infty}}
\prod_{i<j}
\frac{(qx_i^{\pm}x_j^{\mp};q)_{\infty}(qx_i^{\pm}x_j^{\pm};q)_{\infty}}
{(q^{\frac12}t^{2}x_i^{\pm}x_j^{\mp};q)_{\infty}(q^{\frac12}t^{2}x_i^{\pm}x_j^{\pm};q)_{\infty}}. 
\end{align}
It follows \cite{Hatsuda:2024lcc} that 
the $USp(4)$ Nahm pole half-index (\ref{usp4Nahm}) can be obtained from 
the Higgsed Dirichlet half-index with 
\begin{align}
x_1&=q^{\frac14}t, \qquad 
x_2=q^{\frac34}t^3. 
\end{align}
In other words, we have
\begin{align}
&
\mathbb{II}_{\mathcal{D}}^{\textrm{4d $\mathcal{N}=4$ $USp(4)$}}
\left(t,x_1=q^{\frac14}t, x_2=q^{\frac34}t^3;q\right)
\nonumber\\
&=\mathbb{II}_{\textrm{Nahm}}^{\textrm{4d $\mathcal{N}=4$ $USp(4)$}}(t;q)
\mathcal{I}^{\textrm{3d HM}}(t,x=q^{\frac14}t;q)^2 
\nonumber\\
&\times \mathcal{I}^{\textrm{3d HM}}(t,x=q^{\frac34}t^3;q)
\mathcal{I}^{\textrm{3d HM}}(t,x=q^{\frac54}t^5;q). 
\end{align}

\subsubsection{Spinor Wilson line}
For the spinor Wilson line, the one-point function is trivial. 
The two-point function of the boundary Wilson lines in the spinor representation of $Spin(5)$ gauge theory 
obeying the Neumann b.c. takes the from 
\begin{align}
\label{so5N_Wsp}
&
\langle W_{\textrm{sp}} W_{\textrm{sp}}\rangle_{\mathcal{N}}^{\textrm{4d $\mathcal{N}=4$ $Spin(5)$}}(t;q)
\nonumber\\
&=\frac{1}{8}\frac{(q)_{\infty}^2}{(q^{\frac12}t^{-2})^2}
\oint \prod_{i=1}^2 \frac{ds_i}{2\pi is_i}
\frac{(s_i^{\pm};q)_{\infty}}
{(q^{\frac12}t^{-2}s_i^{\pm};q)_{\infty}}
\prod_{i<j}
\frac{(s_i^{\pm}s_j^{\mp};q)_{\infty}(s_i^{\pm}s_j^{\pm};q)_{\infty}}
{(q^{\frac12}t^{-2}s_i^{\pm}s_j^{\mp};q)_{\infty}(q^{\frac12}t^{-2}s_i^{\pm}s_j^{\pm};q)_{\infty}}
\nonumber\\
&\times 
(4+2(s_1+s_2+s_1^{-1}+s_2^{-1})+s_1s_2+s_1^{-1}s_2^{-1}+s_1s_2^{-1}+s_1^{-1}s_2). 
\end{align}
The spinor Wilson line in $Spin(5)$ gauge theory is conjectured to be S-dual to 
the 't Hooft line of magnetic charge magnetic charge $B=(\frac12,\frac12)$ in $USp(4)/\mathbb{Z}_2$ gauge theory. 
Again this corresponds to the minuscule representation. 
We propose that the two-point function of the 't Hooft line operator with the regular Nahm pole b.c. (\ref{NahmBC}) in $USp(4)/\mathbb{Z}_2$ gauge theory 
can be obtained from the $USp(4)$ Dirichlet half-index (\ref{usp4D}) by Higgsing procedure with the following specialization of the global fugacities: 
\begin{align}
x_1&=q^{\frac34}t,\qquad x_2=q^{\frac54}t^3. 
\end{align}
Then the $USp(4)$ Dirichlet half-index (\ref{usp4D}) is factorized as 
\begin{align}
&
\mathbb{II}_{\mathcal{D}}^{\textrm{4d $\mathcal{N}=4$ $USp(4)$}}
\left(t,x_1=q^{\frac34}t, x_2=q^{\frac54}t^3;q\right)
\nonumber\\
&=\langle T_{(\frac12,\frac12)}T_{(\frac12,\frac12)}\rangle_{\textrm{Nahm}}^{\textrm{4d $\mathcal{N}=4$ $USp(4)/\mathbb{Z}_2$}}(t;q)
\mathcal{I}^{\textrm{3d HM}}(t,x=q^{\frac14}t;q)
\mathcal{I}^{\textrm{3d HM}}(t,x=q^{\frac54}t;q)
\nonumber\\
&\times 
\mathcal{I}^{\textrm{3d HM}}(t,x=q^{\frac74}t^3;q)
\mathcal{I}^{\textrm{3d HM}}(t,x=q^{\frac94}t^5;q). 
\end{align}
The resulting two-point function of the dual 't Hooft line operator with the regular Nahm pole b.c. (\ref{NahmBC}) in $USp(4)/\mathbb{Z}_2$ gauge theory takes the form 
\begin{align}
\label{usp4Nahm_T1/2}
\langle T_{(\frac12,\frac12)}T_{(\frac12,\frac12)}\rangle_{\textrm{Nahm}}^{\textrm{4d $\mathcal{N}=4$ $USp(4)/\mathbb{Z}_2$}}(t;q)
&=\frac{(q)_{\infty} (q^{\frac32}t^{2};q)_{\infty}^2 (q^{\frac72}t^{6};q)_{\infty}}
{(q^{\frac12}t^{2};q)_{\infty}(qt^{4};q)_{\infty}(q^2;q)_{\infty}(q^3t^{8};q)_{\infty}}. 
\end{align}
We find that 
\begin{align}
\langle W_{\textrm{sp}} W_{\textrm{sp}}\rangle_{\mathcal{N}}^{\textrm{4d $\mathcal{N}=4$ $Spin(5)$}}(t;q)
&=
\langle T_{(\frac12,\frac12)}T_{(\frac12,\frac12)}\rangle_{\textrm{Nahm}}^{\textrm{4d $\mathcal{N}=4$ $USp(4)/\mathbb{Z}_2$}}(t^{-1};q). 
\end{align}
This provides strong evidence of the duality between  
the boundary Wilson line in the spinor representation of $Spin(5)$ gauge theory with the Neumann b.c. 
and the boundary 't Hooft line of magnetic charge $B=(\frac12,\frac12)$ of $USp(4)/\mathbb{Z}_2$ gauge theory with the Nahm pole b.c. 

\subsubsection{Fundamental Wilson line}
The line defect half-indices for the Wilson lines associated with the non-minuscule representations are not simply obtained by the Higgsing method. 
It would be intriguing to explore the closed-form expressions for such cases. 

For the fundamental Wilson line, 
we find that the one-point function can be expressed as
\begin{align}
\label{so5N_W1_1pt}
\langle W_{\tiny \yng(1)}\rangle_{\mathcal{N}}^{\textrm{4d $\mathcal{N}=4$ $SO(5)$}}(t;q)
&=qt^{-4}\frac{(q^{\frac12}t^2;q)_{\infty}}{(q^{\frac32}t^2;q)_{\infty}}
\frac{(q^{\frac32}t^{-2};q)_{\infty} (q^{\frac72}t^{-6};q)_{\infty}}
{(qt^{-4};q)_{\infty}(q^2t^{-8};q)_{\infty}}. 
\end{align}
Also it follows that the two-point function of a pair of the fundamental Wilson lines can be written as
\begin{align}
\label{so5N_W1_2pt}
&
\langle W_{\tiny \yng(1)} W_{\tiny \yng(1)} \rangle_{\mathcal{N}}^{\textrm{4d $\mathcal{N}=4$ $SO(5)$}}(t;q)
\nonumber\\
&=\frac{(q^{\frac32}t^{-2};q)_{\infty}(q^{\frac52}t^{-6};q)_{\infty}}
{(qt^{-4};q)_{\infty}(q^2t^{-8};q)_{\infty}}
+q^{\frac12}t^{-2}
\frac{(q^{\frac12}t^2;q)_{\infty}(q^{\frac32}t^{-2};q)_{\infty}(q^2t^{-4};q)_{\infty}(q^{\frac72}t^{-6};q)_{\infty}}
{(qt^{-4};q)_{\infty}^2(q^{\frac32}t^2;q)_{\infty}(q^3t^{-8};q)_{\infty}}
\nonumber\\
&+qt^{-4}\frac{(q^{\frac12}t^2;q)_{\infty}(q^{\frac32}t^{-2};q)_{\infty}(q^2t^{-4};q)_{\infty}(q^{\frac52}t^{-2};q)_{\infty}(q^{\frac92}t^{-6};q)_{\infty}}
{(qt^{-4};q)_{\infty}^2(q^{\frac32}t^2;q)_{\infty}(q^3t^{-8};q)_{\infty}(q^{\frac72}t^{-2};q)_{\infty}}. 
\end{align}
Here the first term is the $SO(5)$ Neumann half-index (\ref{so5N}), 
the second the one-point function (\ref{so5N_Wasym2_1pt}) of the rank-$2$ antisymmetric Wilson line 
and the third the one-point function (\ref{so5N_Wsym2_1pt}) of the rank-$2$ symmetric Wilson line. 

\subsubsection{Antisymmetric Wilson line}
For $SO(5)$ gauge theory, there exist the boundary Wilson line transforming in the rank-$2$ antisymmetric representation.
The one-point function of the rank-$2$ antisymmetric Wilson line can be obtained 
from the results (\ref{so5N}), (\ref{so5N_Wsp}) and (\ref{so5N_W1_1pt})
\begin{align}
\label{so5N_Wasym2_1pt}
&
\langle W_{\tiny \yng(1,1)}\rangle_{\mathcal{N}}^{\textrm{4d $\mathcal{N}=4$ $SO(5)$}}(t;q)
\nonumber\\
&=\langle W_{\textrm{sp}} W_{\textrm{sp}}\rangle_{\mathcal{N}}^{\textrm{4d $\mathcal{N}=4$ $Spin(5)$}}(t;q)
-\langle W_{\tiny \yng(1)}\rangle_{\mathcal{N}}^{\textrm{4d $\mathcal{N}=4$ $SO(5)$}}(t;q)
-\mathbb{II}_{\mathcal{N}}^{\textrm{4d $\mathcal{N}=4$ $SO(5)$}}(t;q)
\nonumber\\
&=q^{\frac12}t^{-2}
\frac{(q^{\frac12}t^2;q)_{\infty}(q^{\frac32}t^{-2};q)_{\infty}(q^2t^{-4};q)_{\infty}(q^{\frac72}t^{-6};q)_{\infty}}
{(qt^{-4};q)_{\infty}^2(q^{\frac32}t^2;q)_{\infty}(q^3t^{-8};q)_{\infty}}. 
\end{align}

\subsubsection{Symmetric Wilson lines}
For the one-point function of the boundary Wilson line in the rank-$2$ symmetric Wilson line, 
we find the following closed-form expression 
\begin{align}
&
\label{so5N_Wsym2_1pt}
\langle W_{\tiny \yng(2)}\rangle_{\mathcal{N}}^{\textrm{4d $\mathcal{N}=4$ $SO(5)$}}(t;q)
\nonumber\\
&=qt^{-4}\frac{(q^{\frac12}t^2;q)_{\infty}(q^{\frac32}t^{-2};q)_{\infty}(q^2t^{-4};q)_{\infty}(q^{\frac52}t^{-2};q)_{\infty}(q^{\frac92}t^{-6};q)_{\infty}}
{(qt^{-4};q)_{\infty}^2(q^{\frac32}t^2;q)_{\infty}(q^3t^{-8};q)_{\infty}(q^{\frac72}t^{-2};q)_{\infty}}. 
\end{align}

\subsection{$\mathfrak{so}(2N+1)$}
Now we propose the results for $\mathcal{N}=4$ SYM theories based on the gauge algebra $\mathfrak{so}(2N+1)$ of general rank. 
We start with the half-indices in \cite{Hatsuda:2024lcc} without any insertion of the line operator. 
The half-index of the Neumann b.c. (\ref{NeuBC}) in $SO(2N+1)$ gauge theory is given by
\begin{align}
\label{so2N+1N}
&\mathbb{II}_{\mathcal{N}}^{\textrm{4d $\mathcal{N}=4$ $SO(2N+1)$}}(t;q)
\nonumber\\
&=\frac{1}{2^N N!}\frac{(q)_{\infty}^N}{(q^{\frac12}t^{-2})^N}
\oint \prod_{i=1}^N \frac{ds_i}{2\pi is_i}
\frac{(s_i^{\pm};q)_{\infty}}
{(q^{\frac12}t^{-2}s_i^{\pm};q)_{\infty}}
\prod_{i<j}
\frac{(s_i^{\pm}s_j^{\mp};q)_{\infty}(s_i^{\pm}s_j^{\pm};q)_{\infty}}
{(q^{\frac12}t^{-2}s_i^{\pm}s_j^{\mp};q)_{\infty}(q^{\frac12}t^{-2}s_i^{\pm}s_j^{\pm};q)_{\infty}}. 
\end{align}
The half-index of the dual Nahm pole b.c. in $USp(2N)$ gauge theory is \cite{Hatsuda:2024lcc}
\begin{align}
\label{usp2NNahm}
\mathbb{II}_{\textrm{Nahm}}^{\textrm{4d $\mathcal{N}=4$ $USp(2N)$}}(t;q)
&=\prod_{k=1}^{N}
\frac{(q^{k+\frac12}t^{4k-2};q)_{\infty}}
{(q^k t^{4k};q)_{\infty}}. 
\end{align}
As discussed in \cite{Hatsuda:2024lcc}, 
the $USp(2N)$ Nahm pole half-index (\ref{usp2NNahm}) agrees with the $SO(2N+1)$ Neumann half-index (\ref{so2N+1N}) 
under $t$ $\rightarrow$ $t^{-1}$. 
The half-index of the Dirichlet b.c. in $USp(2N)$ gauge theory is given by 
\begin{align}
\label{usp2ND}
&
\mathbb{II}_{\mathcal{D}}^{\textrm{4d $\mathcal{N}=4$ $USp(2N)$}}(t,x_i;q)
\nonumber\\
&=\frac{(q)_{\infty}^N}{(q^{\frac12}t^2;q)_{\infty}^N}
\prod_{i=1}^N
\frac{(qx_i^{\pm 2};q)_{\infty}}
{(q^{\frac12}t^2 x_i^{\pm 2};q)_{\infty}}
\prod_{i<j}
\frac{(qx_i^{\pm}x_j^{\mp};q)_{\infty}(qx_i^{\pm}x_j^{\pm};q)_{\infty}}
{(q^{\frac12}t^{2}x_i^{\pm}x_j^{\mp};q)_{\infty}(q^{\frac12}t^{2}x_i^{\pm}x_j^{\pm};q)_{\infty}}. 
\end{align}
By fixing the global fugacities as
\begin{align}
x_i&=q^{\frac{2i-1}{4}}t^{2i-1}, 
\end{align}
we get 
\begin{align}
&
\mathbb{II}_{\mathcal{D}}^{\textrm{4d $\mathcal{N}=4$ $USp(2N)$}}
\left(t,x_i=q^{\frac{2i-1}{4}}t^{2i-1};q\right)
\nonumber\\
&=\mathbb{II}_{\textrm{Nahm}}^{\textrm{4d $\mathcal{N}=4$ $USp(2N)$}}(t;q)
\prod_{i=1}^{N}\mathcal{I}^{\textrm{3d HM}}(t,x=q^{\frac{2i-1}{2}}t^{4i-2};q)
\nonumber\\
&\times 
\prod_{i=1}^{N-1}
\mathcal{I}^{\textrm{3d HM}}(t,x=q^{\frac{2i-1}{4}}t^{2i-1};q)^{N-i}
\times 
\prod_{i=1}^{2N-3}
\mathcal{I}^{\textrm{3d HM}}(t,x=q^{\frac{2i+1}{4}}t^{2i+1};q)^{a_N(i)}. 
\end{align}
Here $a_{N}(i)$ is the number of the partitions of $i$  
that fit into a rectangle of size $(N-2) \times 2$. 
It can be obtained from the $q$-binomial coefficients (see e.g. \cite{MR1634067})
\begin{align}
\label{qbinomial}
\left(\begin{matrix}
N\\
2\\
\end{matrix}\right)_q
&=\sum_{i=1}^{2N-3}a_N(i)q^{i-1}, 
\end{align}
where 
\begin{align}
\left(\begin{matrix}
n\\
k\\
\end{matrix}\right)_q
&:=\frac{[n]_q [n-1]_q\cdots [n-k+1]_q}
{[1]_q [2]_q\cdots [k]_q}, \\
[n]_q&:=\frac{1-q^n}{1-q}. 
\end{align}
Taking away the indices of the 3d matters, 
one finds from the Higgsed Dirichlet half-index the Nahm pole half-index (\ref{usp2NNahm}). 

\subsubsection{Spinor representation}
With the Neumann b.c. in $Spin(2N+1)$ gauge theory, 
one finds non-trivial two-point function of the Wilson lines in the spinor representation. 
It is evaluated as the following matrix integral: 
\begin{align}
\label{so2N+1N_Wsp}
&
\langle W_{\textrm{sp}} W_{\textrm{sp}}\rangle_{\mathcal{N}}^{\textrm{4d $\mathcal{N}=4$ $Spin(2N+1)$}}(t;q)
\nonumber\\
&=\frac{1}{2^N N!}\frac{(q)_{\infty}^N}{(q^{\frac12}t^{-2})^N}
\oint \prod_{i=1}^N \frac{ds_i}{2\pi is_i}
\frac{(s_i^{\pm};q)_{\infty}}
{(q^{\frac12}t^{-2}s_i^{\pm};q)_{\infty}}
\prod_{i<j}
\frac{(s_i^{\pm}s_j^{\mp};q)_{\infty}(s_i^{\pm}s_j^{\pm};q)_{\infty}}
{(q^{\frac12}t^{-2}s_i^{\pm}s_j^{\mp};q)_{\infty}(q^{\frac12}t^{-2}s_i^{\pm}s_j^{\pm};q)_{\infty}}
\nonumber\\
&\times 
\prod_{i=1}^N (s_i^{\frac12}+s_i^{-\frac12})^2. 
\end{align}
The boundary Wilson line in the spinor representation of $Spin(2N+1)$ with the Neumann b.c. 
is conjectured to be dual to the boundary 't Hooft line of magnetic charge $B$ $=$ $(\frac12,\cdots, \frac12)$ in $USp(2N)/\mathbb{Z}_2$ gauge theory with the regular Nahm pole b.c. 
We propose that the dual two-point function of the 't Hooft line can be obtained from the $USp(2N)$ Dirichlet half-index (\ref{usp2ND}) via the Higgsing manipulation 
with the following specialization of the global fugacities: 
\begin{align}
x_i&=q^{\frac{2i+1}{4}}t^{2i-1}, \qquad i=1,\cdots, N. 
\end{align}
Accordingly, we find that 
the $USp(2N)$ Dirichlet half-index is Higgsed as
\begin{align}
&
\mathbb{II}_{\mathcal{D}}^{\textrm{4d $\mathcal{N}=4$ $USp(2N)$}}
\left(t,x_i=q^{\frac{2i+1}{4}}t^{2i-1};q\right)
\nonumber\\
&=\langle T_{(\frac12,\cdots,\frac12)}T_{(\frac12,\cdots,\frac12)}\rangle_{\textrm{Nahm}}^{\textrm{4d $\mathcal{N}=4$ $USp(2N)/\mathbb{Z}_2$}}(t;q)
\prod_{i=1}^{N-1}
\mathcal{I}^{\textrm{3d HM}}(t,x=q^{\frac{2i-1}{4}}t^{2i-1};q)^{N-i}
\nonumber\\
&\times 
\prod_{i=1}^{2N-1}
\mathcal{I}^{\textrm{3d HM}}(t,x=q^{\frac{2i+3}{4}}t^{2i-1};q)^{a_{N+1}(i)},
\end{align}
where $a_N(i)$ are the $q$-binomial coefficients defined in (\ref{qbinomial}). 
The resulting two-point function of the boundary 't Hooft line of magnetic charge $B$ $=$ $(\frac12,\cdots, \frac12)$ in $USp(2N)/\mathbb{Z}_2$ gauge theory with the regular Nahm pole b.c. takes the form 
\begin{align}
\label{usp2NNahm_T1/2}
&
\langle T_{(\frac12,\cdots,\frac12)}T_{(\frac12,\cdots,\frac12)}\rangle_{\textrm{Nahm}}^{\textrm{4d $\mathcal{N}=4$ $USp(2N)/\mathbb{Z}_2$}}(t;q)
\nonumber\\
&=\prod_{k=1}^{\lfloor \frac{N+1}{2}\rfloor}
\frac{(q^{\frac{2k+1}{2}}t^{4k-2};q)_{\infty}}
{(q^{k+1}t^{4k-4};q)_{\infty}}
\prod_{k=1}^{N}
\frac{(q^{\frac{k+1}{2}}t^{2(k-1)};q)_{\infty}}
{(q^{\frac{k}{2}}t^{2k};q)_{\infty}}
\prod_{k=1}^{\lfloor \frac{N+1}{2}\rfloor}
\frac{(q^{\frac{2(k+\lfloor \frac{N}{2}\rfloor)+3}{2}}t^{4(k+\lfloor \frac{N}{2}\rfloor)-2};q)_{\infty}}
{(q^{k+\lfloor \frac{N}{2}\rfloor+1}t^{4(k+\lfloor \frac{N}{2}\rfloor)};q)_{\infty}}, 
\end{align}
where
\begin{align}
\lfloor x\rfloor:=\max\{n\in \mathbb{Z}|n\le x\}
\end{align}
is the floor function. 
For example, for $N=3, 4, 5$ we have
\begin{align}
&
\langle T_{(\frac12,\frac12,\frac12)}T_{(\frac12,\frac12,\frac12)}\rangle_{\textrm{Nahm}}^{\textrm{4d $\mathcal{N}=4$ $USp(6)/\mathbb{Z}_2$}}(t;q)
\nonumber\\
&=\frac{(q)_{\infty}(q^{\frac32}t^2;q)_{\infty}^2(q^2t^4;q)_{\infty}(q^{\frac52}t^6;q)_{\infty}(q^{\frac72}t^6;q)_{\infty}(q^{\frac92}t^{10};q)_{\infty}}
{(q^{\frac12}t^2;q)_{\infty}(qt^4;q)_{\infty}(q^{\frac32}t^6;q)_{\infty}(q^2;q)_{\infty}(q^3t^4;q)_{\infty}(q^3t^8;q)_{\infty}(q^4t^{12};q)_{\infty}}, \\
&
\langle T_{(\frac12,\frac12,\frac12,\frac12)}T_{(\frac12,\frac12,\frac12,\frac12)}\rangle_{\textrm{Nahm}}^{\textrm{4d $\mathcal{N}=4$ $USp(8)/\mathbb{Z}_2$}}(t;q)
\nonumber\\
&=\frac{(q)_{\infty}(q^{\frac32}t^2;q)_{\infty}^2(q^2t^4;q)_{\infty}(q^{\frac52}t^6;q)_{\infty}^2(q^{\frac92}t^{10};q)_{\infty}(q^{\frac{11}{2}}t^{14};q)_{\infty}}
{(q^{\frac12}t^2;q)_{\infty}(qt^4;q)_{\infty}(q^{\frac32}t^6;q)_{\infty}(q^2;q)_{\infty}(q^2t^8;q)_{\infty}(q^3t^4;q)_{\infty}(q^4t^{12};q)_{\infty}(q^5t^{16};q)_{\infty}}, \\
&
\langle T_{(\frac12,\frac12,\frac12,\frac12,\frac12)}T_{(\frac12,\frac12,\frac12,\frac12,\frac12)}\rangle_{\textrm{Nahm}}^{\textrm{4d $\mathcal{N}=4$ $USp(10)/\mathbb{Z}_2$}}(t;q)
\nonumber\\
&=\frac{(q)_{\infty}(q^{\frac32}t^2;q)_{\infty}^2(q^2t^4;q)_{\infty}(q^{\frac52}t^6;q)_{\infty}^2}
{(q^{\frac12}t^2;q)_{\infty}(qt^4;q)_{\infty}(q^{\frac32}t^6;q)_{\infty}(q^2;q)_{\infty}(q^2t^8;q)_{\infty}}
\nonumber\\
&\times 
\frac{(q^3t^8;q)_{\infty}(q^{\frac72}t^{10};q)_{\infty}(q^{\frac92}t^{10};q)_{\infty}(q^{\frac{11}{2}}t^{14};q)_{\infty}(q^{\frac{13}{2}}t^{18};q)_{\infty}}
{(q^{\frac52}t^{10};q)_{\infty}(q^3t^4;q)_{\infty}(q^4t^8;q)_{\infty}(q^4t^{12};q)_{\infty}(q^5t^{16};q)_{\infty}(q^6t^{20};q)_{\infty}}. 
\end{align}
It is expected that 
the two-point function (\ref{so2N+1N_Wsp}) of the spinor Wilson line of $Spin(2N+1)$ gauge theory with Neumann b.c. 
becomes equivalent to the two-point function (\ref{usp2NNahm_T1/2}) of the 't Hooft line with $B$ $=$ $(\frac12,\cdots, \frac12)$ in $USp(2N)/\mathbb{Z}_2$ gauge theory with the Nahm pole b.c. under the transformation $t$ $\rightarrow$ $t^{-1}$ 
\begin{align}
\label{dual_BNWsp_CNT1/2}
\langle W_{\textrm{sp}} W_{\textrm{sp}}\rangle_{\mathcal{N}}^{\textrm{4d $\mathcal{N}=4$ $Spin(2N+1)$}}(t;q)
&=
\langle T_{(\frac12,\cdots,\frac12)}T_{(\frac12,\cdots,\frac12)}\rangle_{\textrm{Nahm}}^{\textrm{4d $\mathcal{N}=4$ $USp(2N)/\mathbb{Z}_2$}}(t^{-1};q). 
\end{align}

\subsubsection{Direct evaluation by Macdonald polynomials}
Now let us evaluate the line defect half-indices by means of the norm formula of the Macdonald polynomials of type $B_N$. 
In this case, the root system is given by
\begin{align}
R&=\{ \pm \varepsilon_i| 1\leq i \leq N \} \cup \{ \pm \varepsilon_i \pm \varepsilon_j | 1\leq i<j \leq N \}, \\
R^+&=\{ \varepsilon_i| 1\leq i \leq N \} \cup \{ \varepsilon_i \pm \varepsilon_j | 1\leq i<j \leq N \},
\end{align}
where $\varepsilon_i$ ($i=1,2,\dots, N$) are an orthonormal basis of $\mathbb{R}^N$.
Then,
\begin{align}
\rho=\sum_{i=1}^N \biggl( N-i+\frac{1}{2}\biggr)\varepsilon_i,\qquad \lambda=\sum_{i=1}^N \lambda_i \varepsilon_i,
\end{align}
and
\begin{align}
w_{B_N}(s; \q,\t)=\prod_{i=1}^N \frac{(s_i^{\pm 1};\q)_\infty}{(\t s_i^{\pm 1};\q)_\infty}
\prod_{1\leq i<j \leq N} \frac{(s_i^{\pm 1}s_j^{\pm 1};\q)_\infty}{(\t s_i^{\pm 1}s_j^{\pm 1};\q)_\infty}.
\end{align}

The norm is given by
\begin{align}
&\frac{\langle P_\lambda, P_\lambda \rangle}{\langle 1, 1 \rangle}=\prod_{i=1}^N \frac{(\t^{2N-2i+2};\q)_{2\lambda_i}(\t^{2N-2i}\q;\q)_{2\lambda_i}}{(\t^{2N-2i+1};\q)_{2\lambda_i}(\t^{2N-2i+1}\q;\q)_{2\lambda_i}} \\
&\times \prod_{1\leq i<j\leq N} \frac{(\t^{j-i+1};\q)_{\lambda_i-\lambda_j}(\t^{j-i-1}\q;\q)_{\lambda_i-\lambda_j}}{(\t^{j-i};\q)_{\lambda_i-\lambda_j}(\t^{j-i}\q;\q)_{\lambda_i-\lambda_j}}
\frac{(\t^{2N-i-j+2};\q)_{\lambda_i+\lambda_j}(\t^{2N-i-j}\q;\q)_{\lambda_i+\lambda_j}}{(\t^{2N-i-j+1};\q)_{\lambda_i+\lambda_j}(\t^{2N-i-j+1}\q;\q)_{\lambda_i+\lambda_j}}, \nonumber
\end{align}
and also we have
\begin{align}
\langle 1, 1 \rangle=\frac{(\t;\q)_\infty^N}{(\q;\q)_\infty^N} \prod_{k=1}^N \frac{(\t^{2k-1}\q;\q)_\infty}{(\t^{2k};\q)_\infty}.
\end{align}

According to \cite{MR4139057}, for the type $B_N$ system, the character with one-column Young diagram is related to the Macdonald polynomials by
\begin{align}
&\chi_{(1^r)}(s)=E_r(s)+E_{r-1}(s), \label{eq:chi-AS}\\
&E_r(s)=\sum_{k=0}^{\lfloor \frac{r}{2} \rfloor} 
\frac{(\t^{N-r+1};\t)_{2k}}{(\t^{N-r};\t)_{2k}}\frac{(\t^{2N-2r};\t^2)_{2k}}{(\t^{2N-2r-1}\q;\t^2)_{2k}}
\frac{(\q/\t, \t^{2N-2r-1};\t^2)_{k}}{(\t^2, \t^{2N-2r+2};\t^2)_{k}}\t^k \\
&\times \sum_{j=0}^{r-2k}(-1)^j \frac{(\t^{N-r+2k+1}, -\t^{N-r+2k}\q, -\t^{N-r+2k}\q^{1/2}, \t^{N-r+2k}\q^{1/2};\t)_j}{(\t^{2N-2r+4k}\q;\t)_{2j}}
 P_{(1^{r-2k-j})}(s;\q, \t).\notag
\end{align}
where $E_r(s)$ is generated by the following identity:
\begin{align}
\overline{E}(s|z)=\prod_{i=1}^N (1+z s_i)(1+z s_i^{-1})=\sum_{r=0}^{2N} z^rE_r(s) .
\label{eq:gene-E}
\end{align}
It is obvious to see that $E_{2N-r}(s)=E_r(s)$ for $0 \leq r \leq N$.

\paragraph{Fundamental representation.}
Although the line defect half-indices for the Wilson lines in the fundamental representation do not simply match with 
those for the dual 't Hooft lines which are obtainable by the Higgsing procedure, 
they can be exactly computed by using the Macdonald polynomials. 
For $\lambda=(1)$, we have
\begin{equation}
\begin{aligned}
\chi_{(1)}(s)=E_1(s)+E_0(s)=P_{(1)}(s;\q, \t)+\frac{\t^{N-1}(\t-\q)}{1-\q \t^{2N-1} }.
\end{aligned}
\end{equation}
Therefore
\begin{align}
\langle \chi_{(1)}, 1 \rangle&=\frac{\t^{N-1}(\t-\q)}{1-\q \t^{2N-1} } \langle 1,1 \rangle , \\
\langle \chi_{(1)}, \chi_{(1)} \rangle&=\langle P_{(1)},P_{(1)} \rangle+ \frac{\t^{2N-2}(\t-\q)^2}{(1-\q \t^{2N-1})^2 } \langle 1,1 \rangle,
\end{align}
where
\begin{equation}
\begin{aligned}
\frac{\langle P_{(1)},P_{(1)} \rangle}{\langle 1,1 \rangle}=\frac{(1-\q)(1-\q^2 \t^{2N-2})(1-\t^{2N})(1-\q \t^{2N})}{(1-\t)(1-\q \t^{2N-1})^2 (1-\q^2 \t^{2N-1})}.
\end{aligned}
\end{equation}
Using these results, we finally find
\begin{align}
\frac{\langle W_{\square} \rangle_{\mathcal{N}}^{SO(2N+1)}}{\mathbb{II}_{\mathcal{N}}^{SO(2N+1)}}&=\frac{\t^{N-1}(\t-\q)}{1-\q \t^{2N-1} }, \label{eq:W_fund-B_N}\\ 
\frac{\langle W_{\square}W_{\square} \rangle_{\mathcal{N}}^{SO(2N+1)}}{\mathbb{II}_{\mathcal{N}}^{SO(2N+1)}}&=\frac{(1-\q)(1-\q^2 \t^{2N-2})(1-\t^{2N})(1-\q \t^{2N})}{(1-\t)(1-\q 
\t^{2N-1})^2 (1-\q^2 \t^{2N-1})}+\frac{\t^{2N-2}(\t-\q)^2}{(1-\q \t^{2N-1})^2 }
\label{eq:W_fund-B_N2pt}.
\end{align}
The formula (\ref{eq:W_fund-B_N}) reproduces the expressions (\ref{so3N_W1_1pt}) and (\ref{so5N_W1_1pt}). 
Also one can check that (\ref{eq:W_fund-B_N2pt}) leads to the results (\ref{so3N_W1_2pt}) and (\ref{so5N_W1_2pt}). 
We see that the second term in (\ref{eq:W_fund-B_N}) is the square of the one-point function. 
Thus we can define the connected normalized two-point function of the fundamental Wilson lines by
\begin{align}
\langle \mathcal{W}_{\square}\mathcal{W}_{\square} \rangle_{\mathcal{N},c}^{SO(2N+1)}
&:=\frac{\langle W_{\square}W_{\square} \rangle_{\mathcal{N}}^{SO(2N+1)}}{\mathbb{II}_{\mathcal{N}}^{SO(2N+1)}}
-\left(\frac{\langle W_{\square} \rangle_{\mathcal{N}}^{SO(2N+1)}}{\mathbb{II}_{\mathcal{N}}^{SO(2N+1)}}\right)^2
\nonumber\\
&=\frac{(1-\q)(1-\q^2 \t^{2N-2})(1-\t^{2N})(1-\q \t^{2N})}{(1-\t)(1-\q \t^{2N-1})^2 (1-\q^2 \t^{2N-1})}.
\end{align}

\paragraph{Antisymmetric representations.}
Furthermore, we obtain the exact formulas for the line defect half-indices of the Wilson line in the antisymmetric representations. 
The one-point function vanish for odd rank antisymmetric representations, however it does not for even rank. 
From \eqref{eq:chi-AS}, we have
\begin{align}
\frac{\langle W_{(1^{r})} \rangle_{\mathcal{N}}^{SO(2N+1)}}{\mathbb{II}_{\mathcal{N}}^{SO(2N+1)}}=\frac{\langle \chi_{(1^r)}, 1 \rangle}{\langle 1, 1 \rangle}
=\mathcal{E}_{N,r}+\mathcal{E}_{N, r-1}.
\end{align}
where
\begin{equation}
\begin{aligned}
\label{Er-1_product2}
\mathcal{E}_{N,r}:=\frac{\langle E_{r}, 1 \rangle}{\langle 1, 1 \rangle}
=\sum_{k=0}^{\lfloor \frac{r}{2} \rfloor} (-1)^{r-2k} \frac{(\t^{N-r+1};\t)_{2k}}{(\t^{N-r};\t)_{2k}}\frac{(\t^{2N-2r};\t^2)_{2k}}{(\t^{2N-2r-1}\q;\t^2)_{2k}}
\frac{(\q/\t, \t^{2N-2r-1};\t^2)_{k}}{(\t^2, \t^{2N-2r+2};\t^2)_{k}}\\
\times \frac{(\t^{N-r+2k+1}, -\t^{N-r+2k}\q, -\t^{N-r+2k}\q^{1/2}, \t^{N-r+2k}\q^{1/2};\t)_{r-2k}}{(\t^{2N-2r+4k}\q;\t)_{2r-4k}} \t^k.
\end{aligned}
\end{equation}
Using it, we observe the following simple recursive relations:
\begin{align}
\frac{\langle W_{(1^{2m+1})} \rangle_{\mathcal{N}}^{SO(2N+1)}}{\langle W_{(1^{2m})} \rangle_{\mathcal{N}}^{SO(2N+1)}}
&=\frac{\t^{N-2m}(1-\q \t^{2m-1})}{1-\q \t^{2N-2m-1}},\\
\frac{\langle W_{(1^{2m})} \rangle_{\mathcal{N}}^{SO(2N+1)}}{\langle W_{(1^{2m-1})} \rangle_{\mathcal{N}}^{SO(2N+1)}}
&=\frac{1-\t^{2N-2m+2}}{\t^{N-2m+1}(1-\t^{2m})}.
\end{align}
Combining them and \eqref{eq:W_fund-B_N}, we find
\begin{align}
\label{so2N+1N_Wasym2}
\frac{\langle W_{(1^{2})} \rangle_{\mathcal{N}}^{SO(2N+1)}}{\mathbb{II}_{\mathcal{N}}^{SO(2N+1)}}
&=\frac{(\t-\q)(1-\t^{2N})}{(1-\t^2)(1-\q \t^{2N-1})}, \\
\label{so2N+1N_Wasym3}
\frac{\langle W_{(1^{3})} \rangle_{\mathcal{N}}^{SO(2N+1)}}{\mathbb{II}_{\mathcal{N}}^{SO(2N+1)}}
&=\frac{\t^{N-2}(\t-\q)(1-\q \t)(1-\t^{2N})}{(1-\t^2)(1-\q \t^{2N-1})(1-\q \t^{2N-3})},\\
\label{so2N+1N_Wasym4}
\frac{\langle W_{(1^{4})} \rangle_{\mathcal{N}}^{SO(2N+1)}}{\mathbb{II}_{\mathcal{N}}^{SO(2N+1)}}
&=\frac{\t(\t-\q)(1-\q \t)(1-\t^{2N})(1-\t^{2N-2})}{(1-\t^2)(1-\t^4)(1-\q \t^{2N-1})(1-\q \t^{2N-3})}.
\end{align}
We see that 
the result (\ref{so2N+1N_Wasym2}) is consistent with the formula (\ref{so5N_Wasym2_1pt}). 

\paragraph{Spinor representation.}
Noticing that from \eqref{ch_so2N+1_sp} and \eqref{eq:gene-E}, a product of the character of the spinor representation is given by
\begin{align}
\chi_\text{sp}^2(s)=\prod_{i=1}^N (1+s_i)(1+s_i^{-1})=\overline{E}(s|1)=2\sum_{r=0}^{N-1} E_r(s)+E_N(s), 
\end{align}
we easily obtain
\begin{equation}
\begin{aligned}
\frac{\langle W_\text{sp} W_\text{sp} \rangle_{\mathcal{N}}^{Spin(2N+1)}}{\mathbb{II}_{\mathcal{N}}^{SO(2N+1)}}
=2\sum_{r=0}^{N-1} \mathcal{E}_{N,r}+\mathcal{E}_{N,N}.
\end{aligned}
\end{equation}
We observe the following recursion relation:
\begin{align}
\frac{\langle W_\text{sp} W_\text{sp} \rangle_{\mathcal{N}}^{Spin(2N+1)}/\mathbb{II}_{\mathcal{N}}^{SO(2N+1)}}
{\langle W_\text{sp} W_\text{sp} \rangle_{\mathcal{N}}^{Spin(2N-1)}/\mathbb{II}_{\mathcal{N}}^{SO(2N-1)}}
=\frac{(1+\t^N)(1-\q \t^{N-1})}{1-\q \t^{2N-1}}.
\end{align}
From it, we finally find
\begin{equation}
\begin{aligned}
\frac{\langle W_\text{sp} W_\text{sp} \rangle_{\mathcal{N}}^{Spin(2N+1)}}{\mathbb{II}_{\mathcal{N}}^{SO(2N+1)}}
=\prod_{k=0}^{N-1} \frac{(1+\t^{k+1})(1-\q \t^{k})}{1-\q \t^{2k+1}}
=\frac{(-\t;\t)_N (\q;\t)_N}{(\q \t; \t^2)_N}.
\end{aligned}
\end{equation}
One can check that this reproduces the result (\ref{usp2NNahm_T1/2}) 
obtained from the $USp(2N)$ Dirichlet half-index (\ref{usp2ND}) via the Higgsing manipulation 
according to the duality identity (\ref{dual_BNWsp_CNT1/2}).

\section{$\mathfrak{usp}(2N)$}
\label{sec_usp2N}

\subsection{$\mathfrak{usp}(2)$}
According to the isomorphism $USp(2)$ $\cong$ $Spin(3)$, 
the Wilson line defect half-indices with the Neumann b.c. are comparable with those in subsection \ref{sec_so3}. 
However, $USp(2)'$ gauge theory that is constructed in the background of $\widetilde{\textrm{O3}}^+$-plane 
obeys the modified Neumann b.c. including a single half-hypermultiplet in the fundamental representation \cite{Gaiotto:2008ak}. 
Here we investigate the case with the boundary lines. 
The half-index of the ordinary Neumann b.c. for $USp(2)$ gauge theory is given by
\begin{align}
\label{usp2N}
\mathbb{II}_{\mathcal{N}}^{\textrm{4d $\mathcal{N}=4$ $USp(2)$}}(t;q)
&=\frac12 \frac{(q)_{\infty}}{(q^{\frac12}t^{-2};q)_{\infty}}
\oint \frac{ds}{2\pi is}
\frac{(s^{\pm 2};q)_{\infty}}{(q^{\frac12}t^{-2}s^{\pm 2};q)_{\infty}}. 
\end{align}
The dual $SO(3)$ Nahm pole half-index is \cite{Hatsuda:2024lcc}
\begin{align}
\label{so3Nahm}
\mathbb{II}_{\textrm{Nahm}}^{\textrm{4d $\mathcal{N}=4$ $SO(3)$}}(t;q)
&=\frac{(q^{\frac32}t^{2};q)_{\infty}}{(qt^{4};q)_{\infty}}. 
\end{align}
Under the transformation $t$ $\rightarrow$ $t^{-1}$, the two expressions (\ref{usp2N}) and (\ref{so3Nahm}) coincide. 
The Nahm pole index can be found by Higgsing the $SO(3)$ Dirichlet half-index 
\begin{align}
\label{so3D}
\mathbb{II}_{\mathcal{D}}^{\textrm{4d $\mathcal{N}=4$ $SO(3)$}}(t,x;q)
&=\frac{(q)_{\infty}}{(q^{\frac12}t^2;q)_{\infty}}
\frac{(qx;q)_{\infty}(qx^{-1};q)_{\infty}}
{(q^{\frac12}t^2x;q)_{\infty}(q^{\frac12}t^2x^{-1};q)_{\infty}}
\end{align}
with $x=q^{1/2}t^2$ so that 
\begin{align}
&
\mathbb{II}_{\mathcal{D}}^{\textrm{4d $\mathcal{N}=4$ $SO(3)$}}
\left(t,x=q^{\frac12}t^2;q\right)
\nonumber\\
&=\mathbb{II}_{\textrm{Nahm}}^{\textrm{4d $\mathcal{N}=4$ $SO(3)$}}(t;q)
\mathcal{I}^{\textrm{3d HM}}(t,x=q^{\frac14}t;q). 
\end{align}

For the Neumann b.c. for $USp(2)'$ gauge theory 
the half-index reads
\begin{align}
\label{usp2'N}
\mathbb{II}_{\mathcal{N}+\textrm{hyp}}^{\textrm{4d $\mathcal{N}=4$ $USp(2)'$}}(t;q)
&=\frac12 \frac{(q)_{\infty}}{(q^{\frac12}t^{-2};q)_{\infty}}
\oint \frac{ds}{2\pi is}
\frac{(s^{\pm 2};q)_{\infty}}{(q^{\frac12}t^{-2}s^{\pm 2};q)_{\infty}}
\frac{(q^{\frac34}t^{-1}s^{\mp};q)_{\infty}}{(q^{\frac14}ts^{\pm};q)_{\infty}}. 
\end{align}
S-duality predicts that it is dual to the Nahm pole b.c. for $USp(2)'$ gauge theory. 
In fact, it coincides with the expression (\ref{usp2Nahm}), which can be viewed as the $USp(2)'$ half-index upon the transformation $t$ $\rightarrow$ $t^{-1}$. 

\subsubsection{Fundamental Wilson line}
For $USp(2)$ gauge theory, the one-point function of the boundary Wilson line in the fundamental representation with the Neumann b.c. vanishes 
so we begin with the two-point function of the fundamental Wilson lines of the form 
\begin{align}
\label{usp2N_W1}
\langle W_{\tiny \yng(1)} W_{\tiny \yng(1)}\rangle_{\mathcal{N}}^{\textrm{4d $\mathcal{N}=4$ $USp(2)$}}(t;q)
&=\frac12 \frac{(q)_{\infty}}{(q^{\frac12}t^{-2};q)_{\infty}}
\oint \frac{ds}{2\pi is}
\frac{(s^{\pm 2};q)_{\infty}}{(q^{\frac12}t^{-2}s^{\pm 2};q)_{\infty}}(s+s^{-1})^2. 
\end{align}
The boundary Wilson line in the fundamental representation of $USp(2)$ with the Neumann b.c. is conjecturally dual to 
the boundary 't Hooft line of magnetic charge $B=(1)$ of $SO(3)$ gauge theory with the Nahm pole b.c. 
Let us consider the Higgsing procedure with the specialization of the global fugacity of the $SO(3)$ Dirichlet half-index (\ref{so3D}) as $x=q^{3/2}t^2$. 
It leads to the factorized Dirichlet half-index (\ref{so3D}) of the form
\begin{align}
&
\mathbb{II}_{\mathcal{D}}^{\textrm{4d $\mathcal{N}=4$ $SO(3)$}}
\left(t,x=q^{\frac32}t^2;q\right)
\nonumber\\
&=\langle T_{(1)}T_{(1)}\rangle_{\textrm{Nahm}}^{\textrm{4d $\mathcal{N}=4$ $SO(3)$}}(t;q)
\mathcal{I}^{\textrm{3d HM}}(t,x=q^{\frac54}t;q). 
\end{align}
The resulting two-point function of 
the boundary 't Hooft line of magnetic charge $B=(1)$ of $SO(3)$ gauge theory with the Nahm pole b.c. is 
\begin{align}
\label{so3Nahm_T1}
\langle T_{(1)}T_{(1)}\rangle_{\textrm{Nahm}}^{\textrm{4d $\mathcal{N}=4$ $SO(3)$}}(t;q)
&=\frac{(q)_{\infty} (q^{\frac32}t^{2};q)_{\infty}}{(q^{\frac12}t^{2};q)_{\infty}(q^2;q)_{\infty}}
\frac{(q^{\frac52}t^{2};q)_{\infty}}{(q^2t^{4};q)_{\infty}}. 
\end{align}
In fact, this agrees with the two-point function (\ref{usp2N_W1}) of the boundary Wilson line in the fundamental representation of $USp(2)$ with the Neumann b.c. 
under $t$ $\rightarrow$ $t^{-1}$
\begin{align}
\langle W_{\tiny \yng(1)} W_{\tiny \yng(1)}\rangle_{\mathcal{N}}^{\textrm{4d $\mathcal{N}=4$ $USp(2)$}}(t;q)
&=\langle T_{(1)}T_{(1)}\rangle_{\textrm{Nahm}}^{\textrm{4d $\mathcal{N}=4$ $SO(3)$}}(t^{-1};q). 
\end{align}

For $USp(2)'$ gauge theory the line defect half-indices do not seem to be simply obtained via the Higgsing procedure. 
In this case, the one-point function of the fundamental Wilson line with the Neumann b.c. is non-trivial. 
We find that it can be expressed as
\begin{align}
\label{usp2'N_W1}
\langle W_{\tiny \yng(1)} \rangle_{\mathcal{N}+\textrm{hyp}}^{\textrm{4d $\mathcal{N}=4$ $USp(2)'$}}(t;q)
&=q^{\frac14}t \frac{(q^{\frac52}t^{-2};q)_{\infty}}
{(q^2t^{-4};q)_{\infty}}. 
\end{align}

\subsubsection{Symmetric Wilson lines}
For odd rank symmetric Wilson line in $USp(2)$ gauge theory with the Neumann b.c. the one-point function vanishes. 
However, for even rank symmetric Wilson line, it is non-trivial. 
Since the rank-$2k$ symmetric Wilson line in $USp(2)$ gauge theory is isomorphic to the rank-$k$ symmetric Wilson line in $SO(3)$ gauge theory, 
the correlators can be simply obtained from the previous analysis for $SO(3)$ gauge theory. 
For example, the one-point function is given by
\begin{align}
\label{usp2N_Wsym2k_1pt}
\langle W_{(2k)} \rangle_{\mathcal{N}}^{\textrm{4d $\mathcal{N}=4$ $USp(2)$}}(t;q)
&=\langle W_{(k)}\rangle_{\mathcal{N}}^{\textrm{4d $\mathcal{N}=4$ $SO(3)$}}(t;q)
\nonumber\\
&=q^{\frac{k}{2}}t^{-2k}
\frac{(q^{\frac12}t^2;q)_{\infty}(q^{\frac{2k+3}{2}}t^{-2};q)_{\infty}}
{(q^{\frac{2k+1}{2}}t^2;q)_{\infty}(qt^{-4};q)_{\infty}}. 
\end{align}

\subsection{$\mathfrak{usp}(4)$}
Next consider $USp(4)$ gauge theory. 
Again as we have $USp(4)$ $\cong$ $Spiin(5)$, 
the line defect half-indices akin to the previous analysis, but the modified Neumann b.c. for $USp(4)'$ gauge theory needs to be examined. 
The half-index of the ordinary Neumann b.c. for $USp(4)$ gauge theory is given by
\begin{align}
\label{usp4N}
&
\mathbb{II}_{\mathcal{N}}^{\textrm{4d $\mathcal{N}=4$ $USp(4)$}}(t;q)
\nonumber\\
&=\frac{1}{8} \frac{(q)_{\infty}^2}{(q^{\frac12}t^{-2};q)_{\infty}^2}
\oint \prod_{i=1}^2 \frac{ds_i}{2\pi is_i}
\frac{(s_i^{\pm 2};q)_{\infty}}{(q^{\frac12}t^{-2}s_i^{\pm 2};q)_{\infty}}
\prod_{i<j}
\frac{(s_i^{\pm}s_j^{\mp};q)_{\infty}(s_i^{\pm}s_j^{\pm};q)_{\infty}}
{(q^{\frac12}t^{-2}s_i^{\pm}s_j^{\mp};q)_{\infty}(q^{\frac12}t^{-2}s_i^{\pm}s_j^{\pm};q)_{\infty}}. 
\end{align}
The Neumann b.c. of $USp(4)$ gauge theory is dual to the regular Nahm pole b.c. in $SO(5)$ gauge theory. 
The half-index of the dual Nahm pole b.c. for $SO(5)$ gauge theory is given by \cite{Hatsuda:2024lcc}
\begin{align}
\label{so5Nahm}
\mathbb{II}_{\textrm{Nahm}}^{\textrm{4d $\mathcal{N}=4$ $SO(5)$}}(t;q)
&=\frac{(q^{\frac32}t^{2};q)_{\infty} (q^{\frac52}t^{6};q)_{\infty}}
{(qt^{4};q)_{\infty}(q^2t^{8};q)_{\infty}}. 
\end{align}
It can be shown that 
the half-indices (\ref{so5N}) and (\ref{usp4Nahm}) are equivalent upon the transformation $t$ $\rightarrow$ $t^{-1}$ \cite{Hatsuda:2024lcc}.  
The half-index of the Dirichlet b.c. for $SO(5)$ gauge theory reads
\begin{align}
\label{so5D}
&
\mathbb{II}_{\mathcal{D}}^{\textrm{4d $\mathcal{N}=4$ $SO(5)$}}(t,x_1,x_2;q)
\nonumber\\
&=\frac{(q)_{\infty}^2}{(q^{\frac12}t^2;q)_{\infty}^2}
\prod_{i=1}^2
\frac{(qx_i^{\pm};q)_{\infty}}
{(q^{\frac12}t^2 x_i^{\pm};q)_{\infty}}
\prod_{i<j}
\frac{(qx_i^{\pm}x_j^{\mp};q)_{\infty}(qx_i^{\pm}x_j^{\pm};q)_{\infty}}
{(q^{\frac12}t^2x_i^{\pm}x_j^{\mp};q)_{\infty}(q^{\frac12}t^2x_i^{\pm}x_j^{\pm};q)_{\infty}}. 
\end{align}
The Nahm pole half-index (\ref{so5Nahm}) can be found 
by Higgsing the Dirichlet half-index (\ref{so5D}) with the specialized fugacities 
\begin{align}
x_1&=q^{\frac12}t^2,
\qquad 
x_2=qt^4, 
\end{align}
as one finds the Higgsed Dirichlet half-index
\begin{align}
&
\mathbb{II}_{\mathcal{D}}^{\textrm{4d $\mathcal{N}=4$ $SO(5)$}}
\left(t,x_1=q^{\frac12}t^2,x_2=qt^4;q\right)
\nonumber\\
&=\mathbb{II}_{\textrm{Nahm}}^{\textrm{4d $\mathcal{N}=4$ $SO(5)$}}(t;q)
\mathcal{I}^{\textrm{3d HM}}(t,x=q^{\frac14}t;q)^2
\nonumber\\
&\times 
\mathcal{I}^{\textrm{3d HM}}(t,x=q^{\frac34}t^3;q)
\mathcal{I}^{\textrm{3d HM}}(t,x=q^{\frac54}t^5;q). 
\end{align}

For $USp(4)'$ gauge theory, the modified Neumann b.c. is expected to be dual to the Nahm pole b.c. of $USp(4)'$ theory. 
In fact, it is confirmed in \cite{Hatsuda:2024lcc} that 
the half-index of the modified Neumann b.c. with a fundamental half-hyper agrees with the $USp(4)$ Nahm pole half-index (\ref{usp4Nahm}).

\subsubsection{Fundamental Wilson line}
Consider the boundary Wilson line in the fundamental representation of $USp(4)$ gauge theory with the Neumann b.c. 
The two-point function of the boundary Wilson lines is given by
\begin{align}
\label{usp4N_W1}
&
\langle W_{\tiny \yng(1)} W_{\tiny \yng(1)}\rangle_{\mathcal{N}}^{\textrm{4d $\mathcal{N}=4$ $USp(4)$}}(t;q)
\nonumber\\
&=\frac{1}{8} \frac{(q)_{\infty}^2}{(q^{\frac12}t^{-2};q)_{\infty}^2}
\oint \prod_{i=1}^2 \frac{ds_i}{2\pi is_i}
\frac{(s_i^{\pm 2};q)_{\infty}}{(q^{\frac12}t^{-2}s_i^{\pm 2};q)_{\infty}}
\prod_{i<j}
\frac{(s_i^{\pm}s_j^{\mp};q)_{\infty}(s_i^{\pm}s_j^{\pm};q)_{\infty}}
{(q^{\frac12}t^{-2}s_i^{\pm}s_j^{\mp};q)_{\infty}(q^{\frac12}t^{-2}s_i^{\pm}s_j^{\pm};q)_{\infty}}
\nonumber\\
&\times (s_1+s_1^{-1}+s_2+s_2^{-1}). 
\end{align}
The fundamental Wilson line in $USp(4)$ gauge theory is conjectured to be dual 
to the 't Hooft line of magnetic charge $B=(1,0)$ in $SO(5)$ gauge theory 
which is associated with the minuscule representation. 
To apply the Higgsing procedure to the $SO(5)$ Dirichlet half-index (\ref{so5D}), 
we choose the global fugacities as
\begin{align}
x_1&=q^{\frac12}t^2,\qquad 
x_2=q^2t^4. 
\end{align}
We then find
\begin{align}
&
\mathbb{II}_{\mathcal{D}}^{\textrm{4d $\mathcal{N}=4$ $SO(5)$}}
\left(t,x_1=q^{\frac12}t^2,x_2=q^2t^4;q\right)
\nonumber\\
&=\langle T_{(1,1)}T_{(1,1)}\rangle_{\textrm{Nahm}}^{\textrm{4d $\mathcal{N}=4$ $SO(5)$}}(t;q)
\mathcal{I}^{\textrm{3d HM}}(t,x=q^{\frac14}t;q)
\mathcal{I}^{\textrm{3d HM}}(t,x=q^{\frac54}t;q)
\nonumber\\
&\times 
\mathcal{I}^{\textrm{3d HM}}(t,x=q^{\frac74}t^3;q)
\mathcal{I}^{\textrm{3d HM}}(t,x=q^{\frac94}t^5;q). 
\end{align}
The two-point function of the boundary 't Hooft line of magnetic charge $B=(1,0)$ in $SO(5)$ gauge theory with the Nahm pole b.c. is given by
\begin{align}
\label{so5Nahm_T1}
\langle T_{(1,0)}T_{(1,0)}\rangle_{\textrm{Nahm}}^{\textrm{4d $\mathcal{N}=4$ $SO(5)$}}(t;q)
&=\frac{(q)_{\infty} (q^{\frac32}t^{2};q)_{\infty}^2 (q^{\frac72}t^{6};q)_{\infty}}
{(q^{\frac12}t^{2};q)_{\infty}(qt^{4};q)_{\infty}(q^2;q)_{\infty}(q^3t^{8};q)_{\infty}}. 
\end{align}
Under the transformation $t$ $\rightarrow$ $t^{-1}$, 
this agrees with the two-point functions (\ref{usp4N_W1}) of the boundary Wilson line in the fundamental representation of $USp(4)$ gauge theory with the Neumann b.c. 
\begin{align}
\langle W_{\tiny \yng(1)} W_{\tiny \yng(1)}\rangle_{\mathcal{N}}^{\textrm{4d $\mathcal{N}=4$ $USp(4)$}}(t;q)
&=\langle T_{(1,0)}T_{(1,0)}\rangle_{\textrm{Nahm}}^{\textrm{4d $\mathcal{N}=4$ $SO(5)$}}(t^{-1};q). 
\end{align}
This supports the duality between the boundary Wilson line in the fundamental representation of $USp(4)$ gauge theory with the Neumann b.c. 
and the boundary 't Hooft line of magnetic charge $B=(1,0)$ of $SO(5)$ theory with the regular Nahm pole b.c. 

For $USp(4)'$ gauge theory, the Higgsing procedure does not seem to work simply for the line defect half-indices. 
In particular, there exists non-trivial one-point function of the boundary Wilson line in the fundamental representation. 
We find that it can be expressed as
\begin{align}
\langle W_{\tiny \yng(1)} \rangle_{\mathcal{N}+\textrm{hyp}}^{\textrm{4d $\mathcal{N}=4$ $USp(4)'$}}(t;q)
&=q^{\frac14}t\frac{(q^{\frac32}t^{-2};q)_{\infty}(q^{\frac72}t^{-6};q)_{\infty}}
{(qt^{-4};q)_{\infty}(q^3t^{-8};q)_{\infty}}. 
\end{align}
It would be desirable to explore the way to evaluate other line defect half-indices systematically. 

\subsubsection{Antisymmetric Wilson line}
The rank-$2$ antisymmetric Wilson line in $USp(4)$ gauge theory is equivalent to the fundamental Wilson line in $SO(5)$ gauge theory. 
Thus the one-point function of the boundary Wilson line is equivalent to (\ref{so5N_W1_1pt})
\begin{align}
\label{usp4N_Wasym2_1pt}
\langle W_{\tiny \yng(1,1)} \rangle_{\mathcal{N}}^{\textrm{4d $\mathcal{N}=4$ $USp(4)$}}(t;q)
&=\langle W_{\tiny \yng(1)} \rangle_{\mathcal{N}}^{\textrm{4d $\mathcal{N}=4$ $SO(5)$}}(t;q). 
\end{align}
For $USp(4)'$ gauge theory with the modified Neumann b.c. 
we obtain the same result 
\begin{align}
\langle W_{\tiny \yng(1,1)} \rangle_{\mathcal{N}+\textrm{hyp}}^{\textrm{4d $\mathcal{N}=4$ $USp(4)'$}}(t;q)
&=\langle W_{\tiny \yng(1)} \rangle_{\mathcal{N}}^{\textrm{4d $\mathcal{N}=4$ $SO(5)$}}(t;q). 
\end{align}

\subsubsection{Symmetric Wilson lines}
The rank-$2$ symmetric Wilson line in $USp(4)$ gauge theory is equivalent to the rank-$2$ antisymmetric Wilson line in $SO(5)$ gauge theory. 
Hence the expression for the one-point function is given by (\ref{so5N_Wasym2_1pt})
\begin{align}
\label{usp4N_Wsym2_1pt}
\langle W_{\tiny \yng(2)} \rangle_{\mathcal{N}}^{\textrm{4d $\mathcal{N}=4$ $USp(4)$}}(t;q)
&=\langle W_{\tiny \yng(1,1)} \rangle_{\mathcal{N}}^{\textrm{4d $\mathcal{N}=4$ $SO(5)$}}(t;q). 
\end{align}
On the other hand, for $USp(4)'$ gauge theory with the modified Neumann b.c., 
the one-point function of the rank-$2$ symmetric Wilson line is not equal to (\ref{usp4N_Wsym2_1pt}). 

\subsection{$\mathfrak{usp}(2N)$}
We promote the results so far to the gauge theory based on the gauge algebra $\mathfrak{usp}(2N)$ with general rank. 
The half-index of the Neumann b.c. for $USp(2N)$ gauge theory is evaluated as
\begin{align}
\label{usp2NN}
&
\mathbb{II}_{\mathcal{N}}^{\textrm{4d $\mathcal{N}=4$ $USp(2N)$}}(t;q)
\nonumber\\
&=\frac{1}{2^N N!} \frac{(q)_{\infty}^N}{(q^{\frac12}t^{-2};q)_{\infty}^N}
\oint \prod_{i=1}^N \frac{ds_i}{2\pi is_i}
\frac{(s_i^{\pm 2};q)_{\infty}}{(q^{\frac12}t^{-2}s_i^{\pm 2};q)_{\infty}}
\prod_{i<j}
\frac{(s_i^{\pm}s_j^{\mp};q)_{\infty}(s_i^{\pm}s_j^{\pm};q)_{\infty}}
{(q^{\frac12}t^{-2}s_i^{\pm}s_j^{\mp};q)_{\infty}(q^{\frac12}t^{-2}s_i^{\pm}s_j^{\pm};q)_{\infty}}. 
\end{align}
As the Neumann b.c. of $USp(2N)$ gauge theory is dual to the regular Nahm pole b.c. of $SO(2N+1)$ gauge theory, 
when $t$ is flipped to $t^{-1}$, 
the half-index (\ref{usp2NN}) agrees with the half-index \cite{Hatsuda:2024lcc}
\begin{align}
\label{so2N+1Nahm}
\mathbb{II}_{\textrm{Nahm}}^{\textrm{4d $\mathcal{N}=4$ $SO(2N+1)$}}(t;q)
&=\prod_{k=1}^N
\frac{(q^{k+\frac{1}{2}}t^{4k-2};q)_{\infty}}
{(q^{k}t^{4k};q)_{\infty}}. 
\end{align}
The half-index of the Dirichlet b.c. for $SO(2N+1)$ gauge theory takes the form
\begin{align}
\label{so2N+1D}
&
\mathbb{II}_{\mathcal{D}}^{\textrm{4d $\mathcal{N}=4$ $SO(2N+1)$}}(t,x_i;q)
\nonumber\\
&=\frac{(q)_{\infty}^N}{(q^{\frac12}t^2;q)_{\infty}^N}
\prod_{i=1}^N 
\frac{(qx_i^{\pm};q)_{\infty}}
{(q^{\frac12}t^2 x_i^{\pm};q)_{\infty}}
\prod_{i<j}
\frac{(qx_i^{\pm}x_j^{\mp};q)_{\infty}(qx_i^{\pm}x_j^{\pm};q)_{\infty}}
{(q^{\frac12}t^2x_i^{\pm}x_j^{\mp};q)_{\infty}(q^{\frac12}t^2x_i^{\pm}x_j^{\pm};q)_{\infty}}. 
\end{align}
When we specialize the global fugacities as
\begin{align}
x_i&=q^{\frac{i}{2}}t^{2i}, 
\end{align}
the Dirichlet half-index (\ref{usp2ND}) reduces to
\begin{align}
&
\mathbb{II}_{\mathcal{D}}^{\textrm{4d $\mathcal{N}=4$ $SO(2N+1)$}}
\left(t,x_{i}=q^{\frac{i}{2}}t^{2i};q\right)
\nonumber\\
&=\mathbb{II}_{\textrm{Nahm}}^{\textrm{4d $\mathcal{N}=4$ $SO(2N+1)$}}(t;q)
\prod_{i=1}^{N}
\mathcal{I}^{\textrm{3d HM}}(t,x=q^{\frac{2i-1}{4}}t^{2i-1};q)
\nonumber\\
&\times 
\prod_{i=1}^{N-1}
\mathcal{I}^{\textrm{3d HM}}(t,x=q^{\frac{2i-1}{4}}t^{2i-1};q)^{N-i}
\prod_{i=1}^{2N-3}
\mathcal{I}^{\textrm{3d HM}}(t,x=q^{\frac{2i+3}{4}}t^{2i+3};q)^{a_N(i)}, 
\end{align}
where $a_N(i)$ are the $q$-binomial coefficients defined in (\ref{qbinomial}). 
We can find the Nahm pole half-index (\ref{so2N+1Nahm}) 
by stripping off the extra contributions as the indices of the decoupled 3d matters. 

\subsubsection{Fundamental Wilson line}
For the boundary Wilson line in the fundamental representation with the Neumann b.c. in $USp(2N)$ gauge theory, the one -point function is trivial. 
The two-point function is non-trivial with the following form of the matrix integral: 
\begin{align}
\label{usp2NN_W1}
&
\langle W_{\tiny \yng(1)} W_{\tiny \yng(1)}\rangle_{\mathcal{N}}^{\textrm{4d $\mathcal{N}=4$ $USp(2N)$}}(t;q)
\nonumber\\
&=\frac{1}{2^N N!} \frac{(q)_{\infty}^N}{(q^{\frac12}t^{-2};q)_{\infty}^N}
\oint \prod_{i=1}^N \frac{ds_i}{2\pi is_i}
\frac{(s_i^{\pm 2};q)_{\infty}}{(q^{\frac12}t^{-2}s_i^{\pm 2};q)_{\infty}}
\prod_{i<j}
\frac{(s_i^{\pm}s_j^{\mp};q)_{\infty}(s_i^{\pm}s_j^{\pm};q)_{\infty}}
{(q^{\frac12}t^{-2}s_i^{\pm}s_j^{\mp};q)_{\infty}(q^{\frac12}t^{-2}s_i^{\pm}s_j^{\pm};q)_{\infty}}
\nonumber\\
&\times 
\left[ \sum_{i=1}^N (s_i+s_i^{-1}) \right]^2. 
\end{align}
The fundamental Wilson line in $USp(2N)$ gauge theory is dual to the 't Hooft line of magnetic charge $B=(1,0,\cdots,0)$ in $SO(2N+1)$ gauge theory. 
This corresponds to the minuscule representation. 
In order to get the two-point function of the dual 't Hooft line with the Nahm pole b.c. 
we propose the Higgsing manipulation by fixing the global fugacities of the $SO(2N+1)$ Dirichlet half-index (\ref{so2N+1D}) as
\begin{align}
x_{i}&=q^{\frac{i}{2}}t^{2i}, \quad i=1,\cdots, N-1, \nonumber\\ 
x_{N}&=q^{\frac{N}{2}+1}t^{2N}. 
\end{align}
Accordingly, we find that 
the $SO(2N+1)$ Dirichlet half-index (\ref{so2N+1D}) is factorized as
\begin{align}
&
\mathbb{II}_{\mathcal{D}}^{\textrm{4d $\mathcal{N}=4$ $SO(2N+1)$}}
\left(t,\left\{x_{i}=q^{\frac{i}{2}}t^{2i}\right\}_{i=1}^{N-1},x_{N}=q^{\frac{N}{2}+1}t^{2N};q\right)
\nonumber\\
&=\langle T_{(1,0,\cdots,0)}T_{(1,0,\cdots,0)}\rangle_{\textrm{Nahm}}^{\textrm{4d $\mathcal{N}=4$ $SO(2N+1)$}}(t;q)
\mathcal{I}^{\textrm{3d HM}}(t,x=q^{\frac{1}{4}}t;q)^{N-1}
\nonumber\\
&\times 
\prod_{i=1}^{N-2}
\mathcal{I}^{\textrm{3d HM}}(t,x=q^{\frac{4i-1}{4}}t^{4i-1};q)^{N-i-1}
\prod_{i=1}^{N-2}
\mathcal{I}^{\textrm{3d HM}}(t,x=q^{\frac{4i+1}{4}}t^{4i+1};q)^{N-i-1}
\nonumber\\
&\times 
\prod_{i=1}^{2N-1}
\mathcal{I}^{\textrm{3d HM}}(t,x=q^{\frac{2i+3}{4}}t^{2i-1};q). 
\end{align}
For $N>1$ the two-point function of the boundary 't Hooft lines of magnetic charge $B=(1,0,\cdots,0)$ in $SO(2N+1)$ gauge theory with the Nahm pole b.c. 
takes the form 
\begin{align}
\label{so2N+1Nahm_T1}
&
\langle T_{(1,0,\cdots,0)}T_{(1,0,\cdots,0)}\rangle_{\textrm{Nahm}}^{\textrm{4d $\mathcal{N}=4$ $SO(2N+1)$}}(t;q)
\nonumber\\
&=\frac{(q)_{\infty}(q^{\frac32}t^2;q)_{\infty}^2(q^{\frac{2N+3}{2}}t^{4N-2};q)_{\infty}}
{(q^{\frac12}t^2;q)_{\infty}(q^2;q)_{\infty}(qt^4;q)_{\infty}(q^{N+1}t^{4N};q)_{\infty}}
\prod_{k=1}^{N-2}
\frac{(q^{\frac{2k+3}{2}}t^{4k+2};q)_{\infty}}
{(q^{k+1}t^{4k+4};q)_{\infty}}. 
\end{align}
For example, for $N=3,4,5$ the two-point functions are
\begin{align}
\label{so7Nahm_T1}
&
\langle T_{(1,0,0)}T_{(1,0,0)}\rangle_{\textrm{Nahm}}^{\textrm{4d $\mathcal{N}=4$ $SO(7)$}}(t;q)
\nonumber\\
&=\frac{(q)_{\infty}(q^{\frac32}t^2;q)_{\infty}^2(q^{\frac52}t^6;q)_{\infty}(q^{\frac92}t^{10};q)_{\infty}}
{(q^{\frac12}t^2;q)_{\infty}(qt^4;q)_{\infty}(q^2;q)_{\infty}(q^2t^8;q)_{\infty}(q^4t^{12};q)_{\infty}}, \\
\label{so9Nahm_T1}
&
\langle T_{(1,0,0,0)}T_{(1,0,0,0)}\rangle_{\textrm{Nahm}}^{\textrm{4d $\mathcal{N}=4$ $SO(9)$}}(t;q)
\nonumber\\
&=\frac{(q)_{\infty}(q^{\frac32}t^2;q)_{\infty}^2(q^{\frac52}t^6;q)_{\infty}(q^{\frac72}t^{10};q)_{\infty}(q^{\frac{11}{2}}t^{14};q)_{\infty}}
{(q^{\frac12}t^2;q)_{\infty}(qt^4;q)_{\infty}(q^2;q)_{\infty}(q^2t^8;q)_{\infty}(q^3t^{12};q)_{\infty}(q^5t^{16};q)_{\infty}}, \\
\label{so11Nahm_T1}
&
\langle T_{(1,0,0,0,0)}T_{(1,0,0,0,0)}\rangle_{\textrm{Nahm}}^{\textrm{4d $\mathcal{N}=4$ $SO(11)$}}(t;q)
\nonumber\\
&=\frac{(q)_{\infty}(q^{\frac32}t^2;q)_{\infty}^2(q^{\frac52}t^6;q)_{\infty}(q^{\frac72}t^{10};q)_{\infty}(q^{\frac{9}{2}}t^{14};q)_{\infty}(q^{\frac{13}{2}}t^{18};q)_{\infty}}
{(q^{\frac12}t^2;q)_{\infty}(qt^4;q)_{\infty}(q^2;q)_{\infty}(q^2t^8;q)_{\infty}(q^3t^{12};q)_{\infty}(q^4t^{16};q)_{\infty}(q^{6}t^{20};q)_{\infty}}. 
\end{align}
As a consequence of S-duality of the fundamental Wilson line in $USp(2N)$ gauge theory obeying the Neumann b.c. 
and the 't Hooft line of magnetic charge $B=(1,0,\cdots,0)$ in $SO(2N+1)$ gauge theory satisfying the Nahm pole b.c. 
it is expected that the two-point function (\ref{usp2NN_W1}) agrees with the two-point function (\ref{so2N+1Nahm_T1}) after flipping $t$ to $t^{-1}$ 
\begin{align}
\label{dual_CNW1_BNT1}
\langle W_{\tiny \yng(1)} W_{\tiny \yng(1)}\rangle_{\mathcal{N}}^{\textrm{4d $\mathcal{N}=4$ $USp(2N)$}}(t;q)
&=
\langle T_{(1,0,\cdots,0)}T_{(1,0,\cdots,0)}\rangle_{\textrm{Nahm}}^{\textrm{4d $\mathcal{N}=4$ $SO(2N+1)$}}(t^{-1};q). 
\end{align} 

\subsubsection{Direct evaluation by Macdonald polynomials}
Now we would like to calculate the line defect half-indices by making use of the norm formula of the Macdonald polynomials of type $C_N$. 
In this case, the root system is given by
\begin{align}
R&=\{ \pm 2\varepsilon_i| 1\leq i \leq N \} \cup \{ \pm \varepsilon_i \pm \varepsilon_j | 1\leq i<j \leq N \}, \\
R^+&=\{ 2\varepsilon_i| 1\leq i \leq N \} \cup \{ \varepsilon_i \pm \varepsilon_j | 1\leq i<j \leq N \}.
\end{align}
Then,
\begin{align}
\rho=\sum_{i=1}^N ( N-i+1)\varepsilon_i,\qquad \lambda=\sum_{i=1}^N \lambda_i \varepsilon_i,
\end{align}
and
\begin{align}
w_{C_N}(s;\q,\t)=\prod_{i=1}^N \frac{(s_i^{\pm 2};\q)_\infty}{(\t s_i^{\pm 2};\q)_\infty}
\prod_{1\leq i<j \leq N} \frac{(s_i^{\pm 1}s_j^{\pm 1};\q)_\infty}{(\t s_i^{\pm 1}s_j^{\pm 1};\q)_\infty}.
\end{align}

The norm is given by
\begin{align}
&\frac{\langle P_\lambda, P_\lambda \rangle}{\langle 1, 1 \rangle}=\prod_{i=1}^N \frac{(\t^{N-i+2};\q)_{\lambda_i}(\t^{N-i}\q;\q)_{\lambda_i}}{(\t^{N-i+1};\q)_{\lambda_i}(\t^{N-i+1}\q;\q)_{\lambda_i}}\nonumber \\
&\times \prod_{1\leq i<j\leq N} \frac{(\t^{j-i+1};\q)_{\lambda_i-\lambda_j}(\t^{j-i-1}\q;\q)_{\lambda_i-\lambda_j}}{(\t^{j-i};\q)_{\lambda_i-\lambda_j}(\t^{j-i}\q;\q)_{\lambda_i-\lambda_j}}
\frac{(\t^{2N-i-j+3};\q)_{\lambda_i+\lambda_j}(\t^{2N-i-j+1}\q;\q)_{\lambda_i+\lambda_j}}{(\t^{2N-i-j+2};\q)_{\lambda_i+\lambda_j}(\t^{2N-i-j+2}\q;\q)_{\lambda_i+\lambda_j}},
\end{align}
and also we have
\begin{align}
\langle 1, 1 \rangle=\frac{(\t;\q)_\infty^N}{(\q;\q)_\infty^N} \prod_{k=1}^N \frac{(\t^{2k-1}\q;\q)_\infty}{(\t^{2k};\q)_\infty}.
\end{align}

According to \cite{MR3856220}, for the type $C_N$ system, the character with one-column Young diagram, 
i.e. the character of the antisymmetric representation, is related to the Macdonald polynomials by
\begin{align}
\chi_{(1^r)}(s)&=E_r(s)-E_{r-2}(s),\\
E_r(s)&=\sum_{j=0}^{\lfloor \frac{r}{2} \rfloor} \frac{(\q \t; \t^2)_j (\t^{2N-2r+2j+2};\t^2)_j}{(\t^2; \t^2)_j (\q \t^{2N-2r+2j+1};\t^2)_j} P_{(1^{r-2j})}(s;\q, \t).
\end{align}

\paragraph{Fundamental representation.}
For $\lambda=(1)$, we have
\begin{align}
\chi_{(1)}(s)=E_1(s)=P_{(1)}(s;\q,\t).
\end{align}
Hence
\begin{align}
\langle \chi_{(1)}, \chi_{(1)} \rangle=\langle P_{(1)}, P_{(1)} \rangle=\frac{(1-\q)(1-\t^{2N})}{(1-\t)(1-\q \t^{2N-1})}\langle 1, 1 \rangle,
\end{align}
and
\begin{align}
\frac{\langle W_{\square}W_{\square} \rangle_{\mathcal{N}}^{USp(2N)}}{\mathbb{II}_{\mathcal{N}}^{USp(2N)}}&=\frac{(1-\q)(1-\t^{2N})}{(1-\t)(1-\q \t^{2N-1})}.
\end{align}
This is consistent with the duality identity (\ref{dual_CNW1_BNT1}) 
and the expression (\ref{so2N+1Nahm_T1}) for the two-point function of the 't Hooft lines 
of magnetic charge $B=(1,0,\cdots,0)$ in $SO(2N+1)$ gauge theory with the Nahm pole b.c. 

\paragraph{Antisymmetric representations.}
The Wilson lines transforming in other representations are not dual to the 't Hooft lines whose magnetic charges are associated with the minuscule modules. 
This seems to prevent us from deriving the line defect half-indices for the dual 't Hooft lines via the Higgsing procedure. 
However, we can still obtain the exact formulas for the line defect half-indices according to the inner product of the Macdonald polynomials.

If $r=2m$ is even, the inner product $\langle \chi_{(1^{2m})}, 1 \rangle$ is non-vanishing. It is easy to see
\begin{align}
\langle \chi_{(1^{2m})}, 1 \rangle&=\langle E_{2m}, 1 \rangle-\langle E_{2m-2}, 1 \rangle \\
&=\frac{(\q \t; \t^2)_{m} (\t^{2N-2m+2};\t^2)_{m}}{(\t^2; \t^2)_{m} (\q \t^{2N-2m+1};\t^2)_{m}}\langle 1,1 \rangle
-\frac{(\q \t; \t^2)_{m-1} (\t^{2N-2m};\t^2)_{m-1}}{(\t^2; \t^2)_{m-1} (\q \t^{2N-2m-1};\t^2)_{m-1}}\langle 1,1 \rangle.
\end{align}
Thus we obtain
\begin{align}
\frac{\langle W_{(1^{2m})} \rangle_{\mathcal{N}}^{USp(2N)}}{\mathbb{II}_{\mathcal{N}}^{USp(2N)}}=\frac{(\q \t; \t^2)_{m} (\t^{2N-2m+2};\t^2)_{m}}{(\t^2; \t^2)_{m} (\q \t^{2N-2m+1};\t^2)_{m}}
-\frac{(\q \t; \t^2)_{m-1} (\t^{2N-2m};\t^2)_{m-1}}{(\t^2; \t^2)_{m-1} (\q \t^{2N-2m-1};\t^2)_{m-1}}.
\end{align}
In particular, for $m=1$,  we have
\begin{align}
\frac{\langle W_{(1^2)} \rangle_{\mathcal{N}}^{USp(2N)}}{\mathbb{II}_{\mathcal{N}}^{USp(2N)}}=\frac{(1-\q \t)(1-\t^{2N})}{(1-\t^2)(1-\q \t^{2N-1})}-1
=\frac{\t(\t-\q)(1-\t^{2N-2})}{(1-\t^2)(1-\q \t^{2N-1})}.
\end{align}
For $N=2$, this reproduces the result (\ref{usp4N_Wasym2_1pt}). 

\section{$\mathfrak{so}(2N)$}
\label{sec_so2N}

\subsection{$\mathfrak{so}(4)$}
The half-index of the Neumann b.c. of $SO(4)$ gauge theory reads
\begin{align}
\label{so4N}
\mathbb{II}_{\mathcal{N}}^{\textrm{4d $\mathcal{N}=4$ $SO(4)$}}(t;q)
&=\frac{1}{4}
\frac{(q)_{\infty}^2}{(q^{\frac12}t^{-2};q)_{\infty}^2}
\oint \prod_{i=1}^{2}
\frac{ds_i}{2\pi is_i}
\prod_{i<j}
\frac{(s_i^{\pm}s_j^{\mp};q)_{\infty}(s_i^{\pm}s_j^{\pm};q)_{\infty}}
{(q^{\frac12}t^{-2}s_i^{\pm}s_j^{\mp};q)_{\infty}(q^{\frac12}t^{-2}s_i^{\pm}s_j^{\pm};q)_{\infty}}. 
\end{align}
The Neumann b.c. of $SO(4)$ gauge theory is dual to the Nahm pole b.c. of $SO(4)$ Nahm pole b.c. 
The half-index of the Nahm pole b.c. is given by \cite{Hatsuda:2024lcc}
\begin{align}
\label{so4Nahm}
\mathbb{II}_{\textrm{Nahm}}^{\textrm{4d $\mathcal{N}=4$ $SO(4)$}}(t;q)
&=\frac{(q^{\frac32}t^2;q)_{\infty}^2}
{(qt^4;q)_{\infty}^2}. 
\end{align}
As discussed in \cite{Hatsuda:2024lcc},  
it agrees with the Neumann half-index (\ref{so4N}) under the transformation $t$ $\rightarrow$ $t^{-1}$. 
Besides, we have the Dirichlet half-index for $SO(4)$ gauge theory 
\begin{align}
\label{so4D}
\mathbb{II}_{\mathcal{D}}^{\textrm{4d $\mathcal{N}=4$ $SO(4)$}}(t;q)
&=\frac{(q)_{\infty}^2}{(q^{\frac12}t^2;q)_{\infty}}
\prod_{i<j}
\frac{(qx_i^{\pm}x_j^{\mp};q)_{\infty}(qx_i^{\pm}x_j^{\pm};q)_{\infty}}
{(q^{\frac12}t^2x_i^{\pm}x_j^{\mp};q)_{\infty}(q^{\frac12}t^2x_i^{\pm}x_j^{\pm};q)_{\infty}}. 
\end{align}
Applying the Higgsing procedure to the Dirichlet half-index (\ref{so4D}) with
\begin{align}
x_1&=1,\qquad 
x_2=q^{\frac12}t^2, 
\end{align} 
one finds the Higgsed Dirichlet half-index 
\begin{align}
&
\mathbb{II}_{\mathcal{D}}^{\textrm{4d $\mathcal{N}=4$ $SO(4)$}}
\left(t,x_1=1,x_2=q^{\frac12}t^2;q\right)
\nonumber\\
&=\mathbb{II}_{\textrm{Nahm}}^{\textrm{4d $\mathcal{N}=4$ $SO(4)$}}(t;q)
\mathcal{I}^{\textrm{3d HM}}(t,x=q^{\frac14}t;q)^2
\end{align}
and the Nahm pole half-index (\ref{so4Nahm}). 

For the disconnected component that consists of elements of determinant $-1$, 
the Neumann half-index is given by
\begin{align}
\label{so4-N}
\mathbb{II}_{\mathcal{N}}^{\textrm{4d $\mathcal{N}=4$ $SO(4)^{-}$}}(t;q)
&=\frac12 
\frac{(\pm q;q)_{\infty}}
{(\pm q^{\frac12}t^{-2};q)_{\infty}}
\oint \frac{ds}{2\pi is}
\frac{(s^{\pm};q)_{\infty}(-s^{\pm};q)_{\infty}}
{(q^{\frac12}t^{-2}s^{\pm};q)_{\infty}(-q^{\frac12}t^{-2}s^{\pm};q)_{\infty}}. 
\end{align}
The half-index of the Nahm pole half-index for the disconnected part of orthogonal gauge theory is given by \cite{Hatsuda:2024lcc}
\begin{align}
\label{so4-Nahm}
\mathbb{II}_{\textrm{Nahm}}^{\textrm{4d $\mathcal{N}=4$ $SO(4)^{-}$}}(t;q)
&=\frac{(\pm q^{\frac32}t^2;q)_{\infty}}
{(\pm qt^{4};q)_{\infty}}. 
\end{align}
Under the transformation $t$ $\rightarrow$ $t^{-1}$, 
the Neumann half-index (\ref{so4-N}) and the Nahm pole half-index (\ref{so4-Nahm}) coincide. 
Again it can be obtained from the Dirichlet half-index 
\begin{align}
\label{so4-D}
\mathbb{II}_{\mathcal{D}}^{\textrm{4d $\mathcal{N}=4$ $SO(4)^{-}$}}(t,x;q)
&=\frac{(\pm q;q)_{\infty}}
{(\pm q^{\frac12}t^2;q)_{\infty}}
\frac{(qx^{\pm};q)_{\infty}(-qx^{\pm};q)_{\infty}}
{(q^{\frac12}t^2x^{\pm};q)_{\infty}(-q^{\frac12}t^2x^{\pm};q)_{\infty}}
\end{align}
via the Higgsing manipulation with $x=q^{1/2}t^2$. 
By gauging the $\mathbb{Z}_2$ global symmetry, one gets the half-indices of $O(4)$ gauge theory. 
In particular, the Nahm pole half-index is given by \cite{Hatsuda:2024lcc}
\begin{align}
\mathbb{II}_{\textrm{Nahm}}^{\textrm{4d $\mathcal{N}=4$ $O(4)^{\pm}$}}(t;q)
&=\frac12 
\frac{(q^{\frac32}t^2;q)_{\infty}}{(qt^4;q)_{\infty}}
\left[
\frac{(q^{\frac32}t^2;q)_{\infty}}{(qt^4;q)_{\infty}}
\pm 
\frac{(-q^{\frac32}t^2;q)_{\infty}}{(-qt^4;q)_{\infty}}
\right]. 
\end{align}

\subsubsection{Spinor Wilson line}
For $Spin(4)$ gauge theory, there exist Wilson lines transforming in the chiral spinor representation and the antichiral spinor representation. 
The two-point function of a pair of the boundary Wilson lines in the chiral spinor representation of $Spin(4)$ gauge theory with the Neumann b.c. is given by
\begin{align}
\label{so4N_Wsp}
&
\langle W_{\textrm{sp}} W_{\textrm{sp}}\rangle_{\mathcal{N}}^{\textrm{4d $\mathcal{N}=4$ $Spin(4)$}}(t;q)
\nonumber\\
&=\frac{1}{4}
\frac{(q)_{\infty}^2}{(q^{\frac12}t^{-2};q)_{\infty}^2}
\oint \prod_{i=1}^{2}
\frac{ds_i}{2\pi is_i}
\prod_{i<j}
\frac{(s_i^{\pm}s_j^{\mp};q)_{\infty}(s_i^{\pm}s_j^{\pm};q)_{\infty}}
{(q^{\frac12}t^{-2}s_i^{\pm}s_j^{\mp};q)_{\infty}(q^{\frac12}t^{-2}s_i^{\pm}s_j^{\pm};q)_{\infty}}
\nonumber\\
&\times (2+s_1s_2+s_1^{-1}s_2^{-1}). 
\end{align}
Analogously, we have the two-point function of the antichiral spinor Wilson lines that is equivalent to (\ref{so4N_Wsp}). 
The spinor Wilson line in $Spin(4)$ gauge theory is dual to the 't Hooft line of magnetic charge $(\frac12,\frac12)$ in $SO(4)/\mathbb{Z}_2$ gauge theory. 
We propose that the two-point function of the dual 't Hooft lines can be obtained via the Higgsing procedure 
by specializing the global fugacities of the $SO(4)$ Dirichlet half-index (\ref{so4D}) as
\begin{align}
x_1&=q^{\frac12},\qquad 
x_2=qt^2. 
\end{align}
Then one finds
\begin{align}
&
\mathbb{II}_{\mathcal{D}}^{\textrm{4d $\mathcal{N}=4$ $SO(4)$}}
\left(t,x_1=q^{\frac12},x_2=qt^2;q\right)
\nonumber\\
&=\langle T_{(\frac12,\frac12)}T_{(\frac12,\frac12)}\rangle_{\textrm{Nahm}}^{\textrm{4d $\mathcal{N}=4$ $SO(4)/\mathbb{Z}_2$}}(t;q)
\mathcal{I}^{\textrm{3d HM}}(t,x=q^{\frac14}t;q)
\mathcal{I}^{\textrm{3d HM}}(t,x=q^{\frac54}t;q), 
\end{align}
where the two-point function of the dual 't Hooft lines of magnetic charge $B=(\frac12,\frac12)$ in $SO(4)/\mathbb{Z}_2$ gauge theory with the Nahm pole b.c. is given by
\begin{align}
\label{so4Nahm_T1/2}
\langle T_{(\frac12,\frac12)}T_{(\frac12,\frac12)}\rangle_{\textrm{Nahm}}^{\textrm{4d $\mathcal{N}=4$ $SO(4)/\mathbb{Z}_2$}}(t;q)
&=\frac{(q)_{\infty}(q^{\frac32}t^2;q)_{\infty}^2(q^{\frac52}t^2;q)_{\infty}}
{(q^{\frac12}t^2;q)_{\infty}(qt^4;q)_{\infty}(q^2;q)_{\infty}(q^2t^4;q)_{\infty}}. 
\end{align}
In fact, we find that the two-point functions (\ref{so4N_Wsp}) and (\ref{so4Nahm_T1/2}) coincide under $t$ $\rightarrow$ $t^{-1}$
\begin{align}
\langle W_{\textrm{sp}} W_{\textrm{sp}}\rangle_{\mathcal{N}}^{\textrm{4d $\mathcal{N}=4$ $Spin(4)$}}(t;q)
&=
\langle T_{(\frac12,\frac12)}T_{(\frac12,\frac12)}\rangle_{\textrm{Nahm}}^{\textrm{4d $\mathcal{N}=4$ $SO(4)/\mathbb{Z}_2$}}(t^{-1};q). 
\end{align}
Because of the isomorphism $Spin(4)$ $\cong$ $SU(2)\times SU(2)$ 
and $SO(4)$ $\cong$ $(SU(2)\times SU(2))/\mathbb{Z}_2$, 
they are also expressible in terms of the results for $SU(2)$ gauge theory. 
We have 
\begin{align}
\langle W_{\textrm{sp}} W_{\textrm{sp}}\rangle_{\mathcal{N}}^{\textrm{4d $\mathcal{N}=4$ $Spin(4)$}}(t;q)
&=\mathbb{II}_{\mathcal{N}}^{\textrm{4d $\mathcal{N}=4$ $SU(2)$}}(t;q)
\langle W_{\tiny \yng(1)} W_{\tiny \yng(1)}\rangle_{\mathcal{N}}^{\textrm{4d $\mathcal{N}=4$ $SU(2)$}}(t;q). 
\end{align}

\subsubsection{Fundamental Wilson line}
The one-point function of the boundary Wilson line in the fundamental representation of $SO(4)$ gauge theory with the Neumann b.c. vanishes. 
The two-point function takes the form 
\begin{align}
\label{so4N_W1}
&
\langle W_{\tiny \yng(1)} W_{\tiny \yng(1)}\rangle_{\mathcal{N}}^{\textrm{4d $\mathcal{N}=4$ $SO(4)$}}(t;q)
\nonumber\\
&=\frac{1}{4}
\frac{(q)_{\infty}^2}{(q^{\frac12}t^{-2};q)_{\infty}^2}
\oint \prod_{i=1}^{2}
\frac{ds_i}{2\pi is_i}
\prod_{i<j}
\frac{(s_i^{\pm}s_j^{\mp};q)_{\infty}(s_i^{\pm}s_j^{\pm};q)_{\infty}}
{(q^{\frac12}t^{-2}s_i^{\pm}s_j^{\mp};q)_{\infty}(q^{\frac12}t^{-2}s_i^{\pm}s_j^{\pm};q)_{\infty}}
(s_1+s_2+s_1^{-1}+s_2^{-1})^2. 
\end{align}
The dual of the fundamental Wilson line is conjectured to be the 't Hooft line of magnetic charge $B=(1,0)$. 
We propose that the two-point function of the dual 't Hooft lines with the Nahm pole b.c. 
can be found by deforming the $SO(4)$ Dirichlet half-index (\ref{so4D}) via the Higgsing manipulation. 
Let us specialize global fugacities of the $SO(4)$ Dirichlet half-index (\ref{so4D}) as
\begin{align}
x_1&=1,\qquad 
x_2=q^{\frac32}t^2. 
\end{align}
Consequently, the $SO(4)$ Dirichlet half-index (\ref{so4D}) is Higgsed as
\begin{align}
&
\mathbb{II}_{\mathcal{D}}^{\textrm{4d $\mathcal{N}=4$ $SO(4)$}}
\left(t,x_1=1,x_2=q^{\frac32}t^2;q\right)
\nonumber\\
&=\langle T_{(1,0)}T_{(1,0)}\rangle_{\textrm{Nahm}}^{\textrm{4d $\mathcal{N}=4$ $SO(4)$}}(t;q)
\mathcal{I}^{\textrm{3d HM}}(t,x=q^{\frac54}t;q)^2. 
\end{align}
The resulting two-point function of the dual 't Hooft line of magnetic charge $B=(1,0)$ 
with the Nahm pole b.c. takes the form 
\begin{align}
\label{so4Nahm_T1}
\langle T_{(1,0)}T_{(1,0)}\rangle_{\textrm{Nahm}}^{\textrm{4d $\mathcal{N}=4$ $SO(4)$}}(t;q)
&=\frac{(q)_{\infty}^2 (q^{\frac32}t^{2};q)_{\infty}^2 (q^{\frac52}t^{2};q)_{\infty}^2}
{(q^{\frac12}t^{2};q)_{\infty}^2(q^2;q)_{\infty}^2(q^2t^{4};q)_{\infty}^2}. 
\end{align}
We have confirmed that 
the two-point functions (\ref{so4N_W1}) and (\ref{so4Nahm_T1}) agree with each other under the transformation $t$ $\rightarrow$ $t^{-1}$
\begin{align}
\langle W_{\tiny \yng(1)} W_{\tiny \yng(1)}\rangle_{\mathcal{N}}^{\textrm{4d $\mathcal{N}=4$ $SO(4)$}}(t;q)
&=\langle T_{(1,0)}T_{(1,0)}\rangle_{\textrm{Nahm}}^{\textrm{4d $\mathcal{N}=4$ $SO(4)$}}(t^{-1};q). 
\end{align}
Again the isomorphism $Spin(4)$ $\cong$ $SU(2)\times SU(2)$ 
and $SO(4)$ $\cong$ $(SU(2)\times SU(2))/\mathbb{Z}_2$ allow us to write them 
as the $SU(2)$ correlators
\begin{align}
\langle W_{\tiny \yng(1)} W_{\tiny \yng(1)}\rangle_{\mathcal{N}}^{\textrm{4d $\mathcal{N}=4$ $SO(4)$}}(t;q)
&={\langle W_{\tiny \yng(1)} W_{\tiny \yng(1)}\rangle_{\mathcal{N}}^{\textrm{4d $\mathcal{N}=4$ $SU(2)$}}(t;q)}^2. 
\end{align}

Next consider the disconnected component that consists of elements of determinant $-1$. 
The two-point function of the boundary Wilson line in the fundamental representation satisfying the Neumann b.c. is 
\begin{align}
\label{so4-N_W1}
&
\langle W_{\tiny \yng(1)} W_{\tiny \yng(1)}\rangle_{\mathcal{N}}^{\textrm{4d $\mathcal{N}=4$ $SO(4)^{-}$}}(t;q)
\nonumber\\
&=\frac12 
\frac{(\pm q;q)_{\infty}}
{(\pm q^{\frac12}t^{-2};q)_{\infty}}
\oint \frac{ds}{2\pi is}
\frac{(s^{\pm};q)_{\infty}(-s^{\pm};q)_{\infty}}
{(q^{\frac12}t^{-2}s^{\pm};q)_{\infty}(-q^{\frac12}t^{-2}s^{\pm};q)_{\infty}}. 
\end{align}
The two-point function of the dual 't Hooft line can be also obtained by following the Higgsing procedure. 
By fixing the global fugacities of the $SO(4)^{-}$ Dirichlet half-index (\ref{so4-D}) as $x=q^{3/2}t^2$, 
we find
\begin{align}
&
\mathbb{II}_{\mathcal{D}}^{\textrm{4d $\mathcal{N}=4$ $SO(4)^{-}$}}
\left(t,x=q^{\frac32}t^2;q\right)
\nonumber\\
&=\langle T_{(1)}T_{(1)}\rangle_{\textrm{Nahm}}^{\textrm{4d $\mathcal{N}=4$ $SO(4)^{-}$}}(t;q)
\mathcal{I}^{\textrm{3d HM}}(t,x=q^{\frac54}t;q)
\mathcal{I}^{\textrm{3d HM}}(t,x=-q^{\frac54}t;q). 
\end{align}
The two-point function of the dual 't Hooft line is given by
\begin{align}
\label{so4-Nahm_T1}
\langle T_{(1)}T_{(1)}\rangle_{\textrm{Nahm}}^{\textrm{4d $\mathcal{N}=4$ $SO(4)^{-}$}}(t;q)
&=
\frac{(\pm q;q)_{\infty} (\pm q^{\frac32}t^2;q)_{\infty} (\pm q^{\frac52}t^2;q)_{\infty}}
{(\pm q^{\frac12}t^2;q)_{\infty} (\pm q^{2};q)_{\infty} (\pm q^2t^4;q)_{\infty}}. 
\end{align}
Indeed, upon the flip $t$ to $t^{-1}$, 
this agrees with the two-point function (\ref{so4-N_W1}) of the fundamental Wilson lines subject to the Neumann b.c. 
It is straightforward to get the two-point function of the fundamental Wilson lines of $O(4)$ gauge theory 
with the Neumann b.c. by gauging the $\mathbb{Z}_2$ global symmetry 
\begin{align}
\label{o4N_W1}
&
\langle W_{\tiny \yng(1)} W_{\tiny \yng(1)}\rangle_{\mathcal{N}}^{\textrm{4d $\mathcal{N}=4$ $O(4)^{\pm}$}}(t;q)
\nonumber\\
&=\frac12\left[
\langle W_{\tiny \yng(1)} W_{\tiny \yng(1)}\rangle_{\mathcal{N}}^{\textrm{4d $\mathcal{N}=4$ $SO(4)$}}(t;q)
\pm 
\langle W_{\tiny \yng(1)} W_{\tiny \yng(1)}\rangle_{\mathcal{N}}^{\textrm{4d $\mathcal{N}=4$ $SO(4)^{-}$}}(t;q)
\right]. 
\end{align}
We have 
\begin{align}
\langle W_{\tiny \yng(1)} W_{\tiny \yng(1)}\rangle_{\mathcal{N}}^{\textrm{4d $\mathcal{N}=4$ $O(4)^{\pm}$}}(t;q)
&=
\langle T_{(1,0)}T_{(1,0)}\rangle_{\textrm{Nahm}}^{\textrm{4d $\mathcal{N}=4$ $O(4)$}}(t^{-1};q), 
\end{align}
where 
\begin{align}
&
\langle T_{(1,0)}T_{(1,0)}\rangle_{\textrm{Nahm}}^{\textrm{4d $\mathcal{N}=4$ $O(4)$}}(t;q)
=\frac12
\frac{(q)_{\infty}(q^{\frac32}t^2;q)_{\infty}(q^{\frac52}t^2;q)_{\infty}}
{(q^{\frac12}t^2;q)_{\infty}(q^2;q)_{\infty}(q^{2}t^4;q)_{\infty}}
\nonumber\\
&\times 
\left[
\frac{(q)_{\infty}(q^{\frac32}t^2;q)_{\infty}(q^{\frac52}t^2;q)_{\infty}}
{(q^{\frac12}t^2;q)_{\infty}(q^2;q)_{\infty}(q^{2}t^4;q)_{\infty}}
\pm 
\frac{(-q;q)_{\infty}(-q^{\frac32}t^2;q)_{\infty}(-q^{\frac52}t^2;q)_{\infty}}
{(-q^{\frac12}t^2;q)_{\infty}(-q^2;q)_{\infty}(-q^{2}t^4;q)_{\infty}}
\right]. 
\end{align}

\subsubsection{Antisymmetric Wilson line}
Next consider the Wilson lines in the higher rank representations. 
In this case one-point line defect correlation functions can be non-trivial. 
For the rank-$2$ antisymmetric representation, 
i.e. the adjoint representation, 
the one-point function is given by
\begin{align}
\label{so4N_Wasym2}
\langle W_{\tiny \yng(1,1)} \rangle_{\mathcal{N}}^{\textrm{4d $\mathcal{N}=4$ $SO(4)$}}(t;q)
&=
\langle W_{\overline{\tiny \yng(1,1)}} \rangle_{\mathcal{N}}^{\textrm{4d $\mathcal{N}=4$ $SO(4)$}}(t;q)
\nonumber\\
&=q^{\frac12}t^{-2}
\frac{(q^{\frac12}t^2;q)_{\infty}(q^{\frac32}t^{-2};q)_{\infty}(q^{\frac52}t^{-2};q)_{\infty}}
{(qt^{-4};q)^2(q^{\frac32}t^{2};q)_{\infty}}. 
\end{align}
Similarly to the line defect index without boundary (see \cite{Hatsuda:2025jze}), 
the expression (\ref{so4N_Wasym2}) can be simply a product of the $SU(2)$ half-index (\ref{usp2N}) and the one-point function (\ref{usp2N_Wsym2k_1pt}) 
of the rank-$2$ $SU(2)$ symmetric Wilson line. 

Also we observe that the multi-point function of the antisymmetric Wilson lines has factorization property
\begin{align}
&
\langle (W_{\tiny \yng(1,1)})^k (W_{\overline{\tiny \yng(1,1)}})^m \rangle_{\mathcal{N}}^{\textrm{4d $\mathcal{N}=4$ $SO(4)$}}(t;q)
\nonumber\\
&=\frac{(qt^{-4};q)_{\infty}^2}{(q^{\frac32}t^{-2};q)_{\infty}^2}
\langle (W_{\tiny \yng(1,1)})^k \rangle_{\mathcal{N}}^{\textrm{4d $\mathcal{N}=4$ $SO(4)$}}(t;q)
\langle (W_{\overline{\tiny \yng(1,1)}})^m \rangle_{\mathcal{N}}^{\textrm{4d $\mathcal{N}=4$ $SO(4)$}}(t;q). 
\end{align}
This generalizes the factorization property of the half-BPS $SO(4)$ antisymmetric Wilson line defect indices in \cite{Hatsuda:2025jze}. 

\subsubsection{Symmetric Wilson lines}
For the rank-$2$ symmetric Wilson line, the one-point function can be expressed as
\begin{align}
\label{so4N_Wsym2}
\langle W_{\tiny \yng(2)} \rangle_{\mathcal{N}}^{\textrm{4d $\mathcal{N}=4$ $SO(4)$}}(t;q)
&=qt^{-4}
\frac{(q^{\frac12}t^2;q)_{\infty}^2(q^{\frac52}t^{-2};q)_{\infty}^2}
{(qt^{-4};q)_{\infty}^2(q^{\frac32}t^2;q)_{\infty}^2}. 
\end{align}
This can be identified with the square of the one-point function 
(\ref{usp2N_Wsym2k_1pt}) of the rank-$2$ $SU(2)$ symmetric Wilson line. 

The multi-point function of the symmetric Wilson lines can be factorized as
\begin{align}
\langle \underbrace{W_{(l)}\cdots W_{(l)}}_{k} \rangle_{\mathcal{N}}^{\textrm{4d $\mathcal{N}=4$ $SO(4)$}}(t;q)
&={\langle \underbrace{W_{(l)}\cdots W_{(l)}}_{k} \rangle_{\mathcal{N}}^{\textrm{4d $\mathcal{N}=4$ $SU(2)$}}(t;q)}^2, 
\end{align}
where $\langle \underbrace{W_{(l)}\cdots W_{(l)}}_{k} \rangle_{\mathcal{N}}^{\textrm{4d $\mathcal{N}=4$ $SU(2)$}}(t;q)$ 
is the $k$-point function of the Wilson lines in the rank-$l$ symmetric representation in $\mathcal{N}=4$ $SU(2)$ SYM theory. 
The correlator is non-trivial when either $l$ or $k$ is even. 
This is the same factorization property of the $SO(4)$ symmetric Wilson line defect indices without boundary observed in \cite{Hatsuda:2025jze}.

For the disconnected component the correlation function of the rank-$l$ symmetric Wilson lines 
can be expressed in terms of the $SU(2)$ correlators as
\begin{align}
\langle \underbrace{W_{(l)}\cdots W_{(l)}}_{k} \rangle_{\mathcal{N}}^{\textrm{4d $\mathcal{N}=4$ $SO(4)$}}(t;q)
&=\langle \underbrace{W_{(l)}\cdots W_{(l)}}_{k} \rangle_{\mathcal{N}}^{\textrm{4d $\mathcal{N}=4$ $SU(2)$}}(t^2;q^2), 
\end{align}
Again this takes the same form as the factorization of the line defect indices without boundary in \cite{Hatsuda:2025jze}.

\subsection{$\mathfrak{so}(6)$}
Next consider $\mathcal{N}=4$ SYM theory based on the gauge algebra $\mathfrak{so}(6)$. 
We have the half-index of the Neumann b.c. of $SO(6)$ gauge theory of the form
\begin{align}
\label{so6N}
\mathbb{II}_{\mathcal{N}}^{\textrm{4d $\mathcal{N}=4$ $SO(6)$}}(t;q)
&=\frac{1}{24}
\frac{(q)_{\infty}^3}{(q^{\frac12}t^{-2};q)_{\infty}^3}
\oint \prod_{i=1}^{3}
\frac{ds_i}{2\pi is_i}
\prod_{i<j}
\frac{(s_i^{\pm}s_j^{\mp};q)_{\infty}(s_i^{\pm}s_j^{\pm};q)_{\infty}}
{(q^{\frac12}t^{-2}s_i^{\pm}s_j^{\mp};q)_{\infty}(q^{\frac12}t^{-2}s_i^{\pm}s_j^{\pm};q)_{\infty}}. 
\end{align}
The half-index of the Nahm pole b.c. of $SO(6)$ gauge theory is given by \cite{Hatsuda:2024lcc}
\begin{align}
\label{so6Nahm}
\mathbb{II}_{\textrm{Nahm}}^{\textrm{4d $\mathcal{N}=4$ $SO(6)$}}(t;q)
&=\frac{(q^{\frac32}t^2;q)_{\infty}(q^2t^4;q)_{\infty}(q^{\frac52}t^6;q)_{\infty}}
{(qt^4;q)_{\infty}(q^{\frac32}t^6;q)_{\infty}(q^2t^8;q)_{\infty}}. 
\end{align}
As the Neumann b.c. and the Nahm pole b.c. in $SO(6)$ gauge theory are dual, 
the Neumann half-index (\ref{so6N}) coincide with the Nahm pole half-index \ref{so6Nahm}) under the transformation 
$t$ $\rightarrow$ $t^{-1}$ \cite{Hatsuda:2024lcc}. 
On the other hand, the $SO(6)$ Dirichlet half-index is given by 
\begin{align}
\label{so6D}
\mathbb{II}_{\mathcal{D}}^{\textrm{4d $\mathcal{N}=4$ $SO(6)$}}(t,x_1,x_2,x_3;q)
&=\frac{(q)_{\infty}^3}{(q^{\frac12}t^2;q)_{\infty}^3}
\prod_{i<j}
\frac{(qx_i^{\pm}x_j^{\mp};q)_{\infty}(qx_i^{\pm}x_j^{\pm};q)_{\infty}}
{(q^{\frac12}t^2x_i^{\pm}x_j^{\mp};q)_{\infty}(q^{\frac12}t^2x_i^{\pm}x_j^{\pm};q)_{\infty}}. 
\end{align}
The Nahm pole half-index (\ref{so6Nahm}) can be obtained by specializing the global fugacities 
of the Dirichlet half-index (\ref{so6D}) as
\begin{align}
x_1&=1,\qquad 
x_2=q^{\frac12}t^2,\qquad 
x_3=qt^4, 
\end{align}
and stripping off the extra 3d indices contributed from the decoupled matter fields 
as one find the Higgsed Dirichlet half-index
\begin{align}
&
\mathbb{II}_{\mathcal{D}}^{\textrm{4d $\mathcal{N}=4$ $SO(6)$}}
\left(t,x_1=1,x_2=q^{\frac12}t^2,x_3=qt^4;q\right)
\nonumber\\
&=\mathbb{II}_{\textrm{Nahm}}^{\textrm{4d $\mathcal{N}=4$ $SO(6)$}}(t;q)
\mathcal{I}^{\textrm{3d HM}}(t,x=q^{\frac14}t;q)^3
\nonumber\\
&\times 
\mathcal{I}^{\textrm{3d HM}}(t,x=q^{\frac34}t^3;q)^2
\mathcal{I}^{\textrm{3d HM}}(t,x=q^{\frac54}t^5;q). 
\end{align}

For the disconnected component the Neumann half-index is given by 
\begin{align}
\label{so6-N}
&
\mathbb{II}_{\mathcal{N}}^{\textrm{4d $\mathcal{N}=4$ $SO(6)^{-}$}}(t;q)
\nonumber\\
&=\frac{1}{8}
\frac{(q)_{\infty}^2 (-q;q)_{\infty}}
{(q^{\frac12}t^2;q)_{\infty}^2(-q^{\frac12}t^2;q)_{\infty}}
\oint \prod_{i=1}^2 
\frac{ds_i}{2\pi is_i}
\frac{(s_i^{\pm};q)_{\infty}(-s_i^{\pm};q)_{\infty}}
{(q^{\frac12}t^{-2}s_i^{\pm};q)_{\infty}(-q^{\frac12}t^{-2}s_i^{\pm};q)_{\infty}}
\nonumber\\
&\times 
\prod_{i< j}
\frac{(s_i^{\pm}s_j^{\mp};q)_{\infty}(s_i^{\pm}s_j^{\pm};q)_{\infty}}
{(q^{\frac12}t^{-2}s_i^{\pm}s_j^{\mp};q)_{\infty}(q^{\frac12}t^{-2}s_i^{\pm}s_j^{\pm};q)_{\infty}}. 
\end{align}
Upon the transformation $t$ $\rightarrow$ $t^{-1}$, 
this agrees with the half-index \cite{Hatsuda:2024lcc}
\begin{align}
\label{so6-Nahm}
\mathbb{II}_{\textrm{Nahm}}^{\textrm{4d $\mathcal{N}=4$ $SO(6)^{-}$}}(t;q)
&=\frac{(q^{\frac32}t^2;q)_{\infty}(q^{\frac52}t^6;q)_{\infty}(-q^2t^4;q)_{\infty}}
{(qt^4;q)_{\infty}(q^2t^8;q)_{\infty}(-q^{\frac32}t^6;q)_{\infty}}
\end{align}
of the Nahm pole b.c. for the disconnected part. 
One can get from the Dirichlet half-index (\ref{so6-D}) 
\begin{align}
&
\label{so6-D}
\mathbb{II}_{\mathcal{D}}^{\textrm{4d $\mathcal{N}=4$ $SO(6)^{-}$}}(t;q)
\nonumber\\
&=\frac{(q)_{\infty}^2(-q;q)_{\infty}}
{(q^{\frac12}t^2;q)_{\infty}^2(-q^{\frac12}t^2;q)_{\infty}}
\prod_{i=1}^3
\frac{(qx_i^{\pm};q)_{\infty} (-qx_i^{\pm};q)_{\infty}}
{(q^{\frac12}t^2x_i^{\pm};q)_{\infty} (-q^{\frac12}t^2x_i^{\pm};q)_{\infty}}
\nonumber\\
&\times 
\prod_{i< j}
\frac{(qx_i^{\pm}x_j^{\mp};q)_{\infty}(qx_i^{\pm}x_j^{\pm};q)_{\infty}}
{(q^{\frac12}t^{-2}x_i^{\pm}x_j^{\mp};q)_{\infty}(q^{\frac12}t^{-2}x_i^{\pm}x_j^{\pm};q)_{\infty}}
\end{align}
the Nahm pole half-index (\ref{so6-Nahm}) by means of the Higgsing with 
\begin{align}
x_1&=q^{\frac12}t^2,\qquad 
x_2=qt^4. 
\end{align}

\subsubsection{Spinor Wilson lines}
The $Spin(6)$ gauge theory admits the Wilson lines in the chiral and the antichiral representations which are complex conjugate. 
The two-point function for a pair of the conjugate boundary Wilson line in the spinor representations with the Neumann b.c. is non-trivial. 
It takes the form
\begin{align}
\label{so6N_Wsp}
&
\langle W_{\textrm{sp}} W_{\overline{\textrm{sp}}}\rangle_{\mathcal{N}}^{\textrm{4d $\mathcal{N}=4$ $Spin(6)$}}(t;q)
\nonumber\\
&=\frac{1}{24}
\frac{(q)_{\infty}^3}{(q^{\frac12}t^{-2};q)_{\infty}^3}
\oint \prod_{i=1}^{3}
\frac{ds_i}{2\pi is_i}
\prod_{i<j}
\frac{(s_i^{\pm}s_j^{\mp};q)_{\infty}(s_i^{\pm}s_j^{\pm};q)_{\infty}}
{(q^{\frac12}t^{-2}s_i^{\pm}s_j^{\mp};q)_{\infty}(q^{\frac12}t^{-2}s_i^{\pm}s_j^{\pm};q)_{\infty}}
\nonumber\\
&\times (4+\sum_{i<j}s_is_j+s_i^{-1}sj^{-1}+s_is_j^{-1}+s_i^{-1}s_j). 
\end{align}
The spinor Wilson line is conjecturally dual to the 't Hooft line of magnetic charge $B=(\frac12,\frac12,\frac12)$ of $SO(6)/\mathbb{Z}_2$ gauge theory. 
The two-point function of the dual 't Hooft line with the Nahm pole b.c. can be found 
from the $SO(6)$ Dirichlet half-index (\ref{so6D}) via the Higgsing procedure. 
Let us specialize the global fugacities of the $SO(6)$ Dirichlet half-index (\ref{so6D}) as
\begin{align}
x_1&=q^{\frac12},\qquad 
x_2=qt^2,\qquad 
x_3=q^{\frac32}t^4. 
\end{align}
The Higgsed $SO(6)$ Dirichlet half-index takes the form
\begin{align}
&
\mathbb{II}_{\mathcal{D}}^{\textrm{4d $\mathcal{N}=4$ $SO(6)$}}
\left(t,x_1=q^{\frac12},x_2=qt^2,x_3=q^{\frac32}t^4;q\right)
\nonumber\\
&=\langle T_{(\frac12,\frac12,\frac12)}T_{(\frac12,\frac12,\frac12)}\rangle_{\textrm{Nahm}}^{\textrm{4d $\mathcal{N}=4$ $SO(6)/\mathbb{Z}_2$}}(t;q)
\mathcal{I}^{\textrm{3d HM}}(t,x=q^{\frac14}t;q)^2
\mathcal{I}^{\textrm{3d HM}}(t,x=q^{\frac34}t^3;q)
\nonumber\\
&\times 
\mathcal{I}^{\textrm{3d HM}}(t,x=q^{\frac54}t;q)
\mathcal{I}^{\textrm{3d HM}}(t,x=q^{\frac74}t^3;q)
\mathcal{I}^{\textrm{3d HM}}(t,x=q^{\frac94}t^5;q). 
\end{align}
The two-point function of the boundary 't Hooft line 
of magnetic charge $B=(\frac12,\frac12,\frac12)$ of $SO(6)/\mathbb{Z}_2$ gauge theory with the Nahm pole b.c. is given by
\begin{align}
\label{so6Nahm_T1/2}
&
\langle T_{(\frac12,\frac12,\frac12)}T_{(\frac12,\frac12,\frac12)}\rangle_{\textrm{Nahm}}^{\textrm{4d $\mathcal{N}=4$ $SO(6)/\mathbb{Z}_2$}}(t;q)
\nonumber\\
&=\frac{(q)_{\infty}(q^{\frac32}t^2;q)_{\infty}^2(q^2t^4;q)_{\infty}(q^{\frac72}t^6;q)_{\infty}}
{(q^{\frac12}t^2;q)_{\infty}(qt^4;q)_{\infty}(q^{\frac32}t^6;q)_{\infty}(q^2;q)_{\infty}(q^3t^8;q)_{\infty}}. 
\end{align}
In fact, we find that 
the two expressions (\ref{so6N_Wsp}) and (\ref{so6Nahm_T1/2}) agree with each other after flipping $t$ to $t^{-1}$
\begin{align}
\langle W_{\textrm{sp}} W_{\overline{\textrm{sp}}}\rangle_{\mathcal{N}}^{\textrm{4d $\mathcal{N}=4$ $Spin(6)$}}(t;q)
&=\langle T_{(\frac12,\frac12,\frac12)}T_{(\frac12,\frac12,\frac12)}\rangle_{\textrm{Nahm}}^{\textrm{4d $\mathcal{N}=4$ $SO(6)/\mathbb{Z}_2$}}(t^{-1};q). 
\end{align}
This supports the duality of the spinor Wilson line of $Spin(6)$ gauge theory subject to the Neumann b.c. 
and the 't Hooft line of magnetic charge $B=(\frac12,\frac12,\frac12)$ of $SO(6)/\mathbb{Z}_2$ gauge theory obeying the Nahm pole b.c. 

\subsubsection{Fundamental Wilson line}
The two-point function of the of the boundary Wilson line in the fundamental representation of $SO(6)$ gauge theory with the Neumann b.c. is given by
\begin{align}
\label{so6N_W1}
&
\langle W_{\tiny \yng(1)} W_{\tiny \yng(1)}\rangle_{\mathcal{N}}^{\textrm{4d $\mathcal{N}=4$ $SO(6)$}}(t;q)
\nonumber\\
&=\frac{1}{24}
\frac{(q)_{\infty}^3}{(q^{\frac12}t^{-2};q)_{\infty}^3}
\oint \prod_{i=1}^{3}
\frac{ds_i}{2\pi is_i}
\prod_{i<j}
\frac{(s_i^{\pm}s_j^{\mp};q)_{\infty}(s_i^{\pm}s_j^{\pm};q)_{\infty}}
{(q^{\frac12}t^{-2}s_i^{\pm}s_j^{\mp};q)_{\infty}(q^{\frac12}t^{-2}s_i^{\pm}s_j^{\pm};q)_{\infty}}
\nonumber\\
&\times (s_1+s_2+s_3+s_1^{-1}+s_2^{-1}+s_3^{-1})^2. 
\end{align}
The fundamental Wilson line is dual to the 't Hooft line of magnetic charge $B=(1,0,0)$. 
We propose that the two-point function of the dual 't Hooft lines can be found 
by Higgsing the $SO(6)$ Dirichlet half-index (\ref{so6D}) with the following specialization of the global fugacities: 
\begin{align}
x_1&=1,\qquad 
x_2=q^{\frac12}t^2,\qquad 
x_3=q^2t^4. 
\end{align}
As a result, the $SO(6)$ Dirichlet half-index (\ref{so6D}) is factorized as
\begin{align}
&
\mathbb{II}_{\mathcal{D}}^{\textrm{4d $\mathcal{N}=4$ $SO(6)$}}
\left(t,x_1=1,x_2=q^{\frac12}t^2,x_3=q^{2}t^4;q\right)
\nonumber\\
&=\langle T_{(1,0,0)}T_{(1,0,0)}\rangle_{\textrm{Nahm}}^{\textrm{4d $\mathcal{N}=4$ $SO(6)$}}(t;q)
\mathcal{I}^{\textrm{3d HM}}(t,x=q^{\frac14}t;q)^2
\nonumber\\
&\times 
\mathcal{I}^{\textrm{3d HM}}(t,x=q^{\frac54}t;q)
\mathcal{I}^{\textrm{3d HM}}(t,x=q^{\frac74}t^3;q)^2
\mathcal{I}^{\textrm{3d HM}}(t,x=q^{\frac94}t^5;q). 
\end{align}
We obtain the two-point function of the dual 't Hooft line of magnetic charge $B=(1,0,0)$ of $SO(6)$ gauge theory with the Nahm pole b.c. of the form 
\begin{align}
\label{so6Nahm_T1}
&
\langle T_{(1,0,0)}T_{(1,0,0)}\rangle_{\textrm{Nahm}}^{\textrm{4d $\mathcal{N}=4$ $SO(6)$}}(t;q)
\nonumber\\
&=\frac{(q)_{\infty}(q^{\frac32}t^2;q)_{\infty}^3(q^2t^4;q)_{\infty}(q^3t^4;q)_{\infty}(q^{\frac72}t^6;q)_{\infty}}
{(q^{\frac12}t^2;q)_{\infty}(qt^4;q)_{\infty}(q^2;q)_{\infty}(q^{\frac52}t^2;q)_{\infty}(q^{\frac52}t^6;q)_{\infty}(q^3t^8;q)_{\infty}}. 
\end{align}
In fact, we find the precise matching of the two-point functions (\ref{so6N_W1}) and (\ref{so6Nahm_T1}) under the transformation $t$ $\rightarrow$ $t^{-1}$
as strong evidence of dualities of the line operators along with the boundary conditions
\begin{align}
\langle W_{\tiny \yng(1)} W_{\tiny \yng(1)}\rangle_{\mathcal{N}}^{\textrm{4d $\mathcal{N}=4$ $SO(6)$}}(t;q)
&=\langle T_{(1,0,0)}T_{(1,0,0)}\rangle_{\textrm{Nahm}}^{\textrm{4d $\mathcal{N}=4$ $SO(6)$}}(t^{-1};q). 
\end{align}

Next consider the disconnected component consisting of elements with the determinant $-1$. 
The two-point function of the boundary Wilson lines in the fundamental representation 
obeying the Neumann b.c. takes the form
\begin{align}
\label{so6-N_W1}
&
\langle W_{\tiny \yng(1)} W_{\tiny \yng(1)}\rangle_{\mathcal{N}}^{\textrm{4d $\mathcal{N}=4$ $SO(6)^{-}$}}(t;q)
\nonumber\\
&=\frac{1}{8}
\frac{(q)_{\infty}^2 (-q;q)_{\infty}}
{(q^{\frac12}t^2;q)_{\infty}^2(-q^{\frac12}t^2;q)_{\infty}}
\oint \prod_{i=1}^2 
\frac{ds_i}{2\pi is_i}
\frac{(s_i^{\pm};q)_{\infty}(-s_i^{\pm};q)_{\infty}}
{(q^{\frac12}t^{-2}s_i^{\pm};q)_{\infty}(-q^{\frac12}t^{-2}s_i^{\pm};q)_{\infty}}
\nonumber\\
&\times 
\prod_{i\neq j}
\frac{(s_i^{\pm}s_j^{\mp};q)_{\infty}(s_i^{\pm}s_j^{\pm};q)_{\infty}}
{(q^{\frac12}t^{-2}s_i^{\pm}s_j^{\mp};q)_{\infty}(q^{\frac12}t^{-2}s_i^{\pm}s_j^{\pm};q)_{\infty}}
\nonumber\\
&\times 
(s_1+s_2+s_1^{-1}+s_2^{-1})^2. 
\end{align}
Analogously, the two-point function of the dual 't Hooft line can be derived from the Dirichlet half-index (\ref{so6-D}) 
of the disconnected part of $SO(6)$ gauge theory via the Higgsing manipulation. 
With 
\begin{align}
x_1&=q^{\frac12}t^2,\qquad 
x_2=q^2t^4, 
\end{align}
the Dirichlet half-index (\ref{so6-D}) is factorized as
\begin{align}
&
\mathbb{II}_{\mathcal{D}}^{\textrm{4d $\mathcal{N}=4$ $SO(6)^{-}$}}
\left(t,x_1=q^{\frac12}t^2,x_2=q^2t^4;q\right)
\nonumber\\
&=\langle T_{(1,0)}T_{(1,0)}\rangle_{\textrm{Nahm}}^{\textrm{4d $\mathcal{N}=4$ $SO(6)^{-}$}}(t;q)
\mathcal{I}^{\textrm{3d HM}}(t,x=q^{\frac14}t;q)
\mathcal{I}^{\textrm{3d HM}}(t,x=-q^{\frac14}t;q)
\nonumber\\
&\times 
\mathcal{I}^{\textrm{3d HM}}(t,x=q^{\frac54}t;q)
\mathcal{I}^{\textrm{3d HM}}(t,x=-q^{\frac54}t;q)
\nonumber\\
&\times 
\mathcal{I}^{\textrm{3d HM}}(t,x=q^{\frac74}t^3;q)
\mathcal{I}^{\textrm{3d HM}}(t,x=q^{\frac94}t^5;q), 
\end{align}
where the two-point function of the dual 't Hooft line is 
\begin{align}
\label{so6-N_T1}
&
\langle T_{(1,0)}T_{(1,0)}\rangle_{\textrm{Nahm}}^{\textrm{4d $\mathcal{N}=4$ $SO(6)^{-}$}}(t;q)
\nonumber\\
&=\frac{(q)_{\infty}(q^{\frac32}t^2;q)_{\infty}^2(q^{\frac72}t^6;q)_{\infty}(-q^{\frac32}t^2;q)_{\infty}(-q^2t^4;q)_{\infty}(-q^3t^4;q)_{\infty}}
{(q^{\frac12}t^2;q)_{\infty}(qt^4;q)_{\infty}(q^2;q)_{\infty}(q^3t^8;q)_{\infty}(-qt^4;q)_{\infty}(-q^{\frac52}t^2;q)_{\infty}(-q^{\frac52}t^6;q)_{\infty}}. 
\end{align}
Flipping $t$ to $t^{-1}$, it agrees with the two-point function (\ref{so6-N_W1}) of the fundamental Wilson lines. 
By gauging the $\mathbb{Z}_2$ global symmetry, 
we get the two-point function of $O(6)$ gauge theory
\begin{align}
&
\langle W_{\tiny \yng(1)} W_{\tiny \yng(1)}\rangle_{\mathcal{N}}^{\textrm{4d $\mathcal{N}=4$ $O(6)^{\pm}$}}(t;q)
\nonumber\\
&=
\frac12 
\left[
\langle W_{\tiny \yng(1)} W_{\tiny \yng(1)}\rangle_{\mathcal{N}}^{\textrm{4d $\mathcal{N}=4$ $SO(6)$}}(t;q)
\pm 
\langle W_{\tiny \yng(1)} W_{\tiny \yng(1)}\rangle_{\mathcal{N}}^{\textrm{4d $\mathcal{N}=4$ $SO(6)^{-}$}}(t;q)
\right]. 
\end{align}
The half-index of the dual 't Hooft line of magnetic charge $B=(1,0,0)$ of $O(6)$ gauge theory with the Nahm pole b.c. is given by
\begin{align}
&
\langle T_{(1,0,0)}T_{(1,0,0)}\rangle_{\textrm{Nahm}}^{\textrm{4d $\mathcal{N}=4$ $O(6)^{\pm}$}}(t;q)
=\frac12 \frac{(q)_{\infty}(q^{\frac32}t^2;q)_{\infty}^2(q^{\frac72}t^6;q)_{\infty}}
{(q^{\frac12}t^2;q)_{\infty}(qt^4;q)_{\infty}(q^2;q)_{\infty}(q^3t^8;q)_{\infty}}
\nonumber\\
&\times 
\left[
\frac{(q^{\frac32}t^2;q)_{\infty}(q^2t^4;q)_{\infty}(q^3t^4;q)_{\infty}}
{(qt^4;q)_{\infty}(q^{\frac52}t^2;q)_{\infty}(q^{\frac52}t^6;q)_{\infty}}
\pm 
\frac{(-q^{\frac32}t^2;q)_{\infty}(-q^2t^4;q)_{\infty}(-q^3t^4;q)_{\infty}}
{(-qt^4;q)_{\infty}(-q^{\frac52}t^2;q)_{\infty}(-q^{\frac52}t^6;q)_{\infty}}
\right]. 
\end{align}
As a result of S-duality of line operators and the b.c. for $O(6)$ gauge theories we have
\begin{align}
\langle W_{\tiny \yng(1)} W_{\tiny \yng(1)}\rangle_{\mathcal{N}}^{\textrm{4d $\mathcal{N}=4$ $O(6)^{\pm}$}}(t;q)
&=\langle T_{(1,0,0)}T_{(1,0,0)}\rangle_{\textrm{Nahm}}^{\textrm{4d $\mathcal{N}=4$ $O(6)^{\pm}$}}(t^{-1};q). 
\end{align}

\subsubsection{Antisymmetric Wilson lines}
For the line defect half-indices associated with the non-minuscule representations 
we can not simply obtain the closed-form expressions by applying the Higgsing manipulation. 
It would be intriguing to explore the exact results for such line defect half-indices. 

We remark that the one-point function of the boundary Wilson line in the rank-$2$ antisymmetric representation can be expressed as
\begin{align}
\label{so6N_Wasym2_1pt}
\langle W_{\tiny \yng(1,1)} \rangle_{\mathcal{N}}^{\textrm{4d $\mathcal{N}=4$ $SO(6)$}}(t;q)
&=\langle W_{\tiny \overline{\yng(1,1)}} \rangle_{\mathcal{N}}^{\textrm{4d $\mathcal{N}=4$ $SO(6)$}}(t;q)
\nonumber\\
&=q^{\frac12}t^{-2}
\frac{(q^{\frac12}t^2;q)_{\infty}(q^{\frac32}t^{-2};q)_{\infty}^2(q^2t^{-4};q)_{\infty}(q^{\frac72}t^{-6};q)_{\infty}}
{(q^{\frac12}t^{-2};q)_{\infty}(qt^{-4};q)_{\infty}(q^{\frac32}t^2;q)_{\infty}(q^2t^{-8};q)_{\infty}(q^{\frac52}t^{-6};q)_{\infty}}. 
\end{align}
On the other hand, the one-point function of the boundary Wilson line in the rank-$3$ antisymmetric representation vanishes. 
As we will see, more general formulas can be found by directly evaluating the matrix integral according to the norm of the Macdonald polynomials. 

\subsubsection{Symmetric Wilson lines}
The symmetric Wilson lines of $SO(6)$ gauge theory are dual to the 't Hooft line of magnetic charge which are not minuscule. 
Again the Higgsing procedure does not simply provide us with the closed-form expressions, 
however, we find that the one-point function of the rank-$2$ $SO(6)$ symmetric Wilson line is given by
\begin{align}
\label{so6N_Wsym2_1pt}
&
\langle W_{\tiny \yng(2)} \rangle_{\mathcal{N}}^{\textrm{4d $\mathcal{N}=4$ $SO(6)$}}(t;q)
\nonumber\\
&=qt^{-4}
\frac{(q^{\frac12}t^2;q)_{\infty}(q)_{\infty}(q^{\frac32}t^{-2};q)_{\infty}(q^2t^{-4};q)_{\infty}(q^3t^{-4};q)_{\infty}(q^{\frac72}t^{-6};q)_{\infty}}
{(qt^{-4};q)_{\infty}^2(q^{\frac32}t^2;q)_{\infty}(q^{\frac32}t^{-6};q)_{\infty}(q^2;q)_{\infty}(q^3t^{-8};q)_{\infty}}. 
\end{align}

\subsection{$\mathfrak{so}(2N)$}
Now we propose a generalization of the results so far for gauge theories based on the gauge algebra $\mathfrak{so}(2N)$ with general rank. 
The half-index of the Neumann b.c. of $SO(2N)$ gauge theory is 
\begin{align}
\label{so2NN}
&
\mathbb{II}_{\mathcal{N}}^{\textrm{4d $\mathcal{N}=4$ $SO(2N)$}}(t;q)
\nonumber\\
&=\frac{1}{2^{N-1}N!}
\frac{(q)_{\infty}^N}{(q^{\frac12}t^{-2};q)_{\infty}^N}
\oint \prod_{i=1}^{N}
\frac{ds_i}{2\pi is_i}
\prod_{i<j}
\frac{(s_i^{\pm}s_j^{\mp};q)_{\infty}(s_i^{\pm}s_j^{\pm};q)_{\infty}}
{(q^{\frac12}t^{-2}s_i^{\pm}s_j^{\mp};q)_{\infty}(q^{\frac12}t^{-2}s_i^{\pm}s_j^{\pm};q)_{\infty}}. 
\end{align}
On the other hand, the half-index of the Nahm pole b.c. of $SO(2N)$ gauge theory is given by \cite{Hatsuda:2024lcc}
\begin{align}
\label{so2NNahm}
\mathbb{II}_{\textrm{Nahm}}^{\textrm{4d $\mathcal{N}=4$ $SO(2N)$}}(t;q)
&=\frac{(q^{\frac12+\frac{N}{2}}t^{2N-2};q)_{\infty}}
{(q^{\frac{N}{2}}t^{2N};q)_{\infty}}
\prod_{k=1}^{N-1}
\frac{(q^{\frac12+k}t^{4k-2};q)_{\infty}}
{(q^kt^{4k};q)_{\infty}}. 
\end{align}
Upon the transformation $t$ $\rightarrow$ $t^{-1}$, 
it is demonstrated in \cite{Hatsuda:2024lcc} that 
the half-indices (\ref{so2NN}) and (\ref{so2NNahm}) coincide. 
The half-index of the Dirichlet b.c. for $SO(2N)$ gauge theory is
\begin{align}
\label{so2ND}
&
\mathbb{II}_{\mathcal{D}}^{\textrm{4d $\mathcal{N}=4$ $SO(2N)$}}(t;q)
\nonumber\\
&=\frac{(q)_{\infty}^N}{(q^{\frac12}t^2;q)_{\infty}^N}
\prod_{i<j}
\frac{(qx_i^{\pm}x_j^{\mp};q)_{\infty}(qx_i^{\pm}x_j^{\pm};q)_{\infty}}
{(q^{\frac12}t^2x_i^{\pm}x_j^{\mp};q)_{\infty}(q^{\frac12}t^2x_i^{\pm}x_j^{\pm};q)_{\infty}}. 
\end{align}
By fixing the global fugacities as
\begin{align}
x_i&=q^{\frac{i-1}{2}}t^{2(i-1)}, 
\end{align}
the Dirichlet half-index (\ref{so2ND}) becomes
\begin{align}
&
\mathbb{II}_{\mathcal{D}}^{\textrm{4d $\mathcal{N}=4$ $SO(2N)$}}
\left(t,x_i=q^{\frac{i-1}{2}}t^{2(i-1)};q\right)
\nonumber\\
&=\mathbb{II}_{\textrm{Nahm}}^{\textrm{4d $\mathcal{N}=4$ $SO(2N)$}}(t;q)
\prod_{i=1}^{N-1}
\mathcal{I}^{\textrm{3d HM}}(t,x=q^{\frac{2i-1}{4}}t^{2i-1};q)^{N-i}
\nonumber\\
&\times 
\prod_{i=1}^{2N-3}
\mathcal{I}^{\textrm{3d HM}}(t,x=q^{\frac{2i-1}{4}}t^{2i-1};q)^{a_N(i)}
\end{align}
and removing the extra 3d indices, 
one obtains the Nahm pole half-index (\ref{so2NNahm}). 

For the disconnected component of the orthogonal gauge group, 
the half-index of the Neumann b.c. takes the form
\begin{align}
\label{so2N-N}
&
\mathbb{II}_{\mathcal{N}}^{\textrm{4d $\mathcal{N}=4$ $SO(2N)^{-}$}}(t;q)
\nonumber\\
&=\frac{1}{2^{N-1}(N-1)!}
\frac{(q)_{\infty}^{N-1}(-q;q)_{\infty}}{(q^{\frac12}t^{-2};q)_{\infty}^{N-1}(-q^{\frac12}t^{-2};q)_{\infty}}
\oint \prod_{i=1}^{N-1} \frac{ds_i}{2\pi is_i}
\frac{(s_i^{\pm};q)_{\infty}(-s_i^{\pm};q)_{\infty}}
{(q^{\frac12}t^{-2}s_i^{\pm};q)_{\infty}(-q^{\frac12}t^{-2}s_i^{\pm};q)_{\infty}}
\nonumber\\
&\times 
\prod_{i<j}
\frac{(s_i^{\pm}s_j^{\mp};q)_{\infty}(s_i^{\pm}s_j^{\mp};q)_{\infty}}
{(q^{\frac12}t^{-2}s_i^{\pm}s_j^{\mp};q)_{\infty}(q^{\frac12}t^{-2}s_i^{\pm}s_j^{\mp};q)_{\infty}}. 
\end{align}
The half-index of the Nahm pole b.c. for the disconnected component is given by \cite{Hatsuda:2024lcc}
\begin{align}
\label{so2N-Nahm}
\mathbb{II}_{\textrm{Nahm}}^{\textrm{4d $\mathcal{N}=4$ $SO(2N)^{-}$}}(t;q)
&=\frac{(-q^{\frac{N+1}{2}}t^{2N-2};q)_{\infty}}
{(-q^{\frac{N}{2}}t^{2N};q)_{\infty}}
\prod_{k=1}^{N-1}
\frac{(q^{\frac{2k+1}{2}}t^{4k-2};q)_{\infty}}
{(q^kt^{4k};q)_{\infty}}, 
\end{align}
which agrees with the Neumann half-index (\ref{so2N-N}) under the transformation $t$ $\rightarrow$ $t^{-1}$. 
Similarly, the expression (\ref{so2N-Nahm}) can be found from the Dirichlet half-index 
\begin{align}
\label{so2N-D}
&
\mathbb{II}_{\mathcal{D}}^{\textrm{4d $\mathcal{N}=4$ $SO(2N)^{-}$}}(t;q)
\nonumber\\
&=\frac{(q)_{\infty}^{N-1}(-q;q)_{\infty}}
{(q^{\frac12}t^2;q)_{\infty}^{N-1}(-q^{\frac12}t^2;q)_{\infty}}
\prod_{i=1}^N
\frac{(qx_i^{\pm};q)_{\infty} (-qx_i^{\pm};q)_{\infty}}
{(q^{\frac12}t^2x_i^{\pm};q)_{\infty} (-q^{\frac12}t^2x_i^{\pm};q)_{\infty}}
\nonumber\\
&\times 
\prod_{i< j}
\frac{(qx_i^{\pm}x_j^{\mp};q)_{\infty}(qx_i^{\pm}x_j^{\pm};q)_{\infty}}
{(q^{\frac12}t^{-2}x_i^{\pm}x_j^{\mp};q)_{\infty}(q^{\frac12}t^{-2}x_i^{\pm}x_j^{\pm};q)_{\infty}}
\end{align}
by specializing the global fugacities as
\begin{align}
x_i&=q^{\frac{i}{2}}t^{2i}. 
\end{align}
Upon gauging the $\mathbb{Z}_2$ global symmetry, one finds the half-indices for $O(2N)$ gauge theory. 
The Neumann half-index is 
\begin{align}
\label{o2NN}
\mathbb{II}_{\mathcal{N}}^{\textrm{4d $\mathcal{N}=4$ $O(2N)^{\pm}$}}(t;q)
&=\frac12
\left[
\mathbb{II}_{\mathcal{N}}^{\textrm{4d $\mathcal{N}=4$ $SO(2N)$}}(t;q)
\pm 
\mathbb{II}_{\mathcal{N}}^{\textrm{4d $\mathcal{N}=4$ $SO(2N)^{-}$}}(t;q)
\right]
\end{align}
and the Nahm pole half-index is
\begin{align}
\label{o2NNahm}
\mathbb{II}_{\textrm{Nahm}}^{\textrm{4d $\mathcal{N}=4$ $O(2N)^{\pm}$}}(t;q)
&=\frac12 
\left[
\frac{(q^{\frac{N+1}{2}}t^{2N-2};q)_{\infty}}
{(q^{\frac{N}{2}}t^{2N};q)_{\infty}}
\pm 
\frac{(-q^{\frac{N+1}{2}}t^{2N-2};q)_{\infty}}
{(-q^{\frac{N}{2}}t^{2N};q)_{\infty}}
\right]
\prod_{k=1}^{N-1}
\frac{(q^{\frac{2k+1}{2}}t^{4k-2};q)_{\infty}}
{(q^kt^{4k};q)_{\infty}}. 
\end{align}
The matching of the half-indices (\ref{o2NN}) and (\ref{o2NNahm}) upon the transformation $t$ $\rightarrow$ $t^{-1}$ 
demonstrates 
the duality of the Neumann b.c. and Nahm pole b.c. in $O(2N)$ gauge theory. 

\subsubsection{Spinor Wilson lines}
When $N$ is even, the chiral spinor representation and the antichiral spinor representation are not complex conjugate.  
The non-trivial two-point function of the boundary spinor Wilson lines of $Spin(2N)$ gauge theory with the Neumann b.c. can be found 
for a pair of the chiral spinor Wilson lines or that of the antichiral spinor Wilson lines. 
For a pair of the chiral spinor Wilson lines it is evaluated as
\begin{align}
\label{so2NevenN_Wsp}
&
\langle W_{\textrm{sp}} W_{\textrm{sp}}\rangle_{\mathcal{N}}^{\textrm{4d $\mathcal{N}=4$ $Spin(2N)$}}(t;q)
\nonumber\\
&=\frac{1}{2^{N-1}N!}
\frac{(q)_{\infty}^N}{(q^{\frac12}t^{-2};q)_{\infty}^N}
\oint \prod_{i=1}^{N}
\frac{ds_i}{2\pi is_i}
\prod_{i<j}
\frac{(s_i^{\pm}s_j^{\mp};q)_{\infty}(s_i^{\pm}s_j^{\pm};q)_{\infty}}
{(q^{\frac12}t^{-2}s_i^{\pm}s_j^{\mp};q)_{\infty}(q^{\frac12}t^{-2}s_i^{\pm}s_j^{\pm};q)_{\infty}}
\nonumber\\
&\times 
\frac14 
\left[
\prod_{i=1}^N (s_i^{\frac12}+s_i^{-\frac12})
+\prod_{i=1}^N (s_i^{\frac12}-s_i^{-\frac12})
\right]^2. 
\end{align}
Similarly, one can compute the two-point function of the antichiral spinor Wilson lines, 
which gives the same result as (\ref{so2NevenN_Wsp}). 
For odd $N$ the chiral spinor representation and the antichiral spinor representation are complex conjugate, 
for which the non-trivial two-point function can be found for a pair of the chiral spinor Wilson line and the antichiral spinor Wilson line. 
It takes the form 
\begin{align}
\label{so2NoddN_Wsp}
&
\langle W_{\textrm{sp}} W_{\overline{\textrm{sp}}}\rangle_{\mathcal{N}}^{\textrm{4d $\mathcal{N}=4$ $Spin(2N)$}}(t;q)
\nonumber\\
&=\frac{1}{2^{N-1}N!}
\frac{(q)_{\infty}^N}{(q^{\frac12}t^{-2};q)_{\infty}^N}
\oint \prod_{i=1}^{N}
\frac{ds_i}{2\pi is_i}
\prod_{i<j}
\frac{(s_i^{\pm}s_j^{\mp};q)_{\infty}(s_i^{\pm}s_j^{\pm};q)_{\infty}}
{(q^{\frac12}t^{-2}s_i^{\pm}s_j^{\mp};q)_{\infty}(q^{\frac12}t^{-2}s_i^{\pm}s_j^{\pm};q)_{\infty}}
\nonumber\\
&\times 
\frac14 
\left[
\prod_{i=1}^N (s_i^{\frac12}+s_i^{-\frac12})
+\prod_{i=1}^N (s_i^{\frac12}-s_i^{-\frac12})
\right]
\left[
\prod_{i=1}^N (s_i^{\frac12}+s_i^{-\frac12})
-\prod_{i=1}^N (s_i^{\frac12}-s_i^{-\frac12})
\right]. 
\end{align}
The dual of the spinor Wilson line in $Spin(2N)$ gauge theory 
is the 't Hooft line of magnetic charge $B=(\frac12,\cdots,\frac12)$ for $SO(2N)/\mathbb{Z}_2$ gauge theory. 
This is associated with the minuscule representation. 
We propose that the two-point function can be obtained via the Higgsing process 
by fixing the global fugacities of the $SO(2N)$ Dirichlet half-index (\ref{so2ND}) as
\begin{align}
x_i&=q^{\frac{i}{2}}t^{2(i-1)}, 
\qquad i=1,\cdots, N. 
\end{align}
Accordingly, the $SO(2N)$ Dirichlet half-index (\ref{so2ND}) reduces to 
\begin{align}
&
\mathbb{II}_{\mathcal{D}}^{\textrm{4d $\mathcal{N}=4$ $SO(2N)$}}
\left(t,x_i=q^{\frac{i}{2}}t^{2(i-1)};q\right)
\nonumber\\
&=\langle T_{(\frac12,\cdots,\frac12)}T_{(\frac12,\cdots,\frac12)}\rangle_{\textrm{Nahm}}^{\textrm{4d $\mathcal{N}=4$ $SO(2N)/\mathbb{Z}_2$}}(t;q)
\prod_{i=1}^{N-1}\mathcal{I}^{\textrm{3d HM}}(t,x=q^{\frac{2i-1}{4}}t^{2i-1};q)^{N-i}
\nonumber\\
&\times 
\prod_{i=1}^{2N-3}
\mathcal{I}^{\textrm{3d HM}}(t,x=q^{\frac{2i+3}{4}}t^{2i-1};q)^{a_N(i)}, 
\end{align}
where $a_N(i)$ are the $q$-binomial coefficients defined in (\ref{qbinomial}). 
The resulting two-point function of the dual 't Hooft line of magnetic charge $B=(\frac12,\cdots,\frac12)$ for $SO(2N)/\mathbb{Z}_2$ gauge theory 
subject to the Nahm pole b.c. is
\begin{align}
\label{so2NNahm_T1/2}
&
\langle T_{(\frac12,\cdots,\frac12)}T_{(\frac12,\cdots,\frac12)}\rangle_{\textrm{Nahm}}^{\textrm{4d $\mathcal{N}=4$ $SO(2N)/\mathbb{Z}_2$}}(t;q)
\nonumber\\
&=\prod_{k=1}^{\lfloor \frac{N}{2}\rfloor}
\frac{(q^{\frac{2k+1}{2}}t^{4k-2};q)_{\infty}}
{(q^{k+1}t^{4k-4};q)_{\infty}}
\prod_{k=1}^N
\frac{(q^{\frac{k+1}{2}}t^{2(k-1)};q)_{\infty}}
{(q^{\frac{k}{2}}t^{2k};q)_{\infty}}
\prod_{k=1}^{\lfloor \frac{N}{2}\rfloor}
\frac{(q^{\frac{2(k+\lfloor \frac{N+1}{2}\rfloor)+3}{2}}t^{4(k+\lfloor \frac{N-1}{2})-2};q)_{\infty}}
{(q^{k+\lfloor \frac{N+1}{2}\rfloor+1}t^{4(k+\lfloor \frac{N-1}{2}\rfloor)};q)_{\infty}}. 
\end{align}
For example, for $N=4,5$ we have 
\begin{align}
\label{so8Nahm_T1/2}
&
\langle T_{(\frac12,\frac12,\frac12,\frac12)}T_{(\frac12,\frac12,\frac12,\frac12)}\rangle_{\textrm{Nahm}}^{\textrm{4d $\mathcal{N}=4$ $SO(8)/\mathbb{Z}_2$}}(t;q)
\nonumber\\
&=\frac{(q)_{\infty}(q^{\frac32}t^2;q)_{\infty}^2(q^2t^4;q)_{\infty}(q^{\frac52}t^6;q)_{\infty}^2(q^{\frac72}t^{6};q)_{\infty}(q^{\frac92}t^{10};q)_{\infty}}
{(q^{\frac12}t^2;q)_{\infty}(qt^4;q)_{\infty}(q^{\frac32}t^6;q)_{\infty}(q^2;q)_{\infty}(q^3t^4;q)_{\infty}(q^3t^8;q)_{\infty}(q^4t^{12};q)_{\infty}}, \\
\label{so10Nahm_T1/2}
&
\langle T_{(\frac12,\frac12,\frac12,\frac12,\frac12)}T_{(\frac12,\frac12,\frac12,\frac12,\frac12)}\rangle_{\textrm{Nahm}}^{\textrm{4d $\mathcal{N}=4$ $SO(10)/\mathbb{Z}_2$}}(t;q)
\nonumber\\
&=\frac{(q)_{\infty}(q^{\frac32}t^2;q)_{\infty}^2(q^2t^4;q)_{\infty}(q^{\frac52}t^6;q)_{\infty}^2
(q^3t^8;q)_{\infty}(q^{\frac92}t^{10};q)_{\infty}(q^{\frac{11}{2}}t^{14};q)_{\infty}}
{(q^{\frac12}t^2;q)_{\infty}(qt^4;q)_{\infty}(q^{\frac32}t^6;q)_{\infty}(q^2;q)_{\infty}
(q^2t^8;q)_{\infty}(q^{\frac52}t^{10};q)_{\infty}(q^3t^4;q)_{\infty}(q^4t^{12};q)_{\infty}(q^5t^{16};q)_{\infty}}. 
\end{align}
We claim that 
the expression (\ref{so2NNahm_T1/2}) agrees with 
the two-point functions (\ref{so2NevenN_Wsp}) and (\ref{so2NoddN_Wsp}) of the spinor Wilson lines of $Spin(2N)$ gauge theory 
upon the transformation $t$ $\rightarrow$ $t^{-1}$
\begin{align}
\langle W_{\textrm{sp}} W_{\textrm{sp}}\rangle_{\mathcal{N}}^{\textrm{4d $\mathcal{N}=4$ $Spin(4k)$}}(t;q)
&=
\langle T_{(\frac12,\cdots,\frac12)}T_{(\frac12,\cdots,\frac12)}\rangle_{\textrm{Nahm}}^{\textrm{4d $\mathcal{N}=4$ $SO(4k)/\mathbb{Z}_2$}}(t^{-1};q), \\
\langle W_{\textrm{sp}} W_{\overline{\textrm{sp}}}\rangle_{\mathcal{N}}^{\textrm{4d $\mathcal{N}=4$ $Spin(4k+2)$}}(t;q)
&=
\langle T_{(\frac12,\cdots,\frac12)}T_{(\frac12,\cdots,\frac12)}\rangle_{\textrm{Nahm}}^{\textrm{4d $\mathcal{N}=4$ $SO(4k+2)/\mathbb{Z}_2$}}(t^{-1};q). 
\end{align}

\subsubsection{Fundamental Wilson line}
The two-point function of the boundary Wilson line in the fundamental representation of $SO(2N)$ gauge theory 
obeying the Neumann b.c. takes the form 
\begin{align}
\label{so2NN_W1}
&
\langle W_{\tiny \yng(1)} W_{\tiny \yng(1)}\rangle_{\mathcal{N}}^{\textrm{4d $\mathcal{N}=4$ $SO(2N)$}}(t;q)
\nonumber\\
&=\frac{1}{2^{N-1}N!}
\frac{(q)_{\infty}^N}{(q^{\frac12}t^{-2};q)_{\infty}^N}
\oint \prod_{i=1}^{N}
\frac{ds_i}{2\pi is_i}
\prod_{i<j}
\frac{(s_i^{\pm}s_j^{\mp};q)_{\infty}(s_i^{\pm}s_j^{\pm};q)_{\infty}}
{(q^{\frac12}t^{-2}s_i^{\pm}s_j^{\mp};q)_{\infty}(q^{\frac12}t^{-2}s_i^{\pm}s_j^{\pm};q)_{\infty}}
\nonumber\\
&\times 
\left[
\sum_{i=1}^N 
(s_i+s_i^{-1})
\right]^2. 
\end{align}
It is conjectured that the fundamental Wilson line of $SO(2N)$ gauge theory 
is dual to the 't Hooft line of magnetic charge $B=(1,0,\cdots, 0)$ of $SO(2N)$ gauge theory. 
We propose that 
the two-point function of the dual 't Hooft lines can be derived by applying the Higgsing manipulation 
to the $SO(2N)$ Dirichlet half-index (\ref{so2ND}) by specializing the global fugacities as
\begin{align}
x_i&=q^{\frac{i-1}{2}}t^{2(i-1)}, \qquad i=1,\cdots, N-1, \nonumber\\
x_{N}&=q^{\frac{N+1}{2}}t^{2(N-1)}. 
\end{align}
Then we find that the $SO(2N)$ Dirichlet half-index (\ref{so2ND}) is factorized as
\begin{align}
&
\mathbb{II}_{\mathcal{D}}^{\textrm{4d $\mathcal{N}=4$ $SO(2N)$}}
\left(t,\left\{x_i=q^{\frac{i-1}{2}}t^{2(i-1)}\right\}_{i=1}^{N-1},x_N=q^{\frac{N+1}{2}}t^{2(N-1)};q\right)
\nonumber\\
&=\langle T_{(1,0,\cdots,0)}T_{(1,0,\cdots,0)}\rangle_{\textrm{Nahm}}^{\textrm{4d $\mathcal{N}=4$ $SO(2N)$}}(t;q)
\nonumber\\
&\times 
\prod_{i=1}^{N-2}
\mathcal{I}^{\textrm{3d HM}}(t,x=q^{\frac{2i-1}{4}}t^{2i-1};q)^{N-1-\lfloor \frac{i}{2}\rfloor}
\prod_{i=1}^{N-3}
\mathcal{I}^{\textrm{3d HM}}(t,x=q^{\frac{4N-2i-9}{4}}t^{4N-2i-9};q)^{\lfloor \frac{i}{2}\rfloor}
\nonumber\\
&\times 
\mathcal{I}^{\textrm{3d HM}}(t,x=q^{\frac{2N+1}{4}}t^{2N-3};q)
\prod_{i=1}^{2N-3}
\mathcal{I}^{\textrm{3d HM}}(t,x=q^{\frac{2i+3}{4}}t^{2i-1};q). 
\end{align}
Consequently, we obtain the two-point function 
of the dual 't Hooft line of magnetic charge $B=(1,0,\cdots, 0)$ of $SO(2N)$ gauge theory with the Nahm pole b.c. 
\begin{align}
\label{so2NNahm_T1}
&
\langle T_{(1,0,\cdots,0)}T_{(1,0,\cdots,0)}\rangle_{\textrm{Nahm}}^{\textrm{4d $\mathcal{N}=4$ $SO(2N)$}}(t;q)
\nonumber\\
&=\frac{(q)_{\infty}}{(q^{\frac12}t^2;q)_{\infty}}
\frac{(q^{\frac32}t^2;q)_{\infty}}{(q^2;q)_{\infty}}
\prod_{k=1}^{N-2}
\frac{(q^{\frac{2k+1}{2}}t^{4k-2};q)_{\infty}}
{(q^k t^{4k};q)_{\infty}}
\nonumber\\
&\times 
\frac{(q^{\frac{N}{2}}t^{2N-4};q)_{\infty}}
{(q^{\frac{N-1}{2}}t^{2N-2};q)_{\infty}}
\frac{(q^{\frac{N+1}{2}}t^{2N-2};q)_{\infty}}
{(q^{\frac{N+2}{2}}t^{2N-4};q)_{\infty}}
\frac{(q^{\frac{N+3}{2}}t^{2N-2};q)_{\infty}}
{(q^{\frac{N+2}{2}}t^{2N};q)_{\infty}}
\frac{(q^{\frac{2N+1}{2}}t^{4N-6};q)_{\infty}}
{(q^{N}t^{4N-4};q)_{\infty}}. 
\end{align}
For example, for $N=4, 5$ we have
\begin{align}
\label{so8Nahm_T1}
&
\langle T_{(1,0,0,0)}T_{(1,0,0,0)}\rangle_{\textrm{Nahm}}^{\textrm{4d $\mathcal{N}=4$ $SO(8)$}}(t;q)
\nonumber\\
&=\frac{(q)_{\infty}(q^{\frac32}t^2;q)_{\infty}^2(q^2t^4;q)_{\infty}(q^{\frac52}t^6;q)_{\infty}^2(q^{\frac72}t^{6};q)_{\infty}(q^{\frac92}t^{10};q)_{\infty}}
{(q^{\frac12}t^2;q)_{\infty}(qt^4;q)_{\infty}(q^{\frac32}t^6;q)_{\infty}(q^2;q)_{\infty}(q^3t^4;q)_{\infty}(q^3t^8;q)_{\infty}(q^4t^{12};q)_{\infty}}, \\
\label{so10Nahm_T1}
&
\langle T_{(1,0,0,0,0)}T_{(1,0,0,0,0)}\rangle_{\textrm{Nahm}}^{\textrm{4d $\mathcal{N}=4$ $SO(10)$}}(t;q)
\nonumber\\
&=\frac{(q)_{\infty}(q^{\frac32}t^2;q)_{\infty}^2(q^{\frac52}t^6;q)_{\infty}^2
(q^3t^8;q)_{\infty}(q^{\frac72}t^{10};q)_{\infty}(q^4t^8;q)_{\infty}(q^{\frac{11}{2}}t^{14};q)_{\infty}}
{(q^{\frac12}t^2;q)_{\infty}(qt^4;q)_{\infty}(q^2;q)_{\infty}(q^2t^8;q)_{\infty}^2
(q^{\frac72}t^6;q)_{\infty}(q^{\frac72}t^{10};q)_{\infty}(q^5t^{16};q)_{\infty}}. 
\end{align}
We expect that under the transformation $t$ $\rightarrow$ $t^{-1}$ 
the two-point function (\ref{so2NN_W1}) of the boundary Wilson line in the fundamental representation of $SO(2N)$ 
gives the same result as the two-point function (\ref{so2NNahm_T1}) of the boundary 't Hooft line of magnetic charge $B=(1,0,\cdots,0)$ 
as a consequence of S-duality
\begin{align}
\langle W_{\tiny \yng(1)} W_{\tiny \yng(1)}\rangle_{\mathcal{N}}^{\textrm{4d $\mathcal{N}=4$ $SO(2N)$}}(t;q)&
=\langle T_{(1,0,\cdots,0)}T_{(1,0,\cdots,0)}\rangle_{\textrm{Nahm}}^{\textrm{4d $\mathcal{N}=4$ $SO(2N)$}}(t^{-1};q). 
\end{align}
Note that the two-point functions (\ref{so8Nahm_T1/2}) and (\ref{so8Nahm_T1}) are equal, 
that is the two-point function of the boundary Wilson lines in the spinor representation of $Spin(8)$ 
and that of the boundary Wilson lines in the fundamental representation of $SO(8)$ are equal. 
This stems from the triality of $Spin(8)$ that permutes the chiral spinor, antichiral spinor and the fundamental representations. 

For the disconnected component 
the two-point function of the fundamental Wilson lines with the Neumann b.c. reads
\begin{align}
\label{so2N-N_W1}
&
\langle W_{\tiny \yng(1)} W_{\tiny \yng(1)}\rangle_{\mathcal{N}}^{\textrm{4d $\mathcal{N}=4$ $SO(2N)^{-}$}}(t;q)
\nonumber\\
&=\frac{1}{2^{N-1}(N-1)!}
\frac{(q)_{\infty}^{N-1}(-q;q)_{\infty}}{(q^{\frac12}t^{-2};q)_{\infty}^{N-1}(-q^{\frac12}t^{-2};q)_{\infty}}
\oint \prod_{i=1}^{N-1} \frac{ds_i}{2\pi is_i}
\frac{(s_i^{\pm};q)_{\infty}(-s_i^{\pm};q)_{\infty}}
{(q^{\frac12}t^{-2}s_i^{\pm};q)_{\infty}(-q^{\frac12}t^{-2}s_i^{\pm};q)_{\infty}}
\nonumber\\
&\times 
\prod_{i<j}
\frac{(s_i^{\pm}s_j^{\mp};q)_{\infty}(s_i^{\pm}s_j^{\mp};q)_{\infty}}
{(q^{\frac12}t^{-2}s_i^{\pm}s_j^{\mp};q)_{\infty}(q^{\frac12}t^{-2}s_i^{\pm}s_j^{\mp};q)_{\infty}}
\left[
\sum_{i=1}^{N-1}
(s_i+s_i^{-1})
\right]^2. 
\end{align}
Similarly, we propose that 
the two-point function of the dual 't Hooft lines with the Nahm pole b.c. 
can be found from the $SO(2N)^{-}$ Dirichlet half-index (\ref{so2N-D}) via the Higgsing process after taking away the indices of 3d matters. 
By fixing the global fugacities of the $SO(2N)^{-}$ Dirichlet half-index (\ref{so2N-D}) as
\begin{align}
x_{N-1}&=q^{\frac{N+1}{2}}t^{2N-2},\qquad 
x_i=q^{\frac{i}{2}}t^{2i},\qquad i=1,\cdots, N-2, 
\end{align}
we find that 
\begin{align}
&
\mathbb{II}_{\mathcal{D}}^{\textrm{4d $\mathcal{N}=4$ $SO(2N)^{-}$}}
\left(t,\left\{x_i=q^{\frac{i}{2}}t^{2i}\right\}_{i=1}^{N-2},x_{N-1}=q^{\frac{N+1}{2}}t^{2N-2};q\right)
\nonumber\\
&=\langle T_{(1,0,\cdots,0)}T_{(1,0,\cdots,0)}\rangle_{\textrm{Nahm}}^{\textrm{4d $\mathcal{N}=4$ $SO(2N)^{-}$}}(t;q)
\nonumber\\
&\times 
\prod_{i=1}^{2N-5}
\mathcal{I}^{\textrm{3d HM}}(t,x=q^{\frac{2i-1}{4}}t^{2i-1};q)^{N-2-\lfloor \frac{i}{2}\rfloor}
\prod_{i=1}^{N-2}
\mathcal{I}^{\textrm{3d HM}}(t,x=-q^{\frac{2i-1}{4}}t^{2i-1};q)
\nonumber\\
&\times 
\mathcal{I}^{\textrm{3d HM}}(t,x=-q^{\frac{2N+1}{4}}t^{2N-3};q)
\prod_{i=1}^{2N-3}
\mathcal{I}^{\textrm{3d HM}}(t,x=q^{\frac{2i+3}{4}}t^{2i-1};q), 
\end{align}
where the two-point function of the dual 't Hooft lines is given by
\begin{align}
\label{so2N-Nahm_T1}
&
\langle T_{(1,0,\cdots,0)}T_{(1,0,\cdots,0)}\rangle_{\textrm{Nahm}}^{\textrm{4d $\mathcal{N}=4$ $SO(2N)^{-}$}}(t;q)
\nonumber\\
&=\frac{(q)_{\infty}(q^{\frac32}t^2;q)_{\infty}(q^{\frac{2N+1}{2}}t^{4N-6};q)_{\infty}}
{(q^{\frac12}t^2;q)_{\infty}(q^2;q)_{\infty}(q^N t^{4N-4};q)_{\infty}}
\prod_{k=1}^{N-2}
\frac{(q^{\frac{2k+1}{2}}t^{4k-2};q)_{\infty}}
{(q^k t^{4k};q)_{\infty}}
\nonumber\\
&\times 
\frac{(-q^{\frac{N}{2}}t^{2N-4})_{\infty}(-q^{\frac{N+1}{2}}t^{2N-2};q)_{\infty}(-q^{\frac{N+3}{2}}t^{2N-2};q)_{\infty}}
{(-q^{\frac{N-1}{2}}t^{2N-2};q)_{\infty}(-q^{\frac{N+2}{2}}t^{2N-4};q)_{\infty}(-q^{\frac{N+2}{2}}t^{2N};q)_{\infty}}. 
\end{align}
We conjecture that under the transformation $t$ $\rightarrow$ $t^{-1}$, 
it agrees with the two-point function (\ref{so2N-N_W1}) of the boundary Wilson lines. 
It is now straightforward to obtain the two-point functions for $O(2N)$ gauge theory by gauging the $\mathbb{Z}_2$ global symmetry. 
The two-point function of the boundary Wilson line in the fundamental representation 
of $O(2N)$ gauge theory obeying the Neumann b.c. is given by
\begin{align}
\label{o2NN_W1}
&
\langle W_{\tiny \yng(1)} W_{\tiny \yng(1)}\rangle_{\mathcal{N}}^{\textrm{4d $\mathcal{N}=4$ $O(2N)^{\pm}$}}(t;q)
\nonumber\\
&=\frac12
\left[
\langle W_{\tiny \yng(1)} W_{\tiny \yng(1)}\rangle_{\mathcal{N}}^{\textrm{4d $\mathcal{N}=4$ $SO(2N)$}}(t;q)
\pm 
\langle W_{\tiny \yng(1)} W_{\tiny \yng(1)}\rangle_{\mathcal{N}}^{\textrm{4d $\mathcal{N}=4$ $SO(2N)^{-}$}}(t;q)
\right]. 
\end{align}
Similarly, the two-point function of the boundary 't Hooft line of magnetic charge $B=(1,0,\cdots,0)$ 
of $O(2N)$ gauge theory subject to the Nahm pole b.c. is
\begin{align}
\label{o2NNahm_T1}
&
\langle T_{(1,0,\cdots,0)}T_{(1,0,\cdots,0)}\rangle_{\textrm{Nahm}}^{\textrm{4d $\mathcal{N}=4$ $O(2N)^{\pm}$}}(t;q)
\nonumber\\
&=
\frac12 
\frac{(q)_{\infty}(q^{\frac32}t^2;q)_{\infty}(q^{\frac{2N+1}{2}}t^{4N-6};q)_{\infty}}
{(q^{\frac12}t^2;q)_{\infty}(q^2;q)_{\infty}(q^N t^{4N-4};q)_{\infty}}
\prod_{k=1}^{N-2}
\frac{(q^{\frac{2k+1}{2}}t^{4k-2};q)_{\infty}}
{(q^k t^{4k};q)_{\infty}}
\nonumber\\
&\times 
\Biggl[
\frac{(q^{\frac{N}{2}}t^{2N-4})_{\infty}(q^{\frac{N+1}{2}}t^{2N-2};q)_{\infty}(q^{\frac{N+3}{2}}t^{2N-2};q)_{\infty}}
{(q^{\frac{N-1}{2}}t^{2N-2};q)_{\infty}(q^{\frac{N+2}{2}}t^{2N-4};q)_{\infty}(q^{\frac{N+2}{2}}t^{2N};q)_{\infty}}
\nonumber\\
&\pm 
\frac{(-q^{\frac{N}{2}}t^{2N-4})_{\infty}(-q^{\frac{N+1}{2}}t^{2N-2};q)_{\infty}(-q^{\frac{N+3}{2}}t^{2N-2};q)_{\infty}}
{(-q^{\frac{N-1}{2}}t^{2N-2};q)_{\infty}(-q^{\frac{N+2}{2}}t^{2N-4};q)_{\infty}(-q^{\frac{N+2}{2}}t^{2N};q)_{\infty}}
\Biggr]. 
\end{align}
As a consequence of S-duality, 
we expect that the two-point functions (\ref{o2NN_W1}) and (\ref{o2NNahm_T1}) coincide after flipping $t$ to $t^{-1}$
\begin{align}
\langle W_{\tiny \yng(1)} W_{\tiny \yng(1)}\rangle_{\mathcal{N}}^{\textrm{4d $\mathcal{N}=4$ $O(2N)^{\pm}$}}(t;q)
&=
\langle T_{(1,0,\cdots,0)}T_{(1,0,\cdots,0)}\rangle_{\textrm{Nahm}}^{\textrm{4d $\mathcal{N}=4$ $O(2N)^{\pm}$}}(t^{-1};q). 
\end{align}

\subsubsection{Direct evaluation by Macdonald polynomials}
Let us evaluate the matrix integrals (\ref{so2NevenN_Wsp}) 
and (\ref{so2NoddN_Wsp}) by making use of the norm formula of the Macdonald polynomials of type $D_N$. 
In this case, the root system is given by
\begin{align}
R&=\{ \pm \varepsilon_i \pm \varepsilon_j | 1\leq i<j \leq N \}, \\
R^+&=\{ \varepsilon_i \pm \varepsilon_j | 1\leq i<j \leq N \}.
\end{align}
Then,
\begin{align}
\rho=\sum_{i=1}^N ( N-i)\varepsilon_i,\qquad \lambda=\sum_{i=1}^N \lambda_i \varepsilon_i,
\end{align}
and
\begin{align}
w_{D_N}(s;\q,\t)=\prod_{1\leq i<j \leq N} \frac{(s_i^{\pm 1}s_j^{\pm 1};\q)_\infty}{(\t s_i^{\pm 1}s_j^{\pm 1};\q)_\infty}.
\end{align}

The norm is given by
\begin{equation}
\begin{aligned}
\frac{\langle P_\lambda, P_\lambda \rangle}{\langle 1, 1 \rangle}=\prod_{1\leq i<j\leq N} \frac{(\t^{j-i+1};\q)_{\lambda_i-\lambda_j}(\t^{j-i-1}\q;\q)_{\lambda_i-\lambda_j}}{(\t^{j-i};\q)_{\lambda_i-\lambda_j}(\t^{j-i}\q;\q)_{\lambda_i-\lambda_j}}\\
\times \frac{(\t^{2N-i-j+1};\q)_{\lambda_i+\lambda_j}(\t^{2N-i-j-1}\q;\q)_{\lambda_i+\lambda_j}}{(\t^{2N-i-j};\q)_{\lambda_i+\lambda_j}(\t^{2N-i-j}\q;\q)_{\lambda_i+\lambda_j}},
\end{aligned}
\end{equation}
and also we have
\begin{align}
\langle 1, 1 \rangle=\frac{(\t;\q)_\infty^N}{(\q;\q)_\infty^N} \prod_{k=1}^N \frac{(\t^{2k-1}\q;\q)_\infty}{(\t^{2k};\q)_\infty}.
\end{align}
This reproduces the $SO(2N)$ Neumann half-index (\ref{so2NN}), 
which agrees with the $SO(2N)$ Nahm pole half-index (\ref{so2NNahm}) under $t$ $\rightarrow$ $t^{-1}$. 

According to \cite{MR3856220}, 
for the type $D_N$ system, the character with one-column Young diagram is related to the Macdonald polynomials by
\begin{align}
\chi_{(1^r)}(s)&=E_r(s)\quad (1 \leq r \leq N-1),\qquad \chi_{(1^N)}(s)+\chi_{(1^{N-1}, -1)}(s)=E_N(s)\\
E_r(s)&=\sum_{j=0}^{\lfloor \frac{r}{2} \rfloor} \frac{(\q/\t; \t^2)_j (\t^{2N-2r+2j+2};\t^2)_j}{(\t^2; \t^2)_j (\q \t^{2N-2r+2j-1};\t^2)_j}\cdot \frac{1+\t^{N-r}}{1+\t^{N-r+2j}}\t^j P_{(1^{r-2j})}(s;\q, \t).
\end{align}

\paragraph{Fundamental representation.}
We assume $N \geq 2$. For the fundamental representation $\lambda=(1)$, we have
\begin{align}
\chi_{(1)}(s)=E_1(s)=P_{(1)}(s;\q,\t).
\end{align}
Hence
\begin{align}
\langle \chi_{(1)}, \chi_{(1)} \rangle=\langle P_{(1)}, P_{(1)} \rangle=\frac{(1-\q)(1-\q \t^{N-2})(1-\t^{N})(1-\t^{2N-2})}{(1-\t)(1-\t^{N-1})(1-\q \t^{N-1})(1-\q \t^{2N-3})}\langle 1, 1 \rangle,
\end{align}
and 
\begin{align}
\frac{\langle W_{\square}W_{\square} \rangle_{\mathcal{N}}^{SO(2N)}}{\mathbb{II}_{\mathcal{N}}^{SO(2N)}}&=\frac{(1-\q)(1-\q \t^{N-2})(1-\t^{N})(1-\t^{2N-2})}{(1-\t)(1-\t^{N-1})(1-\q \t^{N-1})(1-\q \t^{2N-3})}.
\end{align}
This gives rise to the exact expression for the matrix integral (\ref{so2NN_W1}) 
and agrees with the expression (\ref{so2NNahm_T1}) obtained from the Higgsing process under the transformation $t$ $\rightarrow$ $t^{-1}$. 

\paragraph{Antisymmetric representations.}
For the non-minuscule module, the Higgsing procedure does not simply work. 
Nevertheless, we can find the exact closed-form expressions for the Wilson line defect correlators 
by means of the inner product of the Macdonald polynomials. 
Let us consider the rank-$k$ antisymmetric Wilson lines. 
If $k=2m$ is even, the inner product $\langle \chi_{(1^{2m})}, 1 \rangle$ is non-vanishing. It is easy to see
\begin{align}
\langle \chi_{(1^{2m})}, 1 \rangle&=\langle E_{2m}, 1 \rangle\\
&=\frac{(\q/\t; \t^2)_{m} (\t^{2N-2m+2};\t^2)_{m}}{(\t^2; \t^2)_{m} (\q \t^{2N-2m-1};\t^2)_{m}} \frac{1+\t^{N-2m}}{1+\t^N}\t^m \langle 1,1 \rangle.
\end{align}
for $m=1,\dots, \lfloor (N-1)/2 \rfloor$.
Thus we obtain
\begin{align}
\frac{\langle W_{(1^{2m})} \rangle_{\mathcal{N}}^{SO(2N)}}{\mathbb{II}_{\mathcal{N}}^{SO(2N)}}=\frac{(\q/\t; \t^2)_{m} (\t^{2N-2m+2};\t^2)_{m}}{(\t^2; \t^2)_{m} (\q \t^{2N-2m-1};\t^2)_{m}} \frac{1+\t^{N-2m}}{1+\t^N}\t^m.
\end{align}
In particular, for $m=1$ and $N \geq 3$,  we have
\begin{align}
\frac{\langle W_{(1^2)} \rangle_{\mathcal{N}}^{SO(2N)}}{\mathbb{II}_{\mathcal{N}}^{SO(2N)}}
=\frac{(\t-\q)(1-\t^N)(1+\t^{N-2})}{(1-\t^2)(1-\q \t^{2N-3})}.
\end{align}

\paragraph{Spinor representations.}
Let us evaluate the two-point functions for the spinor Wilson lines.
Using \eqref{ch_so2N_sp} and \eqref{ch_so2N_antisp}, we find
\begin{align}
\chi_{\text{sp}}^2(s)&=\frac{1}{4} \overline{E}(s|1)+\frac{(-1)^N}{4}\overline{E}(s|{-1})+\frac{1}{2}\prod_{i=1}^N (s_i-s_i^{-1}),\\
\chi_{\overline{\text{sp}}}^2(s)&=\frac{1}{4} \overline{E}(s|1)+\frac{(-1)^N}{4}\overline{E}(s|{-1})-\frac{1}{2}\prod_{i=1}^N (s_i-s_i^{-1}),\\
\chi_{\text{sp}}(s)\chi_{\overline{\text{sp}}}(s)&=\frac{1}{4} \overline{E}(s|1)-\frac{(-1)^N}{4}\overline{E}(s|{-1}).
\end{align}
We insert these characters into the matrix integral. We easily notice that the last term in $\chi_{\text{sp}}^2(s)$ or $\chi_{\overline{\text{sp}}}^2(s)$ does not contribute to the matrix integral due to the symmetry of $s_i \leftrightarrow s_i^{-1}$. 

Let us consider the case of $N=2M$. In this case, since we have
\begin{align}
\frac{1}{4} \overline{E}(s|1)+\frac{(-1)^N}{4}\overline{E}(s|{-1})&=\sum_{m=0}^{M-1} E_{2m}(s)+\frac{1}{2}E_{2M}(s), \\
\frac{1}{4} \overline{E}(s|1)-\frac{(-1)^N}{4}\overline{E}(s|{-1})&=\sum_{m=0}^{M-1} E_{2m+1}(s),
\end{align}
we find
\begin{align}
\label{so2NevenN_Wsp_exact}
\frac{\langle W_{\text{sp}}W_{\text{sp}} \rangle_{\mathcal{N}}^{Spin(4M)}}{\mathbb{II}_{\mathcal{N}}^{SO(4M)}}
&=\frac{\langle W_{\overline{\text{sp}}}W_{\overline{\text{sp}}} \rangle_{\mathcal{N}}^{Spin(4M)}}{\mathbb{II}_{\mathcal{N}}^{SO(4M)}}
=\sum_{m=0}^{M-1} \frac{\langle E_{2m},1\rangle}{\langle 1,1 \rangle}+\frac{1}{2} \frac{\langle E_{2M},1\rangle}{\langle 1,1 \rangle}\notag \\
&=\sum_{m=0}^{M-1} \frac{(\q/\t; \t^2)_{m} (\t^{4M-2m+2};\t^2)_{m}}{(\t^2; \t^2)_{m} (\q \t^{4M-2m-1};\t^2)_{m}} \frac{1+\t^{2M-2m}}{1+\t^{2M}}\t^m\notag \\
&\quad+\frac{1}{2}\frac{(\q/\t; \t^2)_{M} (\t^{2M+1};\t^2)_{M}}{(\t^2; \t^2)_{M} (\q \t^{2M-1};\t^2)_{M}} \frac{1}{1+\t^{2M}}\t^M, \\
\langle W_{\text{sp}}W_{\overline{\text{sp}}} \rangle_{\mathcal{N}}^{Spin(4M)}&=0,
\end{align}
The expression (\ref{so2NevenN_Wsp_exact}) is the closed-form formula for the matrix integral (\ref{so2NevenN_Wsp}). 
We have confirmed that it also agrees with the expression (\ref{so2NNahm_T1/2}) 
evaluated from the Higgsing procedure under the transformation $t\rightarrow t^{-1}$. 

In the case of $N=2M+1$, we have
\begin{align}
\frac{1}{4} \overline{E}(s|1)+\frac{(-1)^N}{4}\overline{E}(s|{-1})&=\sum_{m=0}^{M-1} E_{2m+1}(s)+\frac{1}{2}E_{2M+1}(s), \\
\frac{1}{4} \overline{E}(s|1)-\frac{(-1)^N}{4}\overline{E}(s|{-1})&=\sum_{m=0}^{M} E_{2m}(s).
\end{align}
Therefore we find
\begin{align}
\langle W_{\text{sp}}W_{\text{sp}} \rangle_{\mathcal{N}}^{Spin(4M+2)}&=\langle W_{\overline{\text{sp}}}W_{\overline{\text{sp}}} \rangle_{\mathcal{N}}^{Spin(4M+2)}=0, \\
\label{so2NoddN_Wsp_exact}
\frac{\langle W_{\text{sp}}W_{\overline{\text{sp}}} \rangle_{\mathcal{N}}^{Spin(4M+2)}}{\mathbb{II}_{\mathcal{N}}^{SO(4M+2)}}
&=\sum_{m=0}^M \frac{\langle E_{2m}, 1\rangle}{\langle 1, 1 \rangle}\notag \\
&=\sum_{m=0}^M \frac{(\q/\t; \t^2)_{m} (\t^{4M-2m+4};\t^2)_{m}}{(\t^2; \t^2)_{m} (\q \t^{4M-2m+1};\t^2)_{m}}\frac{1+\t^{2M-2m+1}}{1+\t^{2M+1}}\t^m.
\end{align}
Again the expression (\ref{so2NoddN_Wsp_exact}) is the exact result for the matrix integral (\ref{so2NoddN_Wsp}). 
It coincides with the expression (\ref{so2NNahm_T1/2}) under the transformation $t\rightarrow t^{-1}$. 

\acknowledgments{
The authors would like to thank Yunfeng Jiang, Hai Lin and Masatoshi Noumi for useful discussions and comments. 
The work of Y.H. was supported in part by JSPS KAKENHI Grant Nos. 22K03641 and 23K25790.
The work of T.O. was supported by the Startup Funding no.\ 4007012317 of the Southeast University. 
}

\appendix

\section{Character formulas}
\label{app_ch}

\subsection{$\mathfrak{u}(N)$}
The character of the irreducible representation of $\mathfrak{u}(N)$ 
with the highest weight labeled by the Young diagram $\lambda$ is given by the Schur polynomial
\begin{align}
\chi_{\lambda}^{\mathfrak{u}(N)}
&=s_{\lambda}(s)
=\frac{\det s_j^{\lambda_i+N-i}}{\det s_j^{N-j}}. 
\end{align}
For the representation of $\mathfrak{su}(N)$ the product of the variables $s_i$ is required to be $1$. 

\subsection{$\mathfrak{so}(2N+1)$}
For $\mathfrak{so}(2N+1)$ the character of the irreducible representation 
with the highest weight labeled by $\lambda$ is
\begin{align}
\label{ch_so2N+1_irrep}
\chi_{\lambda}^{\mathfrak{so}(2N+1)}
&=\frac{\det (s_j^{\lambda_i+N-i+1/2} -s_{j}^{-(\lambda_i+N-i+1/2)})}
{\det(s_j^{N-i+1/2}-s_j^{-(N-i+1/2)})}. 
\end{align}
For the spinor representation of $\mathfrak{so}(2N+1)$ 
we have $\lambda=(1/2,\cdots,1/2)$ and 
\begin{align}
\label{ch_so2N+1_sp}
\chi_{\textrm{sp}}^{\mathfrak{so}(2N+1)}
&=\prod_{i=1}^{N}(s_i^{\frac12}+s_i^{-\frac12}). 
\end{align}

\subsection{$\mathfrak{usp}(2N)$}
The character of the irreducible representation of $\mathfrak{usp}(2N)$ 
with the highest weight $\lambda$ is given by 
\begin{align}
\label{ch_usp2N_irrep}
\chi_{\tiny \yng(1)}^{\mathfrak{usp}(2N)}
&=\frac{\det(s_j^{\lambda_i+N-i+1}-s_j^{-\lambda_i-N+i-1})}{\det(s_j^{N-i+1}-s_j^{-N+i-1})}. 
\end{align}

\subsection{$\mathfrak{so}(2N)$}

For $\mathfrak{so}(2N)$ 
the character of the irreducible representation with highest weight labeled by $\lambda=(\lambda_1,\dots, \lambda_N)$ ($\lambda_1 \geq \dots \geq \lambda_{N-1} \geq |\lambda_N| \geq 0$) is \cite{MR1153249}
\begin{align}
\label{ch_so2N_irrep}
\chi_{\lambda}^{\mathfrak{so}(2N)}
&=\frac{\det (s_j^{\lambda_i+N-i}+s_j^{-\lambda_i-N+i})+\det(s_j^{\lambda_i+N-i}-s_j^{-\lambda_i-N+i})
}{\det(s_j^{N-i}+s_j^{-N+i})}. 
\end{align}
The chiral and antichiral spinor representations correspond 
to $\lambda=\text{sp}=(1/2,\dots,1/2, 1/2)$ and $\lambda=\overline{\text{sp}}=(1/2,\dots,1/2, -1/2)$, respectively. 
We have 
\begin{align}
\label{ch_so2N_sp}
\chi_{\textrm{sp}}^{\mathfrak{so}(2N)}
&=\frac12 \left[
\prod_{i=1}^N (s_i^{\frac12}+s_i^{-\frac12})
+ \prod_{i=1}^N (s_i^{\frac12}-s_i^{-\frac12})
\right], 
\end{align}
and 
\begin{align}
\label{ch_so2N_antisp}
\chi_{\overline{\textrm{sp}}}^{\mathfrak{so}(2N)}
&=\frac12 \left[
\prod_{i=1}^N (s_i^{\frac12}+s_i^{-\frac12})
-\prod_{i=1}^N (s_i^{\frac12}-s_i^{-\frac12})
\right]. 
\end{align}
For the disconnected component of orthogonal gauge theory, 
we instead employ the character (\ref{ch_usp2N_irrep}) of the $\mathfrak{usp}(2N-2)$.

\bibliographystyle{utphys}
\bibliography{ref}

\def\polhk#1{\setbox0=\hbox{#1}{\ooalign{\hidewidth
  \lower1.5ex\hbox{`}\hidewidth\crcr\unhbox0}}} \def\cprime{$'$}
\providecommand{\href}[2]{#2}\begingroup\raggedright\begin{thebibliography}{10}

\bibitem{Dimofte:2011py}
T.~Dimofte, D.~Gaiotto, and S.~Gukov, ``{3-Manifolds and 3d Indices},''
  \href{http://dx.doi.org/10.4310/ATMP.2013.v17.n5.a3}{{\em Adv. Theor. Math.
  Phys.} {\bfseries 17} no.~5, (2013) 975--1076},
\href{http://arxiv.org/abs/1112.5179}{{\ttfamily arXiv:1112.5179 [hep-th]}}.

\bibitem{Gang:2012yr}
D.~Gang, E.~Koh, and K.~Lee, ``{Line Operator Index on $S^{1}\times S^{3}$},''
  \href{http://dx.doi.org/10.1007/JHEP05(2012)007}{{\em JHEP} {\bfseries 05}
  (2012) 007},
\href{http://arxiv.org/abs/1201.5539}{{\ttfamily arXiv:1201.5539 [hep-th]}}.

\bibitem{Gaiotto:2008sa}
D.~Gaiotto and E.~Witten, ``{Supersymmetric Boundary Conditions in N=4 Super
  Yang-Mills Theory},'' \href{http://dx.doi.org/10.1007/s10955-009-9687-3}{{\em
  J. Statist. Phys.} {\bfseries 135} (2009) 789--855},
\href{http://arxiv.org/abs/0804.2902}{{\ttfamily arXiv:0804.2902 [hep-th]}}.

\bibitem{Wilson:1974sk}
K.~G. Wilson, ``{Confinement of Quarks},''
  \href{http://dx.doi.org/10.1103/PhysRevD.10.2445}{{\em Phys. Rev. D}
  {\bfseries 10} (1974) 2445--2459}.

\bibitem{Maldacena:1998im}
J.~M. Maldacena, ``{Wilson loops in large N field theories},''
  \href{http://dx.doi.org/10.1103/PhysRevLett.80.4859}{{\em Phys. Rev. Lett.}
  {\bfseries 80} (1998) 4859--4862},
  \href{http://arxiv.org/abs/hep-th/9803002}{{\ttfamily arXiv:hep-th/9803002}}.

\bibitem{Rey:1998ik}
S.-J. Rey and J.-T. Yee, ``{Macroscopic strings as heavy quarks in large N
  gauge theory and anti-de Sitter supergravity},''
  \href{http://dx.doi.org/10.1007/s100520100799}{{\em Eur. Phys. J. C}
  {\bfseries 22} (2001) 379--394},
  \href{http://arxiv.org/abs/hep-th/9803001}{{\ttfamily arXiv:hep-th/9803001}}.

\bibitem{Nahm:1979yw}
W.~Nahm, ``{A Simple Formalism for the BPS Monopole},''
  \href{http://dx.doi.org/10.1016/0370-2693(80)90961-2}{{\em Phys. Lett. B}
  {\bfseries 90} (1980) 413--414}.

\bibitem{Diaconescu:1996rk}
D.-E. Diaconescu, ``{D-branes, monopoles and Nahm equations},''
  \href{http://dx.doi.org/10.1016/S0550-3213(97)00438-0}{{\em Nucl. Phys. B}
  {\bfseries 503} (1997) 220--238},
  \href{http://arxiv.org/abs/hep-th/9608163}{{\ttfamily arXiv:hep-th/9608163}}.

\bibitem{tHooft:1977nqb}
G.~'t~Hooft, ``{On the Phase Transition Towards Permanent Quark Confinement},''
  \href{http://dx.doi.org/10.1016/0550-3213(78)90153-0}{{\em Nucl. Phys. B}
  {\bfseries 138} (1978) 1--25}.

\bibitem{Kapustin:2005py}
A.~Kapustin, ``{Wilson-'t Hooft operators in four-dimensional gauge theories
  and S-duality},'' \href{http://dx.doi.org/10.1103/PhysRevD.74.025005}{{\em
  Phys. Rev. D} {\bfseries 74} (2006) 025005},
  \href{http://arxiv.org/abs/hep-th/0501015}{{\ttfamily arXiv:hep-th/0501015}}.

\bibitem{Kapustin:2006pk}
A.~Kapustin and E.~Witten, ``{Electric-Magnetic Duality And The Geometric
  Langlands Program},''
  \href{http://dx.doi.org/10.4310/CNTP.2007.v1.n1.a1}{{\em Commun. Num. Theor.
  Phys.} {\bfseries 1} (2007) 1--236},
  \href{http://arxiv.org/abs/hep-th/0604151}{{\ttfamily arXiv:hep-th/0604151}}.

\bibitem{Gaiotto:2011nm}
D.~Gaiotto and E.~Witten, ``{Knot Invariants from Four-Dimensional Gauge
  Theory},'' \href{http://dx.doi.org/10.4310/ATMP.2012.v16.n3.a5}{{\em Adv.
  Theor. Math. Phys.} {\bfseries 16} no.~3, (2012) 935--1086},
  \href{http://arxiv.org/abs/1106.4789}{{\ttfamily arXiv:1106.4789 [hep-th]}}.

\bibitem{Witten:2011zz}
E.~Witten, ``{Fivebranes and Knots},''
\href{http://arxiv.org/abs/1101.3216}{{\ttfamily arXiv:1101.3216 [hep-th]}}.

\bibitem{Gaiotto:2008ak}
D.~Gaiotto and E.~Witten, ``{S-Duality of Boundary Conditions In N=4 Super
  Yang-Mills Theory},''
  \href{http://dx.doi.org/10.4310/ATMP.2009.v13.n3.a5}{{\em Adv. Theor. Math.
  Phys.} {\bfseries 13} no.~3, (2009) 721--896},
\href{http://arxiv.org/abs/0807.3720}{{\ttfamily arXiv:0807.3720 [hep-th]}}.

\bibitem{Gomis:2009ir}
J.~Gomis, T.~Okuda, and D.~Trancanelli, ``{Quantum 't Hooft operators and
  S-duality in N=4 super Yang-Mills},''
  \href{http://dx.doi.org/10.4310/ATMP.2009.v13.n6.a9}{{\em Adv. Theor. Math.
  Phys.} {\bfseries 13} no.~6, (2009) 1941--1981},
  \href{http://arxiv.org/abs/0904.4486}{{\ttfamily arXiv:0904.4486 [hep-th]}}.

\bibitem{Aharony:2013hda}
O.~Aharony, N.~Seiberg, and Y.~Tachikawa, ``{Reading between the lines of
  four-dimensional gauge theories},''
  \href{http://dx.doi.org/10.1007/JHEP08(2013)115}{{\em JHEP} {\bfseries 08}
  (2013) 115}, \href{http://arxiv.org/abs/1305.0318}{{\ttfamily arXiv:1305.0318
  [hep-th]}}.

\bibitem{Gaiotto:2019jvo}
D.~Gaiotto and T.~Okazaki, ``{Dualities of Corner Configurations and
  Supersymmetric Indices},''
  \href{http://dx.doi.org/10.1007/JHEP11(2019)056}{{\em JHEP} {\bfseries 11}
  (2019) 056},
\href{http://arxiv.org/abs/1902.05175}{{\ttfamily arXiv:1902.05175 [hep-th]}}.

\bibitem{Gaiotto:2012xa}
D.~Gaiotto, L.~Rastelli, and S.~S. Razamat, ``{Bootstrapping the superconformal
  index with surface defects},''
  \href{http://dx.doi.org/10.1007/JHEP01(2013)022}{{\em JHEP} {\bfseries 01}
  (2013) 022}, \href{http://arxiv.org/abs/1207.3577}{{\ttfamily arXiv:1207.3577
  [hep-th]}}.

\bibitem{Okazaki:2019ony}
T.~Okazaki, ``{Mirror symmetry of 3D $\mathcal{N}=4$ gauge theories and
  supersymmetric indices},''
  \href{http://dx.doi.org/10.1103/PhysRevD.100.066031}{{\em Phys. Rev.}
  {\bfseries D100} no.~6, (2019) 066031},
\href{http://arxiv.org/abs/1905.04608}{{\ttfamily arXiv:1905.04608 [hep-th]}}.

\bibitem{MR2109105}
N.~Bourbaki, {\em Lie groups and {L}ie algebras. {C}hapters 7--9}.
\newblock Elements of Mathematics (Berlin). Springer-Verlag, Berlin, 2005.
\newblock Translated from the 1975 and 1982 French originals by Andrew
  Pressley.

\bibitem{Hatsuda:2024lcc}
Y.~Hatsuda, H.~Lin, and T.~Okazaki, ``{Orbifold ETW brane and half-indices},''
  \href{http://dx.doi.org/10.1007/JHEP12(2024)227}{{\em JHEP} {\bfseries 12}
  (2025) 227}, \href{http://arxiv.org/abs/2409.16841}{{\ttfamily
  arXiv:2409.16841 [hep-th]}}.

\bibitem{Hatsuda:2025jze}
Y.~Hatsuda, H.~Lin, and T.~Okazaki, ``{$\mathcal{N}=4$ line defect correlators
  of type BCD},'' \href{http://arxiv.org/abs/2502.18110}{{\ttfamily
  arXiv:2502.18110 [hep-th]}}.

\bibitem{MR1354144}
I.~G. Macdonald, {\em Symmetric functions and Hall polynomials}.
\newblock Oxford Mathematical Monographs. The Clarendon Press, Oxford
  University Press, New York, second~ed., 1995.
\newblock With contributions by A. Zelevinsky, Oxford Science Publications.

\bibitem{MR1976581}
I.~G. Macdonald, \href{http://dx.doi.org/10.1017/CBO9780511542824}{{\em Affine
  {H}ecke algebras and orthogonal polynomials}}, vol.~157 of {\em Cambridge
  Tracts in Mathematics}.
\newblock Cambridge University Press, Cambridge, 2003.
\newblock \url{https://doi.org/10.1017/CBO9780511542824}.

\bibitem{MR1314036}
I.~Cherednik, ``Double affine {H}ecke algebras and {M}acdonald's conjectures,''
  \href{http://dx.doi.org/10.2307/2118632}{{\em Ann. of Math. (2)} {\bfseries
  141} no.~1, (1995) 191--216}. \url{https://doi.org/10.2307/2118632}.

\bibitem{MR1354956}
I.~Cherednik, ``Macdonald's evaluation conjectures and difference {F}ourier
  transform,'' \href{http://dx.doi.org/10.1007/BF01231441}{{\em Invent. Math.}
  {\bfseries 122} no.~1, (1995) 119--145}.
  \url{https://doi.org/10.1007/BF01231441}.

\bibitem{MR1185831}
I.~Cherednik, ``Double affine {H}ecke algebras, {K}nizhnik-{Z}amolodchikov
  equations, and {M}acdonald's operators,''
  \href{http://dx.doi.org/10.1155/S1073792892000199}{{\em Internat. Math. Res.
  Notices} no.~9, (1992) 171--180}.
  \url{https://doi.org/10.1155/S1073792892000199}.

\bibitem{MR1358032}
I.~Cherednik, ``Nonsymmetric {M}acdonald polynomials,''
  \href{http://dx.doi.org/10.1155/S1073792895000341}{{\em Internat. Math. Res.
  Notices} no.~10, (1995) 483--515}.
  \url{https://doi.org/10.1155/S1073792895000341}.

\bibitem{MR1613515}
I.~Cherednik, ``Intertwining operators of double affine {H}ecke algebras,''
  \href{http://dx.doi.org/10.1007/s000290050017}{{\em Selecta Math. (N.S.)}
  {\bfseries 3} no.~4, (1997) 459--495}.
  \url{https://doi.org/10.1007/s000290050017}.

\bibitem{MR1768938}
S.~Sahi, \href{http://dx.doi.org/10.1090/conm/254/03963}{``Some properties of
  {K}oornwinder polynomials,''} in {\em {$q$}-series from a contemporary
  perspective ({S}outh {H}adley, {MA}, 1998)}, vol.~254 of {\em Contemp.
  Math.}, pp.~395--411.
\newblock Amer. Math. Soc., Providence, RI, 2000.
\newblock \url{https://doi.org/10.1090/conm/254/03963}.

\bibitem{MR1715325}
S.~Sahi, ``Nonsymmetric {K}oornwinder polynomials and duality,''
  \href{http://dx.doi.org/10.2307/121102}{{\em Ann. of Math. (2)} {\bfseries
  150} no.~1, (1999) 267--282}. \url{https://doi.org/10.2307/121102}.

\bibitem{MR1792347}
J.~V. Stokman, ``Koornwinder polynomials and affine {H}ecke algebras,''
  \href{http://dx.doi.org/10.1155/S1073792800000520}{{\em Internat. Math. Res.
  Notices} no.~19, (2000) 1005--1042}.
  \url{https://doi.org/10.1155/S1073792800000520}.

\bibitem{MR1411136}
J.~F. van Diejen, ``Self-dual {K}oornwinder-{M}acdonald polynomials,''
  \href{http://dx.doi.org/10.1007/s002220050102}{{\em Invent. Math.} {\bfseries
  126} no.~2, (1996) 319--339}. \url{https://doi.org/10.1007/s002220050102}.

\bibitem{Ito:2011ea}
Y.~Ito, T.~Okuda, and M.~Taki, ``{Line operators on $S^1 \times \mathbb {R}^3$
  and quantization of the Hitchin moduli space},''
  \href{http://dx.doi.org/10.1007/JHEP03(2016)085}{{\em JHEP} {\bfseries 04}
  (2012) 010}, \href{http://arxiv.org/abs/1111.4221}{{\ttfamily arXiv:1111.4221
  [hep-th]}}. [Erratum: JHEP 03, 085 (2016)].

\bibitem{Brennan:2018yuj}
T.~D. Brennan, A.~Dey, and G.~W. Moore, ``{On \textquoteright{}t Hooft defects,
  monopole bubbling and supersymmetric quantum mechanics},''
  \href{http://dx.doi.org/10.1007/JHEP09(2018)014}{{\em JHEP} {\bfseries 09}
  (2018) 014}, \href{http://arxiv.org/abs/1801.01986}{{\ttfamily
  arXiv:1801.01986 [hep-th]}}.

\bibitem{Brennan:2018rcn}
T.~D. Brennan, A.~Dey, and G.~W. Moore, ``{\textquoteright{}t Hooft defects and
  wall crossing in SQM},''
  \href{http://dx.doi.org/10.1007/JHEP10(2019)173}{{\em JHEP} {\bfseries 10}
  (2019) 173}, \href{http://arxiv.org/abs/1810.07191}{{\ttfamily
  arXiv:1810.07191 [hep-th]}}.

\bibitem{Hayashi:2019rpw}
H.~Hayashi, T.~Okuda, and Y.~Yoshida, ``{Wall-crossing and operator ordering
  for 't Hooft operators in $\mathcal{N} $ = 2 gauge theories},''
  \href{http://dx.doi.org/10.1007/JHEP11(2019)116}{{\em JHEP} {\bfseries 11}
  (2019) 116}, \href{http://arxiv.org/abs/1905.11305}{{\ttfamily
  arXiv:1905.11305 [hep-th]}}.

\bibitem{Hayashi:2020ofu}
H.~Hayashi, T.~Okuda, and Y.~Yoshida, ``{ABCD of \textquoteright{}t Hooft
  operators},'' \href{http://dx.doi.org/10.1007/JHEP04(2021)241}{{\em JHEP}
  {\bfseries 04} (2021) 241}, \href{http://arxiv.org/abs/2012.12275}{{\ttfamily
  arXiv:2012.12275 [hep-th]}}.

\bibitem{Nagasaki:2011ue}
K.~Nagasaki, H.~Tanida, and S.~Yamaguchi, ``{Holographic Interface-Particle
  Potential},'' \href{http://dx.doi.org/10.1007/JHEP01(2012)139}{{\em JHEP}
  {\bfseries 01} (2012) 139}, \href{http://arxiv.org/abs/1109.1927}{{\ttfamily
  arXiv:1109.1927 [hep-th]}}.

\bibitem{Estes:2012nx}
J.~Estes, A.~O'Bannon, E.~Tsatis, and T.~Wrase, ``{Holographic Wilson Loops,
  Dielectric Interfaces, and Topological Insulators},''
  \href{http://dx.doi.org/10.1103/PhysRevD.87.106005}{{\em Phys. Rev. D}
  {\bfseries 87} no.~10, (2013) 106005},
  \href{http://arxiv.org/abs/1210.0534}{{\ttfamily arXiv:1210.0534 [hep-th]}}.

\bibitem{Nagasaki:2013hwa}
K.~Nagasaki and S.~Yamaguchi, ``{\textquoteright{}t Hooft operators on an
  interface and bubbling D5-branes},''
  \href{http://dx.doi.org/10.1103/PhysRevD.89.046002}{{\em Phys. Rev. D}
  {\bfseries 89} no.~4, (2014) 046002},
  \href{http://arxiv.org/abs/1309.3125}{{\ttfamily arXiv:1309.3125 [hep-th]}}.

\bibitem{deLeeuw:2016vgp}
M.~de~Leeuw, A.~C. Ipsen, C.~Kristjansen, and M.~Wilhelm, ``{One-loop Wilson
  loops and the particle-interface potential in AdS/dCFT},''
  \href{http://dx.doi.org/10.1016/j.physletb.2017.02.047}{{\em Phys. Lett. B}
  {\bfseries 768} (2017) 192--197},
  \href{http://arxiv.org/abs/1608.04754}{{\ttfamily arXiv:1608.04754
  [hep-th]}}.

\bibitem{Aguilera-Damia:2016bqv}
J.~Aguilera-Damia, D.~H. Correa, and V.~I. Giraldo-Rivera, ``{Circular Wilson
  loops in defect Conformal Field Theory},''
  \href{http://dx.doi.org/10.1007/JHEP03(2017)023}{{\em JHEP} {\bfseries 03}
  (2017) 023}, \href{http://arxiv.org/abs/1612.07991}{{\ttfamily
  arXiv:1612.07991 [hep-th]}}.

\bibitem{Coccia:2021lpp}
L.~Coccia and C.~F. Uhlemann, ``{Mapping out the internal space in AdS/BCFT
  with Wilson loops},'' \href{http://dx.doi.org/10.1007/JHEP03(2022)127}{{\em
  JHEP} {\bfseries 03} (2022) 127},
  \href{http://arxiv.org/abs/2112.14648}{{\ttfamily arXiv:2112.14648
  [hep-th]}}.

\bibitem{Karch:2022rvr}
A.~Karch, H.~Sun, and C.~F. Uhlemann, ``{Double holography in string theory},''
  \href{http://dx.doi.org/10.1007/JHEP10(2022)012}{{\em JHEP} {\bfseries 10}
  (2022) 012}, \href{http://arxiv.org/abs/2206.11292}{{\ttfamily
  arXiv:2206.11292 [hep-th]}}.

\bibitem{Bergman:2022otk}
O.~Bergman and S.~Hirano, ``{The holography of duality in $ \mathcal{N} $ = 4
  Super-Yang-Mills theory},''
  \href{http://dx.doi.org/10.1007/JHEP11(2022)069}{{\em JHEP} {\bfseries 11}
  (2022) 069}, \href{http://arxiv.org/abs/2208.09396}{{\ttfamily
  arXiv:2208.09396 [hep-th]}}.

\bibitem{Drukker:2015spa}
N.~Drukker, ``{The $ \mathcal{N}=4 $ Schur index with Polyakov loops},''
  \href{http://dx.doi.org/10.1007/JHEP12(2015)012}{{\em JHEP} {\bfseries 12}
  (2015) 012}, \href{http://arxiv.org/abs/1510.02480}{{\ttfamily
  arXiv:1510.02480 [hep-th]}}.

\bibitem{Hatsuda:2023iwi}
Y.~Hatsuda and T.~Okazaki, ``{Exact $ \mathcal{N} $ = 2$^{*}$ Schur line defect
  correlators},'' \href{http://dx.doi.org/10.1007/JHEP06(2023)169}{{\em JHEP}
  {\bfseries 06} (2023) 169}, \href{http://arxiv.org/abs/2303.14887}{{\ttfamily
  arXiv:2303.14887 [hep-th]}}.

\bibitem{Hatsuda:2023imp}
Y.~Hatsuda and T.~Okazaki, ``{Large N and large representations of Schur line
  defect correlators},'' \href{http://dx.doi.org/10.1007/JHEP01(2024)096}{{\em
  JHEP} {\bfseries 01} (2024) 096},
  \href{http://arxiv.org/abs/2309.11712}{{\ttfamily arXiv:2309.11712
  [hep-th]}}.

\bibitem{Hatsuda:2023iof}
Y.~Hatsuda and T.~Okazaki, ``{Excitations of bubbling geometries for line
  defects},'' \href{http://dx.doi.org/10.1103/PhysRevD.109.066013}{{\em Phys.
  Rev. D} {\bfseries 109} no.~6, (2024) 066013},
  \href{http://arxiv.org/abs/2311.13740}{{\ttfamily arXiv:2311.13740
  [hep-th]}}.

\bibitem{Imamura:2024lkw}
Y.~Imamura, ``{Giant Graviton Expansions for the Line Operator Index},''
  \href{http://dx.doi.org/10.1093/ptep/ptae084}{{\em PTEP} {\bfseries 2024}
  no.~6, (2024) 063B03}, \href{http://arxiv.org/abs/2403.11543}{{\ttfamily
  arXiv:2403.11543 [hep-th]}}.

\bibitem{Imamura:2024pgp}
Y.~Imamura and M.~Inoue, ``{Brane expansions for anti-symmetric line operator
  index},'' \href{http://dx.doi.org/10.1007/JHEP08(2024)020}{{\em JHEP}
  {\bfseries 08} (2024) 020}, \href{http://arxiv.org/abs/2404.08302}{{\ttfamily
  arXiv:2404.08302 [hep-th]}}.

\bibitem{Beccaria:2024oif}
M.~Beccaria, ``{Schur line defect correlators and giant graviton expansion},''
  \href{http://dx.doi.org/10.1007/JHEP06(2024)088}{{\em JHEP} {\bfseries 06}
  (2024) 088}, \href{http://arxiv.org/abs/2403.14553}{{\ttfamily
  arXiv:2403.14553 [hep-th]}}.

\bibitem{Beccaria:2024dxi}
M.~Beccaria, ``{Leading large N giant graviton correction to Schur correlators
  in large representations},''
  \href{http://dx.doi.org/10.1016/j.nuclphysb.2024.116638}{{\em Nucl. Phys. B}
  {\bfseries 1006} (2024) 116638},
  \href{http://arxiv.org/abs/2404.12690}{{\ttfamily arXiv:2404.12690
  [hep-th]}}.

\bibitem{Hatsuda:2024uwt}
Y.~Hatsuda, H.~Lin, and T.~Okazaki, ``{Giant graviton expansions and ETW
  brane},'' \href{http://dx.doi.org/10.1007/JHEP09(2024)181}{{\em JHEP}
  {\bfseries 09} (2024) 181}, \href{http://arxiv.org/abs/2405.14564}{{\ttfamily
  arXiv:2405.14564 [hep-th]}}.

\bibitem{Imamura:2024zvw}
Y.~Imamura, A.~Sei, and D.~Yokoyama, ``{Giant graviton expansion for general
  Wilson line operator indices},''
  \href{http://dx.doi.org/10.1007/JHEP09(2024)202}{{\em JHEP} {\bfseries 09}
  (2024) 202}, \href{http://arxiv.org/abs/2406.19777}{{\ttfamily
  arXiv:2406.19777 [hep-th]}}.

\bibitem{Beccaria:2024lbt}
M.~Beccaria, ``{$\mathcal N=4$ SYM line defect Schur index and semiclassical
  string},'' \href{http://arxiv.org/abs/2407.06900}{{\ttfamily arXiv:2407.06900
  [hep-th]}}.

\bibitem{Okazaki:2023kpq}
T.~Okazaki and D.~J. Smith, ``{3d exceptional gauge theories and boundary
  confinement},'' \href{http://dx.doi.org/10.1007/JHEP11(2023)199}{{\em JHEP}
  {\bfseries 11} (2023) 199}, \href{http://arxiv.org/abs/2308.14428}{{\ttfamily
  arXiv:2308.14428 [hep-th]}}.

\bibitem{MR49882}
N.~Jacobson, ``Completely reducible {L}ie algebras of linear transformations,''
  \href{http://dx.doi.org/10.2307/2032629}{{\em Proc. Amer. Math. Soc.}
  {\bfseries 2} (1951) 105--113}. \url{https://doi.org/10.2307/2032629}.

\bibitem{MR125180}
V.~V. Morozov, ``A theorem on the nilpotent element in a semi-simple {L}ie
  algebra,'' {\em Uspehi Mat. Nauk} {\bfseries 15} no.~6(96), (1960) 137--139.

\bibitem{MR114875}
B.~Kostant, ``The principal three-dimensional subgroup and the {B}etti numbers
  of a complex simple {L}ie group,''
  \href{http://dx.doi.org/10.2307/2372999}{{\em Amer. J. Math.} {\bfseries 81}
  (1959) 973--1032}. \url{https://doi.org/10.2307/2372999}.

\bibitem{Montonen:1977sn}
C.~Montonen and D.~I. Olive, ``{Magnetic Monopoles as Gauge Particles?},''
\href{http://dx.doi.org/10.1016/0370-2693(77)90076-4}{{\em Phys. Lett.}
  {\bfseries 72B} (1977) 117--120}.

\bibitem{Assel:2015oxa}
B.~Assel and J.~Gomis, ``{Mirror Symmetry And Loop Operators},''
  \href{http://dx.doi.org/10.1007/JHEP11(2015)055}{{\em JHEP} {\bfseries 11}
  (2015) 055},
\href{http://arxiv.org/abs/1506.01718}{{\ttfamily arXiv:1506.01718 [hep-th]}}.

\bibitem{Hanany:1996ie}
A.~Hanany and E.~Witten, ``{Type IIB superstrings, BPS monopoles, and
  three-dimensional gauge dynamics},''
  \href{http://dx.doi.org/10.1016/S0550-3213(97)00157-0}{{\em Nucl.Phys.}
  {\bfseries B492} (1997) 152--190},
\href{http://arxiv.org/abs/hep-th/9611230}{{\ttfamily arXiv:hep-th/9611230
  [hep-th]}}.

\bibitem{Yamaguchi:2006tq}
S.~Yamaguchi, ``{Wilson loops of anti-symmetric representation and
  D5-branes},'' \href{http://dx.doi.org/10.1088/1126-6708/2006/05/037}{{\em
  JHEP} {\bfseries 05} (2006) 037},
  \href{http://arxiv.org/abs/hep-th/0603208}{{\ttfamily arXiv:hep-th/0603208}}.

\bibitem{Witten:1998xy}
E.~Witten, ``{Baryons and branes in anti-de Sitter space},''
  \href{http://dx.doi.org/10.1088/1126-6708/1998/07/006}{{\em JHEP} {\bfseries
  07} (1998) 006}, \href{http://arxiv.org/abs/hep-th/9805112}{{\ttfamily
  arXiv:hep-th/9805112}}.

\bibitem{Cordova:2016uwk}
C.~Cordova, D.~Gaiotto, and S.-H. Shao, ``{Infrared Computations of Defect
  Schur Indices},'' \href{http://dx.doi.org/10.1007/JHEP11(2016)106}{{\em JHEP}
  {\bfseries 11} (2016) 106}, \href{http://arxiv.org/abs/1606.08429}{{\ttfamily
  arXiv:1606.08429 [hep-th]}}.

\bibitem{MR1153249}
W.~Fulton and J.~Harris,
  \href{http://dx.doi.org/10.1007/978-1-4612-0979-9}{{\em Representation
  theory}}, vol.~129 of {\em Graduate Texts in Mathematics}.
\newblock Springer-Verlag, New York, 1991.
\newblock \url{https://doi.org/10.1007/978-1-4612-0979-9}.
\newblock A first course, Readings in Mathematics.

\bibitem{Hatsuda:2025mvj}
Y.~Hatsuda, ``{Deformed Schur indices and Macdonald polynomials},''
  \href{http://arxiv.org/abs/2503.03952}{{\ttfamily arXiv:2503.03952
  [hep-th]}}.

\bibitem{euler1805disquitiones}
L.~Euler, ``Disquitiones analyticae super evolutione potestatis trinomialis
  $(1+ x+ xx)^n$,'' {\em Novi Comm. Acad. Sci. Petropolitanae} no.~722, (1805)
  75--110.

\bibitem{MR1634067}
G.~E. Andrews, {\em The theory of partitions}.
\newblock Cambridge Mathematical Library. Cambridge University Press,
  Cambridge, 1998.
\newblock Reprint of the 1976 original.

\bibitem{MR4139057}
A.~Hoshino and J.~Shiraishi, ``Branching rules for {K}oornwinder polynomials
  with one column diagrams and matrix inversions,''
  \href{http://dx.doi.org/10.3842/SIGMA.2020.084}{{\em SIGMA Symmetry
  Integrability Geom. Methods Appl.} {\bfseries 16} (2020) Paper No. 084, 28}.
  \url{https://doi.org/10.3842/SIGMA.2020.084}.

\bibitem{MR3856220}
A.~Hoshino and J.~Shiraishi, ``Macdonald polynomials of type {$C_n$} with
  one-column diagrams and deformed {C}atalan numbers,''
  \href{http://dx.doi.org/10.3842/SIGMA.2018.101}{{\em SIGMA Symmetry
  Integrability Geom. Methods Appl.} {\bfseries 14} (2018) Paper No. 101, 33}.
  \url{https://doi.org/10.3842/SIGMA.2018.101}.

\end{thebibliography}\endgroup

\end{document}